\documentclass[a4paper,fleqn,usenatbib]{mnras}

\usepackage{newtxtext,newtxmath}

\usepackage[T1]{fontenc}
\usepackage{ae,aecompl}

\usepackage{graphicx}	
\usepackage{amsmath}	
\usepackage{amssymb}	

\newcommand{\kms}{\,km\,s$^{-1}$} 

\title[Oscillating red giants in eclipsing binaries]{Oscillating red giants in eclipsing binary systems: empirical reference value for asteroseismic scaling relation}

\author[N. Theme\ss l et al.]{
N. Theme\ss l$^{1,2,3}$,\thanks{E-mail: themessl@mps.mpg.de}
S. Hekker$^{1,2}$,
J. Southworth$^{4}$,
P. G. Beck$^{5,6,7}$,
K. Pavlovski$^{8}$,
\newauthor
A. Tkachenko$^{9}$,
G. C. Angelou$^{10,1,2}$,
W. H. Ball$^{11,2}$,
C. Barban$^{12}$,
E. Corsaro$^{13,5,6,7}$,
\newauthor
Y. Elsworth$^{11,2}$,
R. Handberg$^{2}$
and T. Kallinger$^{14}$
\\
$^{1}$Max Planck Institute for Solar System Research, Justus-von-Liebig-Weg 3, 37077 G{\"o}ttingen, DE\\
$^{2}$Stellar Astrophysics Centre, Department of Physics and Astronomy, Aarhus University, Ny Munkegade 120, 8000 Aarhus C, DK\\
$^{3}$Institute for Astrophysics, Georg-August University G{\"o}ttingen, Friedrich-Hund-Platz 1, 37077 G{\"o}ttingen, DE\\
$^{4}$Astrophysics Group, Keele University, Staffordshire ST5 5BG, UK\\
$^{5}$Instituto de Astrof\'{i}sica de Canarias, E-38200 La Laguna, Tenerife, ES\\
$^{6}$Departamento de Astrof\'{i}sica, Universidad de La Laguna, 38205 La Laguna, Tenerife, ES\\
$^{7}$Laboratoire AIM Paris-Saclay, CEA/DRF --- CNRS --- Universit\'e Paris Diderot, IRFU/SAp Centre de Saclay, 91191 Gif-sur-Yvette Cedex, FR\\
$^{8}$Department of Physics, University of Zagreb, Bijeni\v{c}ka cesta 32, 10000 Zagreb, HR\\
$^{9}$Instituut voor Sterrenkunde, KU Leuven, Celestijnenlaan 200D, 3001 Leuven, BE\\
$^{10}$Max-Planck Institut f{\"u}r Astrophysik, Karl-Schwarzschild-Str. 1, 85741 Garching, DE\\
$^{11}$School of Physics and Astronomy, U. Birmingham, Edgbaston, Birmingham B15 2TT, UK\\
$^{12}$LESIA, PSL Research University, CNRS, Univ. Pierre et Marie Curie, Univ. Denis Diderot, Observatoire de Paris, Meudon Cedex, FR\\
$^{13}$INAF -- Osservatorio Astrofisico di Catania, Via S. Sofia 78, 95123 Catania, IT\\
$^{14}$Institute for Astrophysics, University of Vienna, T{\"u}rkenschanzstra\ss e 17, 1180 Vienna, AT\\
}

\date{Accepted XXX. Received YYY; in original form ZZZ}

\pubyear{2018}

\begin{document}
\label{firstpage}
\pagerange{\pageref{firstpage}--\pageref{lastpage}}
\maketitle

\begin{abstract}
The internal structures and properties of oscillating red-giant stars can be accurately inferred through their global oscillation modes (asteroseismology). Based on 1460~days of {\it Kepler} observations we perform a thorough asteroseismic study to probe the stellar parameters and evolutionary stages of three red giants in eclipsing binary systems. We present the first detailed analysis of individual oscillation modes of the red-giant components of KIC\,8410637, KIC\,5640750 and KIC\,9540226. We obtain estimates of their asteroseismic masses, radii, mean densities and logarithmic surface gravities by using the asteroseismic scaling relations as well as grid-based modelling.
As these red giants are in double-lined eclipsing binaries, it is possible to derive their independent dynamical masses and radii from the orbital solution and compare it with the seismically inferred values.
For KIC\,5640750 we compute the first spectroscopic orbit based on both components of this system. We use high-resolution spectroscopic data and light curves of the three systems to determine up-to-date values of the dynamical stellar parameters. With our comprehensive set of stellar parameters we explore consistencies between binary analysis and asteroseismic methods, and test the reliability of the well-known scaling relations.
For the three red giants under study, we find agreement between dynamical and asteroseismic stellar parameters in cases where the asteroseismic methods account for metallicity, temperature and mass dependence as well as surface effects. We are able to attain agreement from the scaling laws in all three systems if we use $\Delta\nu_{\rm ref,emp} = 130.8 \pm 0.9 \,\mu$Hz instead of the usual solar reference value.
\end{abstract}




\section{Introduction}
\label{sec:intro}

Asteroseismology is the study of stellar oscillations with the aim of unravelling the structure and dynamics of stellar interiors. In-depth asteroseismic studies require either high-precision photometric time-series observations or time series of accurate radial velocity measurements (RVs). The former has been obtained by space missions such as MOST \citep[e.g.][]{2007bar,2008kal}, CoRoT \citep[e.g.][]{2007bag,2009der} and {\it Kepler} \citep[e.g.][]{2010bor}.
From 2009 to 2013, the nominal {\it Kepler} mission provided nearly continuous photometric time-series data for more than $100\,000$ stars.
These data are suitable for asteroseismic analyses and led to many discoveries in the field of red-giant seismology: determination of evolutionary stages \citep[e.g.][]{2011bed,2014mos,2017els}, rotation studies \citep[e.g.][]{2012bec,2012bmos}, stellar parameter determinations \citep{2010kal,2010hub,2013hek}, ensemble studies and galactic archaeology \citep[e.g.][]{2012cor,2013mig,2016cas}, amongst others. For recent overviews see \citet[][]{2013bhek} and \citet{2016hek}.

Pulsating red giants exhibit solar-like oscillations that are driven by the turbulent convection in the stellar envelope. The physical properties of red giants, such as mean density and surface gravity and thus stellar mass and radius, can be determined through the study of their oscillations. 
The most commonly used asteroseismic method is based on scaling relations \citep[e.g.][]{1986ulr,1991bro,1995kje} that use direct observables from the oscillation spectrum as input. These so-called global oscillation parameters can be measured in a large number of red giants for which high-precision photometric data are available. However, the asteroseismic scaling relations assume that all stars have an internal structure homologous to the Sun \citep[e.g.][]{2013bel}. 
Since evolved G and K giants span a wide range of masses, metallicities and evolutionary stages different than that of the Sun, the validity of these scaling relations, based on the principle of homology to the Sun, has to be tested. One possibility is to use eclipsing binary systems with a pulsating red-giant component. For double-lined eclipsing binaries, the stellar mass and radius of the red-giant component can be derived independently of asteroseismology through the binary orbit analysis using Kepler's laws. The binary analysis is limited to the cases in which the orbital parameters can be resolved and require spectra covering the full orbital period of the system.

So far a number of eclipsing binary systems with a red-giant component were detected in {\it Kepler} data \citep[e.g.][]{2010hek,2013gau}. The first such system, KIC\,8410637, was identified by \citet{2010hek}, who carried out a preliminary asteroseismic study based on a month long photometric time series of data in which only one eclipse was detected. The stellar parameters of the red-giant star could be measured from both the solar-like oscillations and from spectroscopy. A detailed comparison between the asteroseismic and dynamical stellar mass and radius of the red giant was performed by \citet{2013fra}, who found agreement between the binary and asteroseismic results within uncertainties. When \citet{2014hub} repeated the asteroseismic analysis of KIC\,8410637 with a longer {\it Kepler} dataset, he contested the findings of \citet{2013fra} and reported large discrepancies between the asteroseismic and dynamical stellar parameters. 

\citet{2014bec} carried out a seismic and binary analysis of 18 red-giant stars among which was KIC\,9540226. The red giant was not only found to be in an eccentric eclipsing binary, but also to exhibit an increase in flux during the actual periastron passage \citep{1995kum,2012rem}. These stars are colloquially referred to as ``heartbeat stars'' \citep{2012tho}. \citeauthor{2014bec} calculated the orbital parameters of the system from high-resolution spectroscopy and estimated the stellar parameters of the red giant from the asteroseismic scaling relations. In a more recent study, the mass and the radius of the red-giant component of KIC\,9540226 could be constrained from two consecutive binary analyses\footnote{Note that we only provide the updated dynamical values of \citet{2018bro} in Table~\ref{tab:kic_par} and Figure~\ref{fig:fig11}.} \citep[][]{2016bro,2018bro}. Moreover, \citet{2018bro} computed several estimates of its asteroseismic mass and radius based on different methodologies and by using the asteroseismic observables presented by \citet{2016gau}.

KIC\,8410637, KIC\,5640750 and KIC\,9540226 were also part of several ensemble studies\footnote{Here we only consider the updated values of \citet{2014gau} and not the results by \citet{2013gau}.} \citep[][hereafter G16]{2013gau, 2014gau,2016gau}. In these surveys, eclipse modelling and modelling of the radial velocities were used to derive the orbital and dynamical stellar parameters. In addition, masses and radii of the red-giant components were computed by using the asteroseismic scaling relations. In an extensive comparison between the results from detailed binary modelling and asteroseismology, they showed that the stellar masses and radii are systematically overestimated when the asteroseismic scaling relations are used.

In Table~\ref{tab:kic_par} we summarize the orbital and stellar parameters for the three red-giant stars (KIC\,8410637, KIC\,5640750 and KIC\,9540226) that are the subject of this study.

For a number of red-giant components in eclipsing binary systems it has been found that the dynamical and asteroseismic stellar parameters differ significantly. This leads us to investigate three such systems in detail, both from the binary point of view including a dedicated spectral disentangling analysis as well as by obtaining individual frequencies. In addition to the observational analysis, we use an asteroseismic grid-based approach to model the three red-giant components. KIC\,8410637, KIC\,5640750, and KIC\,9540226 belong to wide eclipsing binary systems where the components are not expected to be strongly influenced by tidal effects and/or mass transfer. All three systems were observed during the nominal four year long {\it Kepler} mission providing a large photometric dataset of unprecedented accuracy and supplemented with additional high-resolution spectra from ground-based observatories.
We analyze these spectroscopic and photometric data and derive up-to-date values of the stellar parameters from both the asteroseismic and orbital analysis. Since the stellar parameters determined using Kepler's laws are considered to be both accurate and precise, they provide a means to test the reliability of the asteroseismic mass and radius from the scaling laws.

For the current in-depth study we obtained orbital solutions and physical properties of three eclipsing binary systems from {\it Kepler} light curves and phase-resolved spectroscopy (Section~\ref{sec:spec}). In addition, we analyzed the Fourier spectra of the red-giant components in these systems to derive both global oscillation parameters as well as individual frequencies (Section~\ref{sec:osc}). We studied their asteroseismic stellar parameters and evolutionary states (Section~\ref{sec:stellarparameters}). 
In Section~\ref{sec:comp} we discuss and compare stellar parameters obtained from different asteroseismic methods and from the binary orbit. In the same section we provide an overview of tests that we performed to investigate the importance of different observables that are used for the determination of the asteroseismic stellar parameters and we present the conclusions of our study in Section~\ref{sec:con}.

\begin{table*}
\centering
\caption{Stellar and orbital parameters of the red giants studied here obtained from the literature (as per the rightmost column). The orbital periods ($P$) and eccentricities ($e$) are determined from orbital analysis of these binary systems. Stellar parameters ($M,R,\log g$) are based on either asteroseismic scaling relations or binary analysis. The latter are indicated with asterisks. Effective temperatures ($T_{\rm eff}$) and logarithmic surface gravities ($\log g$) are mostly derived from spectra within the study referred to. We indicate the cases where they were adopted from the original \citep[\textsuperscript{a},][]{2011bro} and revised \citep[\textsuperscript{b},][]{2014bhub} {\it Kepler} input catalogs (KIC).}
\label{tab:kic_par}
\begin{tabular}{lrllllrr}
\hline\hline
\multicolumn{1}{c}{{$P$} [days]} & \multicolumn{1}{c}{$e$} & \multicolumn{1}{c}{$R$ [$\rm R_\odot$]} & \multicolumn{1}{c}{$M$ [$\rm M_\odot$]} & \multicolumn{1}{c}{$\log g$} & \multicolumn{1}{c}{$T_{\rm eff}$ [K]} & \multicolumn{1}{c}{Evol. phase} & \multicolumn{1}{c}{Publication}\\ \hline
				KIC\,8410637 & & & & & & & \\ \hline
$>75$ &  & $11.80 \pm 0.60$ & $1.70 \pm 0.30$ & $2.700\pm 0.150$ & $4650 \pm 80$ &  & \citet{2010hek}\\
$408.32$ & $0.69$ & $10.74\pm 0.11$* & $1.56 \pm 0.03$* & $2.569 \pm 0.009$* & $4800 \pm 80$ & RC & \citet{2013fra}\\
&  & $11.58 \pm 0.30$ & $1.83\pm 0.14$ & $2.572 \pm 0.011$ & $4800 \pm 80$ &  & \citet{2014hub}\\
$408.32$& $0.69$  & $11.01 \pm 0.26$ & $1.61 \pm 0.11$ & $2.760\pm 0.400$\textsuperscript{b} & $4872\pm 139$\textsuperscript{b} & RC & \citet{2014gau}\\
&  & $11.20\pm 0.20$ & $1.70\pm 0.07$ & $2.569 \pm 0.005$ & $4800 \pm 100$ & & G16$_{\rm seis}$\\
&  &  $10.75\pm 0.20$  &  $1.51\pm 0.07$ & $2.555\pm 0.005$ & $4605 \pm 80$ & RGB & This work$_{\rm seis}$\\
408.32 & 0.686 &  $10.60\pm 0.05$*  &  $1.47\pm 0.02$* & $2.556\pm 0.003$* & $4605 \pm 80$ &  & This work$_{\rm dyn}$\\ \hline
				KIC\,5640750 & & & & & & & \\ \hline
$987.40$ & $0.32$ & $14.27 \pm 0.31$ & $1.45\pm 0.09$ & $2.561\pm 0.400$\textsuperscript{b} & $4727\pm 142$\textsuperscript{b} & RGB/AGB & \citet{2014gau}\\
&  & $13.08\pm 0.26$ & $1.15 \pm 0.06$ & $2.267\pm 0.005$ & $4525 \pm 75$ & RGB & This work$_{\rm seis}$\\
987.40& 0.323 & $13.12\pm 0.09$* & $1.16 \pm 0.01$* & $2.266\pm 0.006$* & $4525 \pm 75$ &  & This work$_{\rm dyn}$\\
				\hline
				KIC\,9540226 & & & & & & & \\ \hline
$175.43$ & $0.39$ & $14.10\pm 0.30$ & $1.60\pm 0.10$ & $2.370\pm 0.010$ & $4600\pm 150$\textsuperscript{a} & RGB & \citet{2014bec}\\
$175.46$ & $0.39$ & $14.01\pm 0.26$ & $1.59 \pm 0.08$ & $2.346\pm 0.030$\textsuperscript{b} & $4761\pm 143$\textsuperscript{b} & RGB & \citet{2014gau}\\
&  & $13.60\pm 0.20$ & $1.45\pm 0.05$ & $2.334\pm 0.004$ & $4692 \pm 65$ &  & G16$_{\rm seis}$\\
$175.44$ & $0.388$  & $12.80\pm 0.10$* & $1.33\pm 0.05$* & $2.349\pm 0.008$* & $4692\pm 65$ & & G16$_{\rm dyn}$\\
&  &  $13.06\pm 0.16$*  &  $1.38\pm 0.04$* & $2.345\pm 0.010$* & $4680\pm 80$ &   & \citet{2018bro}\\
&  &  $12.94\pm 0.25$  &  $1.26\pm 0.06$ & $2.314\pm 0.006$ & $4585\pm 75$ &  RGB & This work$_{\rm seis}$\\ 
175.44 & 0.388  &  $13.43\pm 0.17$*  &  $1.39\pm 0.03$* & $2.326\pm 0.010$* & $4585\pm 75$ &   & This work$_{\rm dyn}$\\ \hline
\end{tabular}
\end{table*}

\section{Physical properties of the systems from light curves and radial velocity time series}
\label{sec:spec}

\subsection{{\it Kepler} light curves and ground-based spectroscopic data}

For the eclipse modelling, we extracted the light curves of each eclipse from the {\it Kepler} datasets. In this case, we retained all data obtained within three eclipse durations of the eclipse. The data were then converted from flux to magnitude units and a low-order polynomial was fitted to normalise the out-of-transit data to zero relative magnitude. This step removes any slow trends due to instrumental effects and stellar activity. We tested the effects of different treatment of the light curve normalisation (e.g. polynomial order), and found that it does not have a significant impact on the best-fitting parameters.

By definition the primary eclipse is deeper than the secondary eclipse, and occurs when the hotter star is eclipsed by the cooler star. For all three objects, the dwarf star is smaller and hotter than the giant, so the primary eclipse is an occultation and the secondary eclipse is a transit. This also means that according to standard terminology \citep[e.g.][]{2001hil} the dwarf is the primary star and the giant is the secondary star. To avoid possible confusion, we instead refer to the stellar components as the ``dwarf'' (denoted as A) and ``giant'' (denoted as B).

Complementary to {\it Kepler} photometry we use spectroscopic data for the binary systems KIC\,8410637, KIC\,5640750 and KIC\,9540226, which were obtained with the \textsc{Hermes} spectrograph \citep{2011ras,2011bras} mounted on the 1.2m \textsc{Mercator} telescope in La Palma, Canary Islands, Spain. These spectra cover the wavelength range from $3\,750-9\,000\,\AA$ with a resolution of $R\simeq 85\,000$. Emission spectra of thorium-argon-neon reference lamps are provided in close proximity to each exposure to allow the most accurate wavelength calibration of the spectra possible. Some \textsc{Hermes} spectra for KIC\,8410637 and KIC\,9540226 were already used in previous studies by \citet{2013fra} and \citet{2014bec}. Observations were continued to extend the number of spectra and time base of the spectroscopic data. Moreover, the long-period system KIC\,5640750 has been monitored spectroscopically by members of our team since its discovery as a binary.

\subsection{Spectroscopic orbital elements from cross-correlation function and spectral disentangling}

\subsubsection{Cross-correlation function ({\sc ccf})}

For the three red giants under study we reanalyzed the archived \textsc{Hermes} data and obtained radial velocities by using the cross-correlation method \citep[e.g.][]{1979ton}. Based on this approach each wavelength-calibrated spectrum in the range from $4\,780-6\,530\,\AA$ was cross-correlated with a line mask optimized for \textsc{Hermes} spectra \citep{2011ras}. In this case a red-giant-star template was used that contains spectral lines corresponding to the spectrum of Arcturus. This method provides excellent precision for deriving the RVs of red-giant stars showing solar-like oscillations \citep{2014bec}. 
For KIC\,8410637 those RVs with large measurement uncertainties were not included in the further analysis. This leaves 43 RVs for the giant, with a root mean square ($rms$) scatter of 0.23{\,km\,s$^{-1}$} around the best fit, and 20 for the dwarf with a scatter of 0.92{\,km\,s$^{-1}$} (Table~\ref{tab:rv84}). 
In the case of KIC\,5640750 we only have RV data of the giant star (22 observations with a scatter of 0.08{\,km\,s$^{-1}$}, Table~\ref{tab:rv56}), since we were not able to detect the signature of the dwarf component with {\sc ccf}. As a further attempt to obtain its RVs we applied the least-squares deconvolution (LSD) method developed by \citet{2013tka}. This technique is similar to a cross-correlation with a set of $\delta$ functions. It is sensitive to small contributions and thus more suitable for the detection of faint components in double-lined spectroscopic binary systems. Although the overall signal-to-noise ratio (S/N) was high, the contribution from the dwarf star was very weak and therefore difficult to detect. With LSD we were not able to measure sufficiently precise RVs for the dwarf component that could be used to further constrain the orbital parameters for the system KIC\,5640750.
For KIC\,9540226 we derived 32 RVs for the giant with a scatter of 0.33{\,km\,s$^{-1}$} that we present in Table~\ref{tab:rv95}. These were supplemented by RV data for the dwarf star recently published by \citet{2016gau} (7 RVs with a scatter of 0.91{\,km\,s$^{-1}$}).

Based on the radial velocities determined for the stars in these binary systems we obtained orbital elements by using Kepler's laws. The lack of RVs for the dwarf star of KIC\,5640750 means we cannot measure the masses and radii of the component stars without additional constraints. As these parameters are important for our current study, we extended the spectroscopic analysis to detect the dwarf component of KIC\,5640750 by using spectral disentangling.

\newcommand{\HT}{\hspace*{-1em}}
\begin{table*}
\centering \caption{\label{tab:orbital_elements} Spectroscopic orbital elements for KIC\,8410637 (columns 2--3), KIC\,5640750 (columns 4--6) and KIC\,9540226 (columns 7--8) determined using cross-correlation ({\sc ccf}) and spectral disentangling ({\sc spd}). We adopted the solutions based on {\sc spd} in the further analysis of these eclipsing binary systems. See Section~\ref{sec:orbital} for parameter definitions. We note that $T_0$ is given in heliocentric julian date (HJD).}
\begin{tabular}{l r l r l r l r l r l r l r l} \hline\hline
Parameter &  \multicolumn{4}{c}{KIC\,8410637} &  \multicolumn{6}{c}{KIC\,5640750} & \multicolumn{4}{c}{KIC\,9540226} \\ \hline
& \multicolumn{2}{c}{{\sc ccf}} & \multicolumn{2}{c}{{\sc spd}} & \multicolumn{2}{c}{{\sc ccf}} & \multicolumn{2}{c}{{\sc spd 1}} & \multicolumn{2}{c}{{\sc spd 2}} & \multicolumn{2}{c}{{\sc ccf}} & \multicolumn{2}{c}{{\sc spd}} \\ \hline
 $P$ [d]		&		408.3248 & \HT $\pm$ 0.0004		& \multicolumn{2}{c}{-}		& 987.398 & \HT $\pm$ 0.006		& \multicolumn{2}{c}{-}	& \multicolumn{2}{c}{-}		&	175.4438 & \HT $\pm$ 0.0008		&  \multicolumn{2}{c}{-}   \\
 $T_0$  [d]		&  398.9449 & \HT $\pm$ 0.0007		&  403.53 & \HT $\pm$ 0.06		& 269.215 & \HT $\pm$ 0.004		& 188.7 & \HT $\pm$ 1.1	&	188.5 & \HT $\pm$ 1.1	& 817.289 & \HT $\pm$ 0.002		& 841.71 & \HT $\pm$ 0.08  \\
 $e$		&  0.686 & \HT $\pm$ 0.001		&  0.694 & \HT $\pm$ 0.004		& 0.326 & \HT $\pm$ 0.002		&  0.323 & \HT $\pm$ 0.008	& 0.322 & \HT $\pm$ 0.008		& 0.3877 & \HT $\pm$ 0.0003		&  0.387 & \HT $\pm$ 0.003 \\
$\omega$ [deg]		&  120.9 & \HT $\pm$ 0.1		&  120.7 & \HT $\pm$ 0.2		& 34.3 & \HT $\pm$ 0.7		&  34.0 & \HT $\pm$ 0.7	& 33.6 & \HT $\pm$ 0.7		& 183.5 & \HT $\pm$ 0.6		&  184.2 & \HT $\pm$ 0.7 \\
 $K_{\rm A}$ [km\,s$^{-1}$]		&   30.33 & \HT $\pm$ 0.22		&  29.37 & \HT $\pm$ 0.12		&  \multicolumn{2}{c}{-}		&  17.21 & \HT $\pm$ 0.18	& 15.10 & \HT $\pm$ 0.19		& 31.48 & \HT $\pm$ 0.40		&  31.94 & \HT $\pm$ 0.32  \\
 $K_{\rm B}$ [km\,s$^{-1}$]		&   25.76 & \HT $\pm$ 0.09		&  26.13 & \HT $\pm$ 0.08		& 14.64 & \HT $\pm$ 0.03		 &  14.68 & \HT $\pm$ 0.05	& 14.66 & \HT $\pm$ 0.06		& 23.24 & \HT $\pm$ 0.21       &  23.33 & \HT $\pm$ 0.14  \\
 $q$		&   0.849 & \HT $\pm$ 0.008		&  0.890 & \HT $\pm$ 0.005		& \multicolumn{2}{c}{-}		&  0.853 & \HT $\pm$ 0.011	& 0.971 & \HT $\pm$ 0.012		& 0.738 & \HT $\pm$ 0.016		&  0.730 & \HT $\pm$ 0.032  \\
 \hline  \\
\end{tabular}
\end{table*}

\subsubsection{Spectral disentangling ({\sc spd})}
\label{sec:spd}

The spectra of the binary stars under study are dominated by the spectra of the red-giant components since they contribute the prevailing fraction of the total light of the systems. From the light curve analysis (see light ratio between components in Table~\ref{tab:sysprop}, Section~\ref{sec:ecl_model}) it was found that the dwarf companions contribute only about 9.2, 6.5, and 2.0~per cent to the total light of the system for KIC\,8410637, KIC\,5640750, and KIC\,9540226, respectively. This makes the RVs of the Doppler shifts of the faint companions more difficult to detect, i.e. the $rms$ scatter of the dwarfs is about three times more uncertain than for the giants for KIC\,8410637 and KIC\,9540226 and undetectable for KIC\,5640750. The spectral lines of both components are, however, present in the spectra and to extract both we apply spectral disentangling ({\sc spd}).

The method of {\sc spd} was developed by \citet{1994sim}. In this method, the individual spectra of the components as well as a set of orbital elements can be optimised simultaneously. During this process the fluxes of the observed spectra are effectively co-added. This results in disentangled spectra that have a higher S/N compared to the observed spectra. There is no need for template spectra like in the cross-correlation method. This is highly beneficial in the case of barely visible components' spectrum, like in our case \citep[see][for other examples]{2013may,2014tor,2015kolb}. With the method of {\sc spd} the spectra of the faint dwarf companions were successfully reconstructed with the fractional light in the visual spectral region at the extreme values of barely $\sim1-2$ per cent.

For the present work, we used the spectral disentangling code {\sc fdbinary} \citep{2004ili}, which operates in Fourier space based on the prescription of \citet{1995had} including some numerical improvements. In particular, the Discrete Fourier Transform is implemented in {\sc fdbinary}, which gives more flexibility in selecting spectral segments for {\sc spd} while still keeping the original spectral resolution. We used the wavelength range of the spectra from $5\,000-6\,000\,\AA$ for both the determination of the orbital elements and the isolation of the individual spectra of the components.

\begin{figure}
\includegraphics[width=\columnwidth]{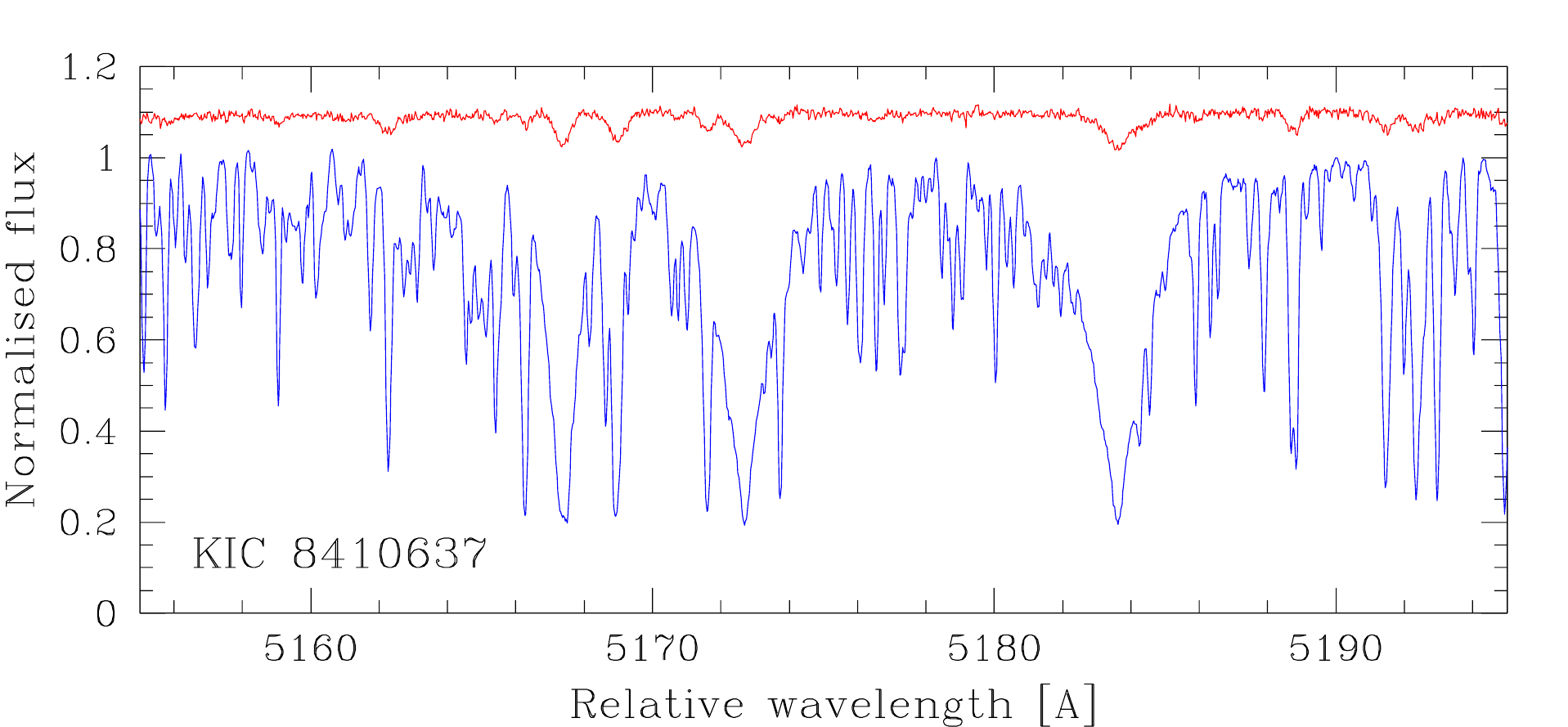} \\
\includegraphics[width=\columnwidth]{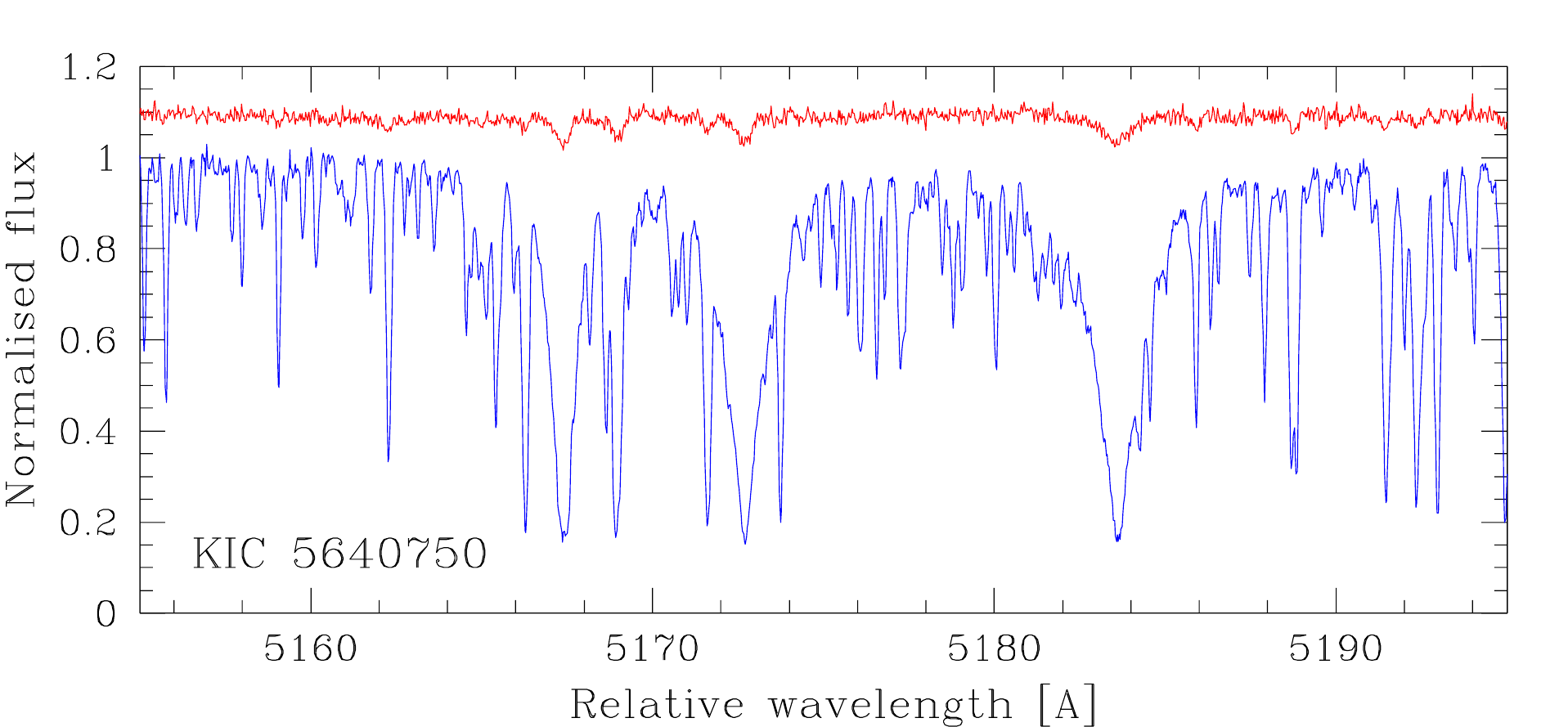} \\
\includegraphics[width=\columnwidth]{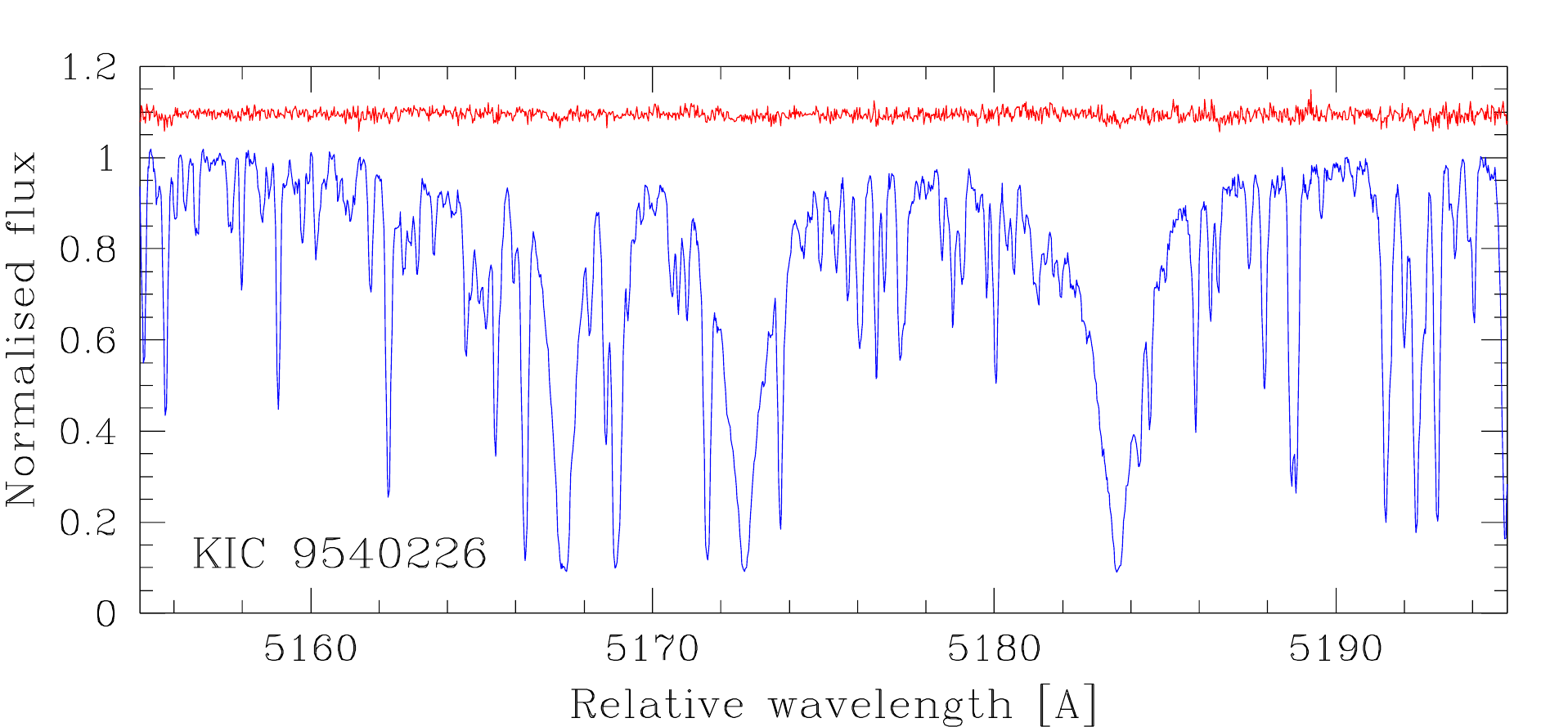} \\
  \caption{Disentangled spectra for the giant (blue) and dwarf (red) component of the eclipsing binary systems KIC\,8410637 (top panel), KIC\,5640750 (middle), and KIC\,9540226 (bottom) centered on \ion{Mg}{i} triplet at $\lambda= 5168-5185$ {\AA}. The spectra are normalized with respect to the composite continuum and for better visibility we use an arbitrary offset between the individual spectra of the binary components.}
    \label{fig:fig01}
\end{figure}

\begin{figure}
\includegraphics[width=\columnwidth]{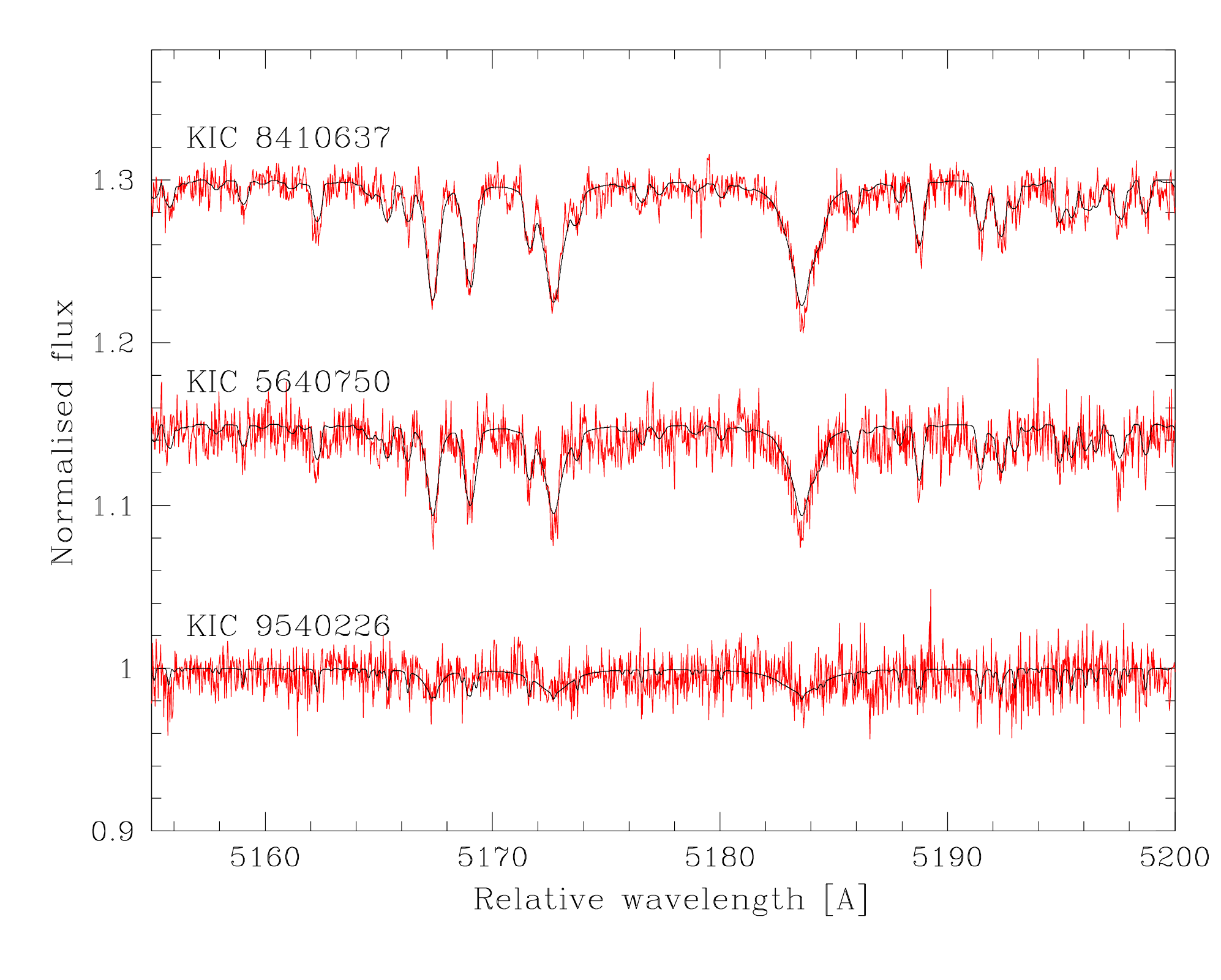} \\
  \caption{Close-ups of the disentangled spectra for the dwarf components (red) of the eclipsing binary systems KIC\,8410637 (top panel), KIC\,5640750 (middle), and KIC\,9540226 (bottom) centered on \ion{Mg}{i} triplet at $\lambda= 5168-5185$ {\AA}. Synthetic spectra are over-plotted in black. The spectra are normalized with respect to the composite continuum and for better visibility we use an arbitrary offset between the individual spectra of the dwarf components.}
    \label{fig:fig01d}
\end{figure}

In {\sc fdbinary} the optimisation is performed with a simplex routine \citep[cf.][]{1989pre}. We performed 100 runs, each with 1000 iterations, examing a relatively wide parameter space around an initial set of parameters. In most cases of high S/N spectra, that are well distributed in the orbital phases, the convergence is achieved quite fast. The uncertainties in the determination of the orbital elements were then calculated with a novel approach using a bootstrapping method (Pavlovski et al. in prep.). The faint companion's spectra for all three systems were extracted (see Figures~\ref{fig:fig01} and \ref{fig:fig01d}). 

\subsubsection{Orbital elements}
\label{sec:orbital}

For the three binary systems under study we report the spectroscopic orbital elements obtained from {\sc ccf} and {\sc spd} analysis in Table~\ref{tab:orbital_elements}. These include the orbital period $P$, the time of periastron $T_0$, the eccentricity $e$, the longitude of periastron $\omega$, the radial velocity semi-amplitudes of the dwarf and giant component $K_{\rm A}$, $K_{\rm B}$ and the mass ratio $q=K_{\rm A}/K_{\rm B}$. The comparison of the results shows agreement between both methods. We note, however, that $T_0$ is different from {\sc spd} and {\sc ccf} for KIC\,5640750 since about one third of the orbital phase is not covered by spectroscopic observations, which results in ambiguities regarding its orbital parameters (Figure~\ref{fig:fig03}).
From {\sc spd} we derived RV semi-amplitudes for all components in the three binary systems making the determination of the dynamical masses for all stars possible. Hence, we adopted these solutions for the further analysis.

{\em KIC\,5640750}:
In the current study we present the first spectroscopic orbit for this binary system based on both components. The {\sc ccf} nor LSD analysis did reveal the radial velocities of the dwarf spectrum. According to our light curve analysis (Section~\ref{sec:ecl_model}) the companion star contributes only $\sim6.5$\,per cent to the total light in the visual passband. In addition, the long orbital period of 987~days makes the detection of the dwarf spectrum difficult since for such small Doppler shifts the spectral lines are along the whole cycle close to the prominent lines of the red-giant component. 
From the {\sc spd} analysis we find two statistically significant solutions for this system which are indistinguishable and whose difference is barely visible in the disentangled spectra.
This ambiguity arises due to an insufficient coverage of the orbital phase which lacks spectroscopic observations between 0 and 0.35 (see bottom left in Figure~\ref{fig:fig03}). Thus, only one extremum in the RV curve is covered by spectroscopic observations. As a result we obtain more than one local minimum in the {\sc spd} analysis due to spurious patterns in the reconstructed spectra of the individual components, which can also affect the quality of the orbital solution \citep{2008hen}. As a further attempt to lift the ambiguity between the two orbital solutions, we rerun the {\sc spd} with fixed $e$ and $\omega$ without success.
In any case, follow-up observations would be required to resolve this ambiguity by filling the gap in the orbital phases. In the current study, we use both solutions of this system to infer the stellar parameters of its components and we check these results for consistencies with asteroseismic stellar parameters. It should be noted that the RV semi-amplitudes for the giant are within 1$\sigma$ confidence level for all solutions.

{\em KIC\,8410637}:
In a comprehensive study by \citet{2013fra} the first spectroscopic orbit was determined for this binary system. Even with about 10~per cent contribution to the total light, the dwarf companion is barely detectable due to a long orbital period of $P \sim 408$~days. \citeauthor{2013fra} used several methods to measure the radial velocities for both components; the line broadening function \citep{2002ruc}, the two-dimensional cross-correlation \citep[{\sc 2D-ccf},][]{1994zuc}, and the Fourier spectral disentangling \citep{1995had}. These three sets of measurements gave consistent orbital parameters within 1$\sigma$ errors. Their final orbital solution is a mean of the results determined from the line broadening function and 2D-{\sc ccf}, and reads, $K_{\rm A} = 30.17\pm0.39$\,\kms, and $K_{\rm B} = 25.85\pm0.07$\,\kms, with the mass ratio, $q = 0.857\pm0.011$. 
Comparing \citeauthor{2013fra} spectroscopic solution with our {\sc ccf} and {\sc spd} results, the agreement is only at a 3$\sigma$ confidence level for the RV semi-amplitudes, and at a 1$\sigma$ level for the geometric orbital parameters, i.e. the eccentricity, and the longitude of periastron. It is difficult to trace the source of these differences. Some systematics could arise because of the different methodology and different datasets that were used. \citeauthor{2013fra} worked with three spectroscopic datasets that were collected with different spectrographs of comparable spectral resolution, {\sc fies} at the Nordic Optical Telescope, {\sc Hermes} at the {\sc Mercator} Telescope, and {\sc ces} at the Th\"{u}ringer Landessternwarte. We used {\sc Hermes} spectra exclusively, hence our dataset is homogeneous, yet less extensive. Since there is no need for template spectra in the {\sc spd} technique, this method is not liable to mismatch problems as the methods used by \citet{2013fra}, as shown in numerical experiments by \citet{2007hen}. 

{\em KIC\,9540226}:
The first attempt to determine the spectroscopic orbit for this binary system was made by \citet{2014bec}. The cross-correlation method applied on 31 {\sc Hermes} spectra did not reveal the dwarf's spectrum. Hence, only the giant's RV semi-amplitude was determined, $K_{\rm B} = 23.32\pm0.04$\,\kms, and the geometric orbital parameters, the eccentricity $e = 0.39\pm0.01$, and the longitude of periastron $\omega = 4.0\pm0.6$\,deg. The {\it Kepler} light curve solution published by \citet{2016gau} shows that the dwarf component contributes barely $\sim2$ per cent to the total light. Despite the low secondary contribution to the total flux, \citeauthor{2016gau} report a detection of the dwarf spectra in 7 out of 12 of their observations by using {\sc ccf}. They used a new series of spectra secured with the 3.5m ARC telescope at Apache Point Observatory. It is encouraging that the spectroscopic orbital elements derived by \citet{2016gau} and ours based on {\sc spd} (Table~\ref{tab:orbital_elements}) agree within 1$\sigma$ uncertainties.

\subsubsection{Individual components' spectra from {\sc spd}}

Spectral disentangling was performed in pure ``separation'' mode \citep{2010pav} since the light curves do not show any significant light variations outside the eclipses. This is also true for the eccentric eclipsing binary system KIC\,9540226 which shows flux modulations at periastron. However, these so-called heartbeat effects are extremely small amplitude that is why they only became widely known through the {\it Kepler} mission. Hence it is justified to use the pure separation mode for all three binary systems.

The disentangled spectra of the components still have a common continuum of a binary system. For the renormalisation of the separated spectra from a common continuum of the binary system to the components' spectra with their individual continua we followed the prescription by \citet{2005pav}. First, an additive correction was made due to different line-blocking of the components. Then these spectra were multiplied for the dilution factor. This multiplicative factor is determined from the light ratio. Since {\it Kepler} photometry is very precise, we preferred the light ratio determined in the light curve analysis (Section~\ref{sec:ecl_model}), rather than the spectroscopically determined one. Disentangled spectra of all binary components could be extracted and are shown in Figures~\ref{fig:fig01} and \ref{fig:fig01d}. The latter presents close-ups of the disentangled spectra for the dwarfs with decreasing S/N from top (KIC\,8410637) to bottom (KIC\,9540226). For the synthetic spectra we used the atmospheric parameters from Table~\ref{tab:atmos} and the light ratios from Table~\ref{tab:sysprop}. Since the dwarf component of KIC\,5640750 is at the limit of detection, we did not obtain its atmospheric parameters and therefore we adjusted the projected rotational velocity to 10\,\kms\ for its synthetic spectrum in Figure~\ref{fig:fig01d}.

\subsection{Eclipse modelling}
\label{sec:ecl_model}

The available light curves of the three systems were modelled with the {\sc jktebop} code \citep[][and references therein]{2013sou} in order to determine their physical properties. {\sc jktebop} parameterises the light curve using the sum and ratio of the fractional radii of the components, $r_{\rm A}+r_{\rm B}$ and $k = r_{\rm B}/r_{\rm A}$. The fractional radii are defined as $r_{\rm A} = R_{\rm A}/a$ and $r_{\rm B} = R_{\rm B}/a$, where $R_{\rm A}$ and $R_{\rm B}$ are the true radii of the stars and $a$ is the orbital semimajor axis. The parameters $r_{\rm A}+r_{\rm B}$ and $k$ were included as fitted parameters, as was the orbital inclination $i$. We fitted for the combination terms $e\cos\omega$ and $e\sin\omega$ where $e$ is the orbital eccentricity and $\omega$ is the argument of periastron. The orbital period, $P$, and midpoint of primary eclipse, $T_0$, were also fitted.

The radiative properties of the stars were modelled using the quadratic limb darkening law \citep{1950kop}, with linear coefficients denoted $u_{\rm A}$ and $u_{\rm B}$ and quadratic coefficients $v_{\rm A}$ and $v_{\rm B}$. We fitted for $u_{\rm B}$, which is well constrained by the shape of the light curve during totality. We fixed $v_{\rm B}$ to theoretical values interpolated from the tabulations of \citet{2010sin}, as it is strongly correlated with $u_{\rm A}$ \citep[e.g.][]{2007sou,2008car}. Both limb darkening coefficients for the dwarf stars ($u_{\rm A}$ and $v_{\rm A}$) were fixed to theoretical values because they are not well constrained by the available data. We also fitted for the central surface brightness ratio of the two stars, $J$.

According to the {\it Kepler} Input Catalog \citep{2011bro}, all three systems have a small but non-zero flux contamination from nearby stars (0.001 for KIC\,8410637, 0.021 for KIC\,5640750 and 0.012 for KIC\,9540226). We obtained solutions with third light, $L_3$, as a fitted parameter but found that they were not significantly different from solutions with $L_3 = 0$. In each case, the best-fitting value of $L_3$ was small and its inclusion had a negligible effect on the other fitted parameters.

We included measured RVs for the stars in the {\sc jktebop} fit, and fitted for the velocity amplitudes of the two stars, $K_{\rm A}$ and $K_{\rm B}$. This was done to include constraints on $e\cos\omega$ and $e\sin\omega$ provided by the RVs and we found that the measured values of $K_{\rm A}$ and $K_{\rm B}$ were in agreement with the input values. However, we did not use them in the subsequent analysis because we prefer the homogeneous set for all dwarfs and giants from {\sc spd} (Section~\ref{sec:spd}). Note that RVs are not available for the dwarf component of KIC\,5640750. We also fitted for the systemic velocities of the stars, $\gamma_{\rm A}$ and $\gamma_{\rm B}$, but did not require $\gamma_{\rm A} = \gamma_{\rm B}$ because the gravitational redshifts of the giants are significantly different to those of the dwarfs. The systemic velocities are formally measured to high precision, but have significantly larger systematic errors due to the intrinsic uncertainty in the stellar RV scale.

As we analysed the {\it Kepler} long-cadence data for each system, the {\sc jktebop} model was numerically integrated to match the 1765\,s sampling rate of these data \citep{2012sou}. This is one point of difference between the current analysis and the study of KIC\,8410637 by \citet{2013fra}. We note that short-cadence data are available for KIC\,9540226 but that we did not use them because the long-cadence data already provide a sufficient sampling rate for both the eclipses and pulsations (Section~\ref{sec:lc}).

\begin{figure*} \includegraphics[width=\textwidth,angle=0]{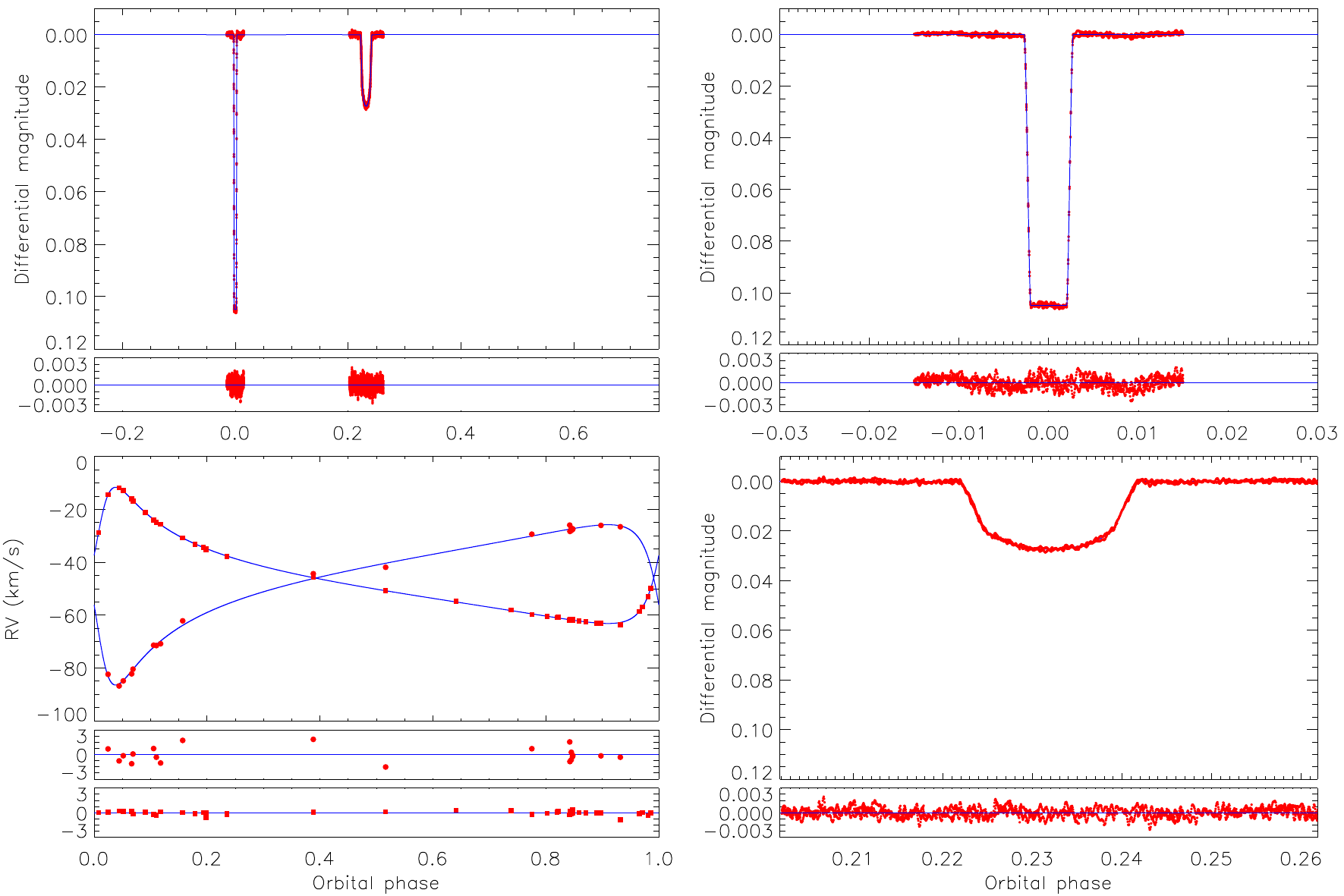}
\caption{\label{fig:fig02} Observational data for KIC\,8410637 and the best-fitting model from {\sc jktebop}. The red points give the data and the blue lines the best fits. The four panels show the phase folded light curve (top left), primary eclipse (top right), the RV curve as derived from {\sc ccf} (left bottom), and secondary eclipse (right bottom). Each panel is accompanied by a plot of the residuals in the lower panel.} \end{figure*}

\begin{figure*} \includegraphics[width=\textwidth,angle=0]{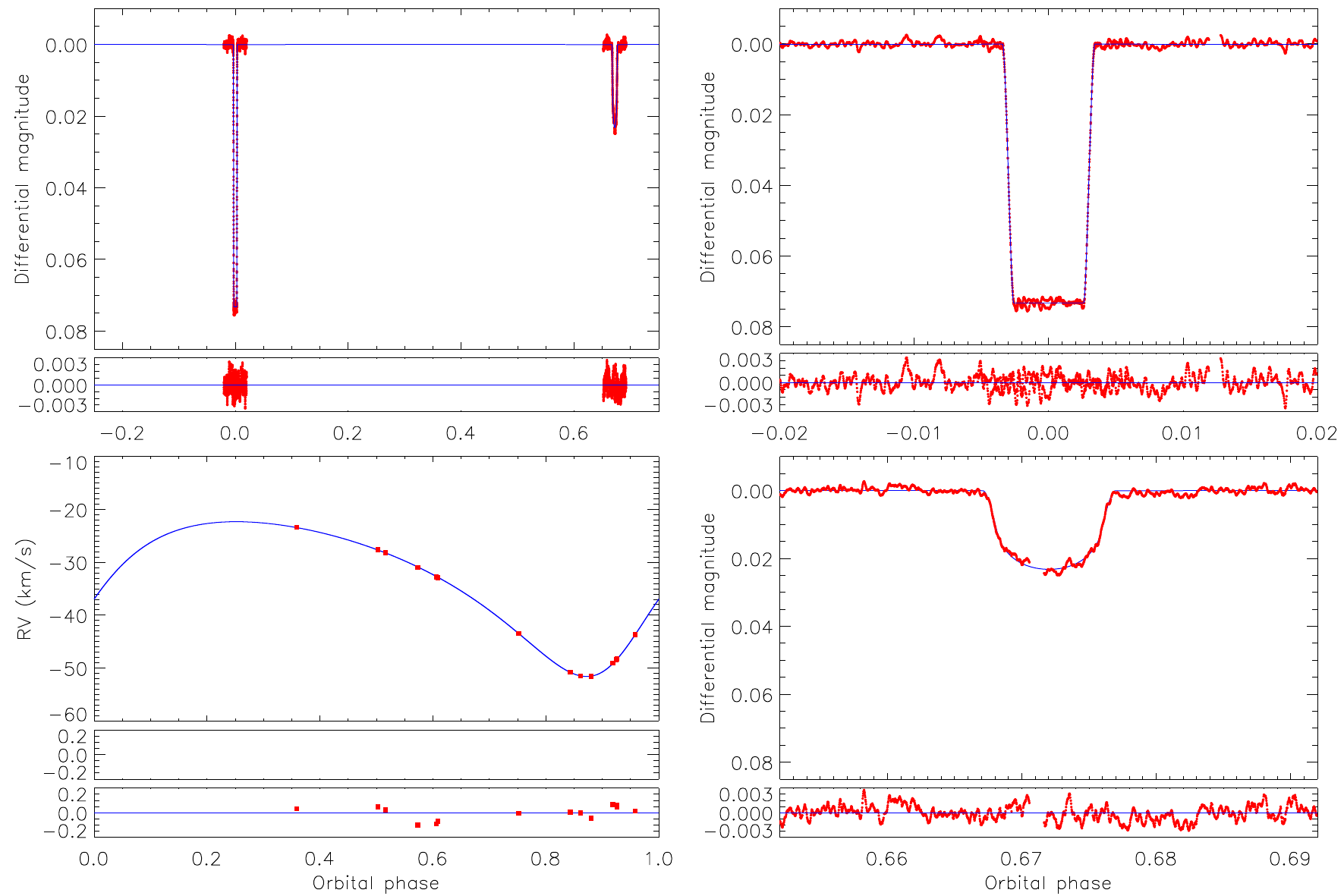}
\caption{\label{fig:fig03} Same as Fig.\,\ref{fig:fig02} now for
KIC\,5640750.} \end{figure*}

\begin{figure*} \includegraphics[width=\textwidth,angle=0]{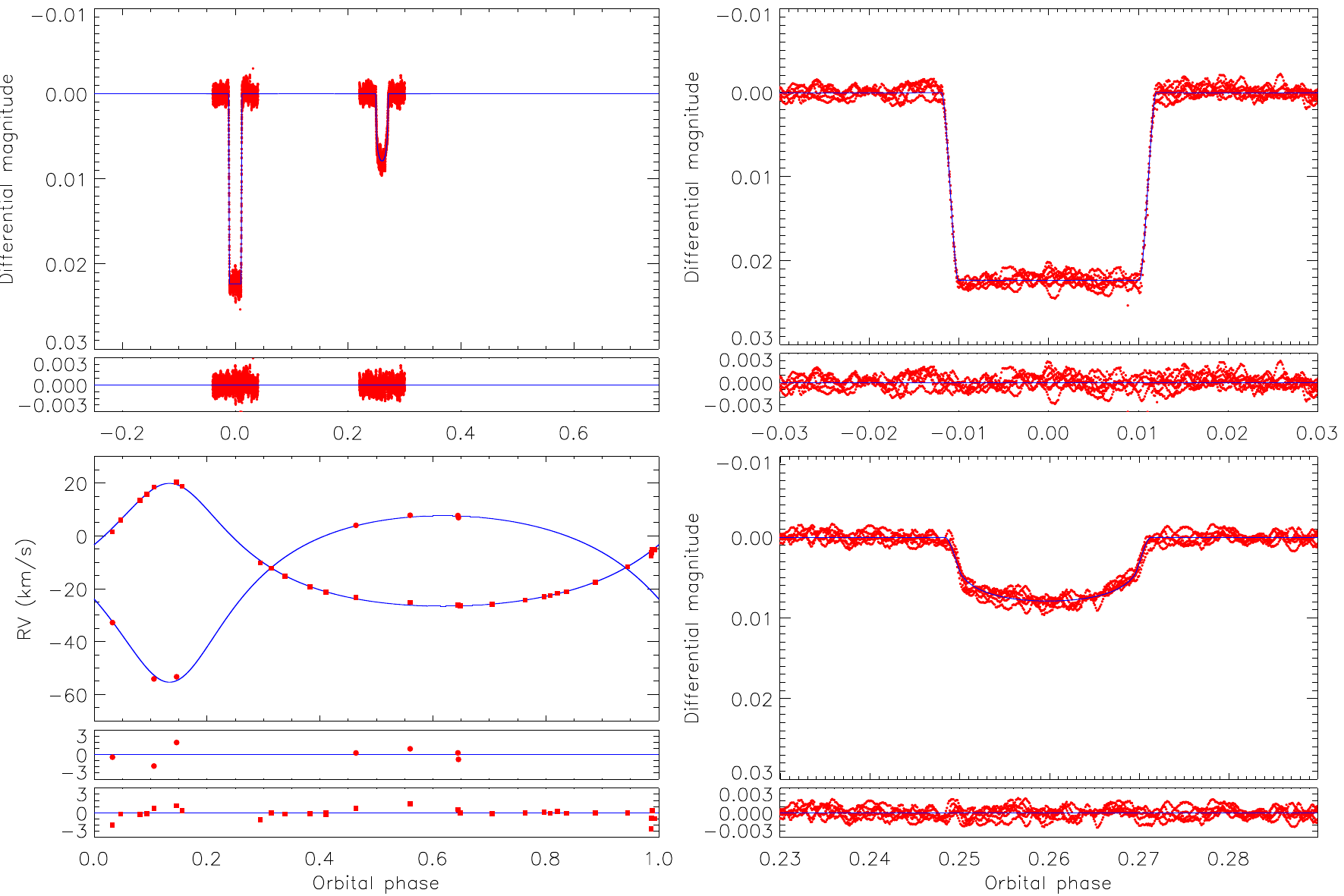}
\caption{\label{fig:fig04} Same as Fig.\,\ref{fig:fig02} now for
KIC\,9540226.} \end{figure*}

The best-fitting values of the fitted parameters for the three systems are listed in Table~\ref{tab:sysprop}, where $M_{\rm A,B}$ are the masses, $R_{\rm A,B}$ the radii, $\log g_{\rm A,B}$ the surface gravities, $L_{\rm A,B}$ the luminosities and $a$ the orbital separation of the two stars. The light ratio $\ell_{\rm B}/\ell_{\rm A}$ of the giant to the dwarf is computed in the {\it Kepler} passband. The light curves and RV data for the three systems are shown in Figures~\ref{fig:fig02}, \ref{fig:fig03} and \ref{fig:fig04}, superimposed on the best-fitting models from {\sc jktebop}. Uncertainty estimates for each parameter were obtained via both Monte Carlo and residual-permutation algorithms \citep[see][]{2008sou}, and the larger of the two uncertainty estimates is reported for each parameter. In most cases we found that the residual-permutation algorithm yielded uncertainties two to three times larger than those from the Monte Carlo algorithm. This is due to the presence of pulsations, which for the purposes of eclipse modelling are simply a source of correlated (red) noise.

\begin{table*} \centering \caption{\label{tab:sysprop}
\setlength{\tabcolsep}{4pt}
Physical properties of the systems measured from the {\it Kepler} light curves and phase-resolved spectroscopy.}
\begin{tabular}{l r l r l p{2pt} r l r l} \hline\hline
\multicolumn{1}{l}{Parameter} & \multicolumn{2}{c}{KIC\,8410637} & \multicolumn{5}{c}{KIC\,5640750} & \multicolumn{2}{c}{KIC\,9540226} \\
\hline
\multicolumn{3}{l}{Parameters fitted using {\sc jktebop}:} & \multicolumn{2}{c}{$1^{\rm st}$ solution}  &     & \multicolumn{2}{c}{$2^{\rm nd}$ solution} & \multicolumn{2}{l}{}\\
$T_0$ [BJD]                       & 2454990.6201 & \HT $\pm$ 0.0007  & \multicolumn{2}{r}{2455269.2144} &$\pm$& \multicolumn{2}{l}{0.0042 } & 2455817.2890 & \HT $\pm$ 0.0024  \\
$P$ [d]                           & 408.32476    & \HT $\pm$ 0.00035 & \multicolumn{2}{r}{987.3981    } &$\pm$& \multicolumn{2}{l}{0.0060 } & 175.44381    & \HT $\pm$ 0.00082 \\
$e\cos\omega$                     & $-$0.35204   & \HT $\pm$ 0.00054 & \multicolumn{2}{r}{0.26916     } &$\pm$& \multicolumn{2}{l}{0.00017} & $-$0.38702   & \HT $\pm$ 0.00011 \\
$e\sin\omega$                     & 0.5884       & \HT $\pm$ 0.0017  & \multicolumn{2}{r}{0.1808      } &$\pm$& \multicolumn{2}{l}{0.0029 } & $-$0.0235    & \HT $\pm$ 0.0042  \\
$r_{\rm A}+r_{\rm B}$             & 0.03730      & \HT $\pm$ 0.00012 & \multicolumn{2}{r}{0.02701     } &$\pm$& \multicolumn{2}{l}{0.00016} & 0.08180      & \HT $\pm$ 0.00087 \\
$k$                               & 6.811        & \HT $\pm$ 0.027   & \multicolumn{2}{r}{7.584       } &$\pm$& \multicolumn{2}{l}{0.066  } & 12.98        & \HT $\pm$ 0.11    \\
$J$                               & 0.2556       & \HT $\pm$ 0.0018  & \multicolumn{2}{r}{0.2695      } &$\pm$& \multicolumn{2}{l}{0.0045 } & 0.2974       & \HT $\pm$ 0.0046  \\
$i$ [degrees]                     & 89.614       & \HT $\pm$ 0.032   & \multicolumn{2}{r}{89.761      } &$\pm$& \multicolumn{2}{l}{0.055  } & 88.73        & \HT $\pm$ 0.19    \\
$u_{\rm B}$                       & 0.528        & \HT $\pm$ 0.024   & \multicolumn{2}{r}{0.573       } &$\pm$& \multicolumn{2}{l}{0.047  } & 0.466        & \HT $\pm$ 0.058   \\
$K_{\rm A}$ [\,km\,s$^{-1}$]      & 30.33        & \HT $\pm$ 0.22    &           17.21 & \HT $\pm$ 0.18 &     & 15.098 & \HT $\pm$ 0.086    & 31.48        & \HT $\pm$ 0.40    \\
$K_{\rm B}$ [\,km\,s$^{-1}$]      & 25.763       & \HT $\pm$ 0.090   &         14.676 & \HT $\pm$ 0.051 &     & 14.664 & \HT $\pm$ 0.056    & 23.24        & \HT $\pm$ 0.21    \\
$\gamma_{\rm A}$ [\,km\,s$^{-1}$] & $-$45.42     & \HT $\pm$ 0.16    & \multicolumn{5}{r}{                                                } & $-$11.70     & \HT $\pm$ 0.22    \\
$\gamma_{\rm B}$ [\,km\,s$^{-1}$] & $-$46.445    & \HT $\pm$ 0.013   & \multicolumn{2}{r}{$-$32.993   } &$\pm$& \multicolumn{2}{l}{0.013  } & $-$12.36     & \HT $\pm$ 0.11    \\
\hline
\multicolumn{7}{l}{Derived parameters:} \\
$T_{\rm eff}$ of dwarf [K]        & 6066         & \HT $\pm$ 200      & \multicolumn{2}{r}{5844     } &$\pm$& \multicolumn{2}{l}{200      } & 5822         & \HT $\pm$ 200     \\
$r_{\rm A}$                       & 0.004775     & \HT $\pm$ 0.000027 & \multicolumn{2}{r}{0.003147 } &$\pm$& \multicolumn{2}{l}{0.000034 } & 0.005850     & \HT $\pm$ 0.000067\\
$r_{\rm B}$                       & 0.032522     & \HT $\pm$ 0.000094 & \multicolumn{2}{r}{0.02387  } &$\pm$& \multicolumn{2}{l}{0.00014  } & 0.07595      & \HT $\pm$ 0.00082 \\
$\ell_{\rm B}/\ell_{\rm A}$       & 9.860        & \HT $\pm$ 0.017    & \multicolumn{2}{r}{14.342   } &$\pm$& \multicolumn{2}{l}{0.060    } & 48.13        & \HT $\pm$ 0.16    \\
Mass ratio                        & 1.124        & \HT $\pm$ 0.006    &       1.173 & \HT $\pm$ 0.013 &     & 1.030 & \HT $\pm$ 0.007       & 1.369        & \HT $\pm$ 0.016   \\
$M_{\rm A}$ [M$_\odot$]           & 1.309        & \HT $\pm$ 0.014    &       1.292 & \HT $\pm$ 0.017 &     & 1.125 & \HT $\pm$ 0.011       & 1.015        & \HT $\pm$ 0.016   \\
$M_{\rm B}$ [M$_\odot$]           & 1.472        & \HT $\pm$ 0.017    &       1.515 & \HT $\pm$ 0.033 &     & 1.158 & \HT $\pm$ 0.014       & 1.390        & \HT $\pm$ 0.031   \\
$R_{\rm A}$ [R$_\odot$]           & 1.556        & \HT $\pm$ 0.010    &       1.853 & \HT $\pm$ 0.023 &     & 1.730 & \HT $\pm$ 0.020       & 1.034        & \HT $\pm$ 0.014   \\
$R_{\rm B}$ [R$_\odot$]           & 10.596       & \HT $\pm$ 0.049    &       14.06 & \HT $\pm$ 0.12  &     & 13.12 & \HT $\pm$ 0.09        & 13.43        & \HT $\pm$ 0.17    \\
$\log g_{\rm A}$ (cgs)            & 4.171        & \HT $\pm$ 0.005    &       4.014 & \HT $\pm$ 0.010 &     & 4.014 & \HT $\pm$ 0.010       & 4.416        & \HT $\pm$ 0.010   \\
$\log g_{\rm B}$ (cgs)            & 2.556        & \HT $\pm$ 0.003    &       2.323 & \HT $\pm$ 0.007 &     & 2.266 & \HT $\pm$ 0.006       & 2.326        & \HT $\pm$ 0.010   \\
$\log L_{\rm A}$ [L$_\odot$]      & 0.468        & \HT $\pm$ 0.058    &       0.555 & \HT $\pm$ 0.060 &     & 0.495 & \HT $\pm$ 0.060       & 0.042        & \HT $\pm$ 0.061   \\
$\log L_{\rm B}$ [L$_\odot$]      & 1.656        & \HT $\pm$ 0.030    &       1.871 & \HT $\pm$ 0.029 &     & 1.811 & \HT $\pm$ 0.029       & 1.854        & \HT $\pm$ 0.030   \\
$a$ [AU]                          & 1.5148       & \HT $\pm$ 0.0054   &       2.738 & \HT $\pm$ 0.016 &     & 2.557 & \HT $\pm$ 0.009       & 0.8218       & \HT $\pm$ 0.0052  \\
$E(B-V)$                          & 0.07         & \HT $\pm$ 0.02     &       0.16  & \HT $\pm$ 0.03  &     & 0.16  & \HT $\pm$ 0.03        & 0.16         & \HT $\pm$ 0.03    \\
Distance [pc]            	    & 1005         & \HT $\pm$ 29       &       1569  & \HT $\pm$ 55    &     & 1464  & \HT $\pm$ 50          & 1667         & \HT $\pm$ 63      \\
\hline \end{tabular} \end{table*}

\subsubsection{Physical properties of the systems}

In Table~\ref{tab:sysprop} we list the physical properties of the systems derived from the spectral disentangling analysis and the {\sc jktebop} analyses. These were calculated using the {\sc jktabsdim} code \citep{2005sou}, and the uncertainties were propagated via a perturbation approach. We emphasise that the velocity amplitudes from the spectral disentangling analysis were preferred over those from the RV measurements because they are available for all six stars.

We also determined the distances to the systems using published optical and near-IR photometry \citep{2011bro,2012hen,2006skr} and the bolometric corrections provided by \citet{2002gir}. Values of $E(B-V)$ were obtained by requiring agreement between the distances at optical and near-IR wavelengths, being $0.16 \pm 0.03$\,mag for KIC\,5640750, $0.07 \pm 0.02$\,mag for KIC\,8410637 and $0.16 \pm 0.03$\,mag for KIC\,9540226. We finally quote the distances determined from the 2MASS $K$-band apparent magnitudes, as these are the least affected by uncertainties in the effective temperatures and $E(B-V)$ values. We conservatively doubled the uncertainties in these measurement to account for some inconsistency in optical apparent magnitudes quoted by different sources. Our distance estimates (see Table~\ref{tab:sysprop}) are much more precise than those from {\it Gaia} Data Release 1 \citep{2016gaib}; future data releases from the {\it Gaia} satellite will significantly improve the distance measurements to these three binary systems.

{\em KIC\,5640750}: We are the first to determine dynamical stellar parameters for this long-period binary system. By using the first set of orbital parameters, denoted as {\sc spd 1} in Table~\ref{tab:orbital_elements}, we obtained $M_{\rm B} = 1.52 \pm 0.03\rm\, M_{\sun}$ and $R_{\rm B} = 14.06 \pm 0.12\rm\, R_{\sun}$ for the red giant component in this system. The second orbital solution ({\sc spd 2}) provided significantly lower stellar parameters with $M_{\rm B} = 1.16 \pm 0.01\rm\, M_{\sun}$ and $R_{\rm B} = 13.12 \pm 0.09\rm\, R_{\sun}$) for the same red-giant star, which results in a relative difference of $\sim 0.4 \,\rm M_{\sun}$ in stellar mass and $\sim 0.9 \rm \,R_{\sun}$ in stellar radius, respectively.

{\em KIC\,8410637}: We found that the velocity amplitudes were different at a 3$\sigma$ level when measured from the RVs compared to the results from spectral disentangling. Our adoption of the velocity amplitudes from spectral disentangling means that we find significantly lower masses for the two components of this system compared to those found by \citet{2013fra} and \citet{2016gau}. However, the discrepancy between the results found by \citet{2013fra} and those from asteroseismic studies led us to investigate this system further. As the dominant source of noise is pulsations in the light curve, we investigated whether the measured radius of the giant was sensitive to which eclipses were included in the analysis. We did this by obtaining eight best fits with each of the eclipses (four primary and four secondary) omitted in turn. The standard deviation of the $R_{\rm B}$ values was 0.047, which is slightly smaller than the error estimate for this quantity in Table~\ref{tab:sysprop}. We therefore conclude that our measured $R_{\rm B}$ is robust against the omission of parts of the input data.

{\em KIC\,9540226}: For this star our measurements of the system parameters can be compared to those found by \citet{2016gau}, who worked with similar data and analysis codes. We find that the agreement between the two sets of results is reasonable but not perfect. Our value of $R_{\rm A}$ and $R_{\rm B}$ are larger by 2$\sigma$ and 2.6$\sigma$, respectively, and the mass measurements agree to within 1$\sigma$. Finally, the mass and the radius of the giant found by \citet{2016bro} are somewhat larger (by 2.4$\sigma$ and 1.9$\sigma$ respectively). In their most recent study, \citet{2018bro} re-analysed this system and obtained considerably lower values for both, the radius and the mass of the red giant. Compared to their latest measurements, our values of $M_{\rm B}$ and $R_{\rm B}$ agree to within 1$\sigma$ and 2$\sigma$ respectively.

\begin{figure*}
		\includegraphics[width=2\columnwidth]{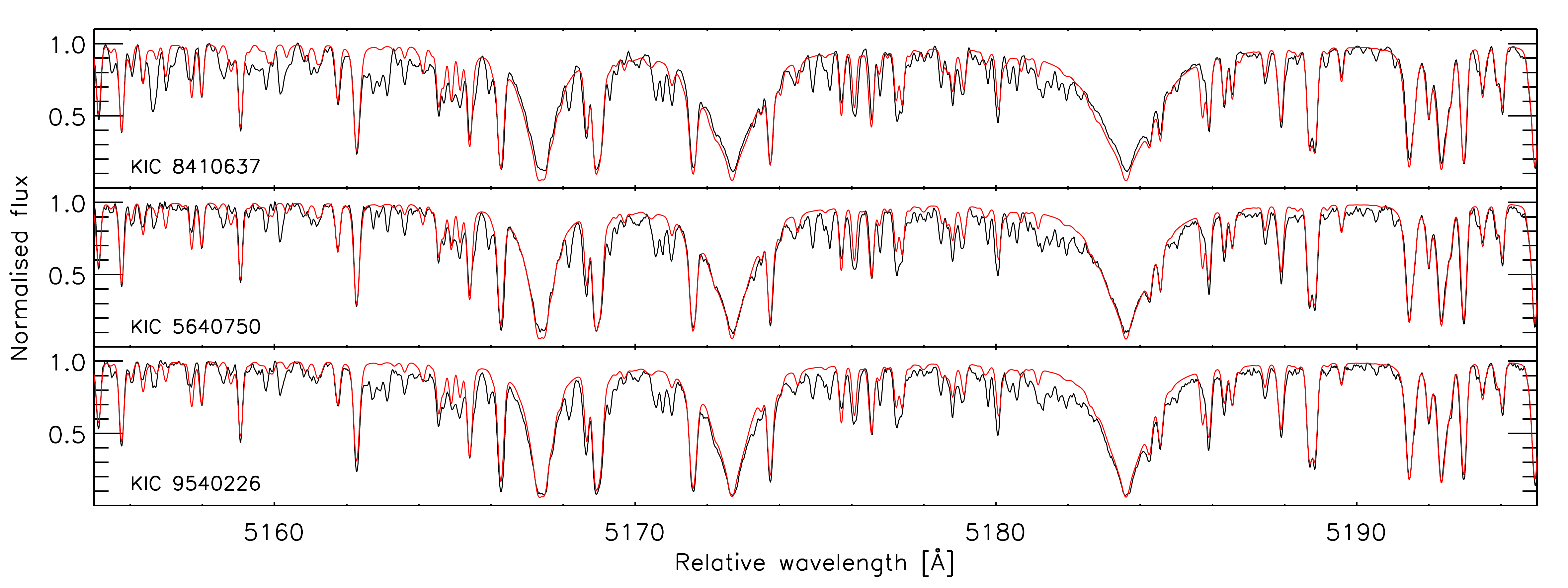}
    \caption{Observed (in black) and best-fitting synthetic (in red) spectra for the giant components in the binary systems in the wavelength range between $5\,155-5\,195$~\AA\ around the \ion{Mg}{i} triplet.}
    \label{fig:fig05}
\end{figure*}

\subsection{Atmospheric parameters}
\label{sec:atmos}

For the extraction of the atmospheric parameters we used the Grid Search in Stellar Parameters \citep[{\sc gssp};][]{2015tka} software package to analyse the disentangled spectra of the evolved components of each of the eclipsing binary systems. 
{\sc gssp} is a LTE-based software package that uses the {\sc SynthV} \citep{1996tsy} radiative transfer code to compute grids of synthetic spectra in an arbitrary wavelength range based on a precomputed grid of plane-parallel atmosphere models from the {\sc LLmodels} code \citep{2004shu}. The atomic data were retrieved from the Vienna Atomic Lines Database \citep[{\sc vald};][]{2000kup}. The optimisation was performed simultaneously for six atmospheric parameters: effective temperature ($T_{\rm eff}$), surface gravity ($\log g$; if not fixed to the value obtained from the light curve solution), micro- and macro-turbulent velocities ($v_{\rm micro}$, $v_{\rm macro}$), projected rotational velocity ($v \sin i$), and global metallicity ([M/H]). The grid of synthetic spectra was built from all possible combinations of the above-mentioned atmospheric parameters and the best-fit solution was obtained by minimising the $\chi^2$ merit function. The 1$\sigma$ errors were derived from $\chi^2$ statistics taking into account possible correlations between the parameters in question. 
In general, {\sc gssp} allows for the analysis of single and binary star spectra, where both composite and disentangled spectra can be analysed for atmospheric parameters and elemental abundances of the individual binary components in the latter case.
We refer the reader to \citet{2015tka} for details on the method implemented in {\sc gssp} and for several methodology tests on the simulated and real spectra of single and binary stars.
In this work, we used the {\sc gssp}-{\sc single module}, where the spectra were treated as those of single stars. By doing so we take advantage of the fact that the light dilution effect could be corrected for based on the a priori knowledge of light factors from the light curve solution. 

Figure~\ref{fig:fig05} shows the best-fitting solutions to a short segment of each observed red-giant spectrum. The atmospheric parameters for KIC\,8410637, KIC\,5640750, and KIC\,9540226 are reported in Table~\ref{tab:atmos} except for the dwarf component of KIC\,9540226. Due to the high noise level in the disentangled spectrum of this dwarf companion (see bottom panel in Figures~\ref{fig:fig01} and \ref{fig:fig01d}), we are not able to obtain precise estimates of its atmospheric parameters from spectral fitting. From its mass and by assuming solar metallicity we can only infer that it is a dwarf star of early to intermediate G spectral type.

{\em KIC\,5640750}: We are the first to determine the atmospheric parameters of the binary components of KIC\,5640750. For the red-giant star we derived $T_{\rm eff}=4525\pm75$\,K and $\rm [M/H]=-0.29\pm0.09$\,dex and for its companion we obtained $T_{\rm eff}=6050\pm350$\,K  and $\rm [M/H]=0.08\pm0.25$\,dex.

{\em KIC\,8410637}: The atmospheric parameters for the stars in this binary system were also determined by \citet{2013fra} from the disentangled spectra of the components. They used the Versatile Wavelength Analysis ({\sc vwa}) package \citep{2004bru}.
The effective temperatures that they determined for the giant and dwarf component, $T_{\rm eff} = 4800\pm80$\,K, and $T_{\rm eff} = 6490\pm160$\,K, respectively, agree with our results (Table~\ref{tab:atmos}) at the 2$\sigma$, and 1$\sigma$ confidence level. The somewhat worse agreement in the effective temperature determinations could be explained as a metallicity effect.
Whilst we found almost solar metallicity for the red giant component, \citeauthor{2013fra} determined [Fe/H] = $0.24\pm0.15$~dex, which was based on numerous \ion{Fe}{i} lines. Since in this temperature range the metal lines become deeper for lower $T_{\rm eff}$ both results could agree in case the degeneracy between the $T_{\rm eff}$ and metallicity can be lifted. This might also explain a better agreement for the $T_{\rm eff}$ of the dwarf companion. \citeauthor{2013fra} fixed the metallicity to [Fe/H] = 0.1\,dex, which is closer to the value we derived, although the uncertainties in the determination of the $T_{\rm eff}$ of the dwarf star are considerably larger than in the case of the red giant component, due to the faintness of the dwarf companion.
The fractional light dilution factor for the RG component is $l_{\rm RG} = 0.9085$, and 0.9080, from the light curve analysis in \citeauthor{2013fra}, and our present study, respectively. The light ratio used in both studies could be another source of slight discrepancies, however it seems unlikely given the small difference between these values. 

{\em KIC\,9540226}: For the red giant component in this binary system, \citet{2016gau} determined the atmospheric parameters through spectroscopic analysis of \ion{Fe}{i} and \ion{Fe}{ii} lines. They used the MOOG spectral synthesis code \citep{2012sne}. It is not clear how they deal with the dilution effect of the secondary component, yet with its contribution of barely $\sim2$~per cent its influence is very small if not negligible. Based on the ARCES spectra they adopted the following principal atmospheric parameters as final results of their work: $T_{\rm eff} = 4692\pm65$\,K, $\log g = 2.2\pm0.2$, and [Fe/H] = $-0.33\pm0.04$\,dex. It is very encouraging that the result from \citet{2016gau} and the analysis in this work agree within 1$\sigma$ uncertainties. \citet{2016bro} first announced preliminary spectroscopic analysis results based mostly on previously published data, which was later on followed by a revised analysis of the same system \citep{2018bro}. In both studies, they derived a lower metallicity ($\rm [Fe/H] = -0.21\pm0.10$ and $\rm [Fe/H]_{rev} = -0.23\pm 0.10$) for the red-giant component. Moreover, their effective temperature measurements for the giant are considerably higher than ours with $T_{\rm eff} = 4780\pm55$\,K and $T_{\rm eff,rev}=4680\pm 80$\,K. This may again be due to the $T_{\rm eff}$ and metallicity degeneracy mentioned earlier.

The effective temperatures of the dwarf components were barely measurable during the spectroscopic analysis due to the low S/N of the disentangled spectra. Thus as a further test we obtained estimates by interpolating between theoretical spectra from the {\sc atlas9} model atmosphere code \citep{1993kur} and using the passband response function of the {\it Kepler} satellite\footnote{{\tt https://keplergo.arc.nasa.gov/kepler\_response\_hires1.txt}}. We determined the effective temperatures of the synthetic spectra which reproduced the central surface brightness ratios measured using {\sc jktebop} versus synthetic spectra for the effective temperatures of the giant stars. The formal uncertainties on these parameters are similar to the uncertainties in the effective temperature measurements for the giants. We instead quote a uniform uncertainty of $\pm$200\,K to account for systematic errors in this method such as dependence on theoretical calculations; the measured {\it Kepler} passband response function; and the metallicities of the stars. 
We report the effective temperatures of each dwarf component of the three eclipsing binary systems in Table~\ref{tab:sysprop}. These results agree with $T_{\rm eff}$ measurements from {\sc spd} for the dwarf companions of KIC\,8410637 and KIC\,5640750 albeit lower by $\sim 300$\,K and $\sim 200$\,K.

\begin{table*}
\centering \caption{\label{tab:atmos} Atmospheric parameters
determined with the {\sc gssp} code. The stellar surface gravities, $\log g$ were kept
fixed to the values coming from the mass and the radius of the stars obtained through the combined photometric-spectroscopic solution (Table~\ref{tab:sysprop}).}
\begin{tabular}{l r l r l c r l r l c r l} \hline\hline
Parameter &  \multicolumn{4}{c}{KIC\,8410637} &&  \multicolumn{4}{c}{KIC\,5640750} && \multicolumn{2}{c}{KIC\,9540226} \\ \hline
& \multicolumn{2}{c}{giant} & \multicolumn{2}{c}{dwarf} && \multicolumn{2}{c}{giant} & \multicolumn{2}{c}{dwarf} && \multicolumn{2}{c}{giant} \\ \hline
$T_{\rm eff}$ [K]		&  4605 & \HT $\pm$ 80		& 6380 & \HT $\pm$ 250		&& 4525 & \HT $\pm$ 75		& 6050 & \HT $\pm$ 350		&& 4585 & \HT $\pm$ 75 \\
 $\log g$ (cgs)		&   \multicolumn{2}{c}{2.56 [fixed]	}	&   \multicolumn{2}{c}{4.17 [fixed]}		&&  \multicolumn{2}{c}{2.32 [fixed]}		&   \multicolumn{2}{c}{4.01 [fixed]	}	&&  \multicolumn{2}{c}{2.33 [fixed]}  \\
$v_{\rm micro}$ [km\,s$^{-1}$]		&  1.14 & \HT $\pm$ 0.18		&  1.95 & \HT $\pm$ 0.45		&& 1.16 & \HT $\pm$ 0.17		&  0.55 & \HT $^{+\,1.05}_{-\,0.55}  $		&& 1.22 & \HT $\pm$ 0.17  \\
$v_{\rm macro}$ [km\,s$^{-1}$]		&  5.0 & \HT $\pm$ 2.5		&  \multicolumn{2}{c}{0.0 [fixed]}		&& 4.7 & \HT $\pm$ 2.5		&  \multicolumn{2}{c}{0.0 [fixed]}		&& 5.1 & \HT $\pm$ 1.5  \\
$v\sin i$ [km\,s$^{-1}$]		&  2 & \HT $^{+\,3}_{-\,2}$		&  17.4 & \HT $\pm$ 1.2		&& 2 & \HT $^{+\,3}_{-\,2}$		&  13.9 & \HT $\pm$ 1.8		&& 1 & \HT $^{+\,3}_{-\,1}$  \\
$[{\rm M/H}]$ (dex)		&  0.02 & \HT $\pm$ 0.08		&  0.01 & \HT $\pm$ 0.14		&& -0.29 & \HT $\pm$ 0.09		&  0.08 & \HT $\pm$ 0.25		&& -0.31 & \HT $\pm$ 0.09  \\
\hline
\end{tabular}
\end{table*}

\section{Stellar properties of oscillating red-giant stars from asteroseismology}
\label{sec:astero}

We complement the binary analysis with a comprehensive study of the stellar oscillations of the systems' red-giant components. For a star showing solar-like oscillations we can infer its asteroseismic mass and radius and thus study consistencies between asteroseismic and dynamical stellar parameters. The asteroseismic approach leads to a more complete description of red giants by revealing their evolutionary stages and ages. In our study, we use well-defined and consistent methods to obtain reliable seismic ($\nu_{\rm max}$ and $\Delta\nu$) and stellar parameters ($M,R,\bar{\rho}$ and $\log g$) for the three stars under study, which we describe here in detail.

\subsection{{\it Kepler} corrected time-series data}
\label{sec:lc}

For the asteroseismic analysis we use {\it Kepler} datasets that have been prepared according to \cite{2014han}. During this procedure long-term variations, outliers, drifts and jumps were removed together with the primary and secondary eclipses. This is a necessary step as the presence of eclipses would interfere with the study of the global oscillations. Figure~\ref{fig:fig06} shows the corrected {\it Kepler} light curves of KIC\,8410637, KIC\,5640750 and KIC\,9540226. The light curve of KIC\,9540226 contains large gaps due to its location on a broken CCD module for three months every year. The Fourier spectra of pulsating red-giant stars reveal a rich set of information consisting of both a granulation as well as an oscillation signal. 

\subsection{The background model}
\label{sec:bg}

Some power in the red-giant Fourier spectrum originates from sources other than the pulsations, such as activity, granulation and photon noise. These signals together form a background on which the oscillations are superimposed. In order to fully exploit the oscillations, we first need to assemble a background model consisting of a constant white noise level and granulation components. Here we use two granulation components with different timescales and a fixed exponent of four. This was shown to be appropriate for describing the granulation background of red-giant stars and provides a global background fit similar to model F proposed by \citet{2014kal}:

\begin{figure}
	\includegraphics[width=\columnwidth]{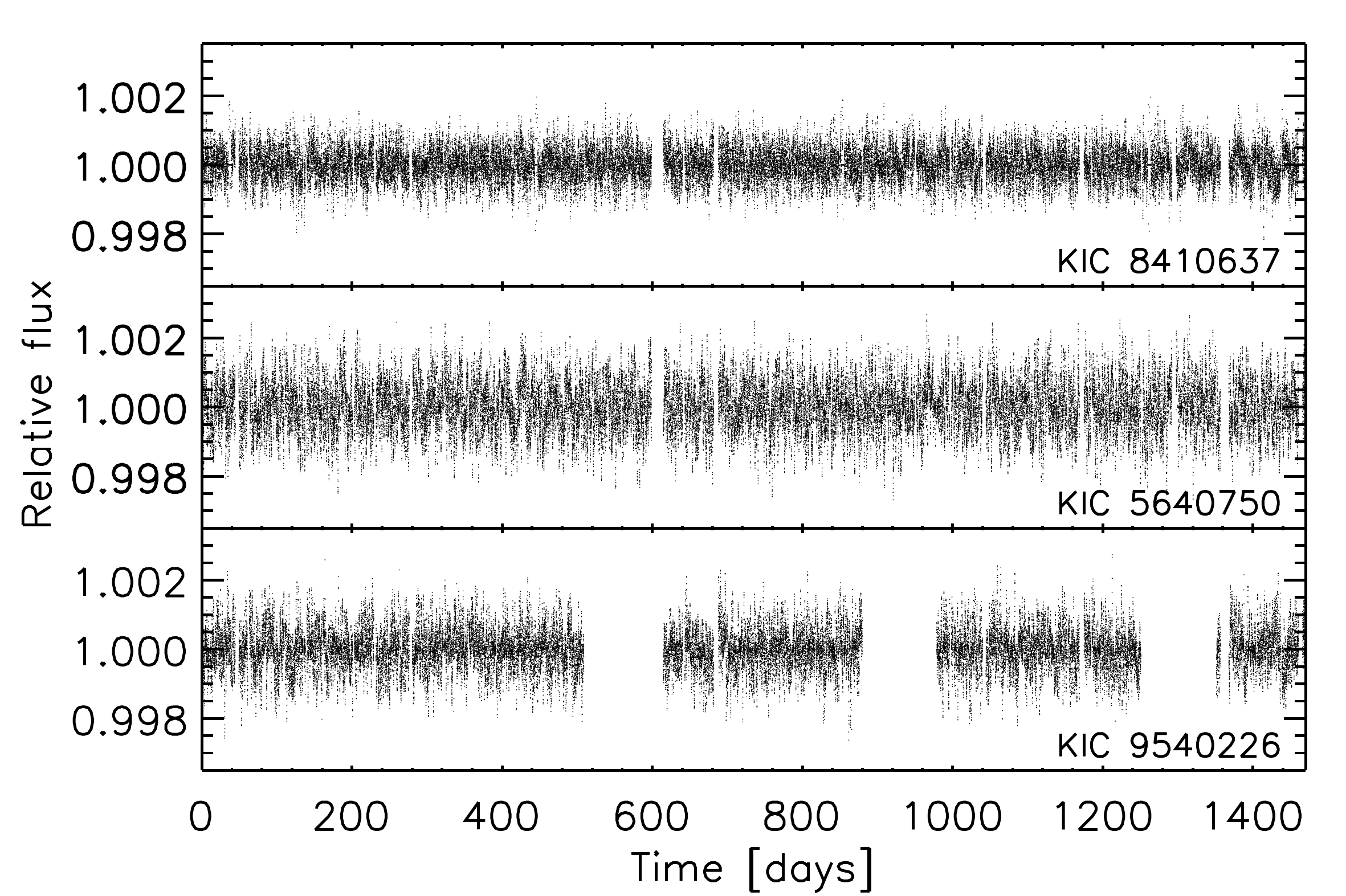}
    \caption{Corrected concatenated {\it Kepler} light curves of the three red-giant stars. The KIC numbers are indicated in each panel. The observations span 1470~days.}
    \label{fig:fig06}
\end{figure}

\begin{equation}
    P_{\rm bg}(\nu)=w_{\rm noise}+\eta(\nu)^2\left[\sum^{2}_{\rm i=1}\frac{\frac{2\sqrt{2}}{\pi}\frac{a_{\rm i}^2}{b_{\rm i}}}{1+(\nu/b_{\rm i})^4}\right].
	\label{eq:granbg}
\end{equation}

Here, $w_{\rm noise}$ describes the white noise contribution to model the photon noise, $a_{\rm i}$ and $b_{\rm i}$ correspond to the root-mean-square ($rms$) amplitude and characteristic frequency of the granulation background component. The stellar granulation and oscillation signals are also influenced by an attenuation $\eta=\rm{sinc}\left(\frac{\pi}{2}\frac{\nu}{\nu_{\rm nyq}}\right)$ due to the integration of the intrinsic signal over discrete time stamps, which increases with higher frequencies approaching the Nyquist frequency $\rm \nu_{nyq}$. 

We fitted the background model over a frequency range from $1\,\mu$Hz up to $283\,\mu$Hz, which is the Nyquist frequency for {\it Kepler} long-cadence data. To avoid influences of the oscillation modes, we excluded the frequency range of the oscillations during this procedure. We note here that we also checked the background fit by taking the oscillations into account simultaneously with a Gaussian-shaped envelope, for which we found agreeing results.
To sample the parameter space of the variables given by Equation~\ref{eq:granbg} we used our own implementation of a Bayesian Markov Chain Monte Carlo (MCMC) framework \citep[e.g.][and references therein]{2011han,2016dav} that employs a Metropolis-Hastings algorithm. In this approach we draw random samples from a probabilistic distribution by using a likelihood function and a proposal distribution (priors) for each of the parameters of interest. We used the exponential log-likelihood introduced by \citet{1986duv} that is suitable for describing the Fourier power density spectrum (PDS) of a solar-like oscillator that has a $\chi^2$ distribution with two degrees of freedom \citep{2003app}. As priors we considered uniform distributions for the global background parameters that are given in the top part of Table~\ref{tab:bg_par}. 
The Metropolis-Hastings MCMC algorithm was run with multiple chains from different initial conditions. From trace plots we assessed the initial burn-in period and we checked that the chains are well mixed and that they explore the relevant parameter space. The initial values of the burn-in phase were then discarded and we ran the algorithm for another 150\,000 iterations before we assessed the convergence of the chains to the posterior distributions. For each distribution we adopted the median as the best-fit parameter value and calculated its 68~per cent credible interval (Table~\ref{tab:bg_results}). 

\begin{figure*}
	\includegraphics[width=2\columnwidth]{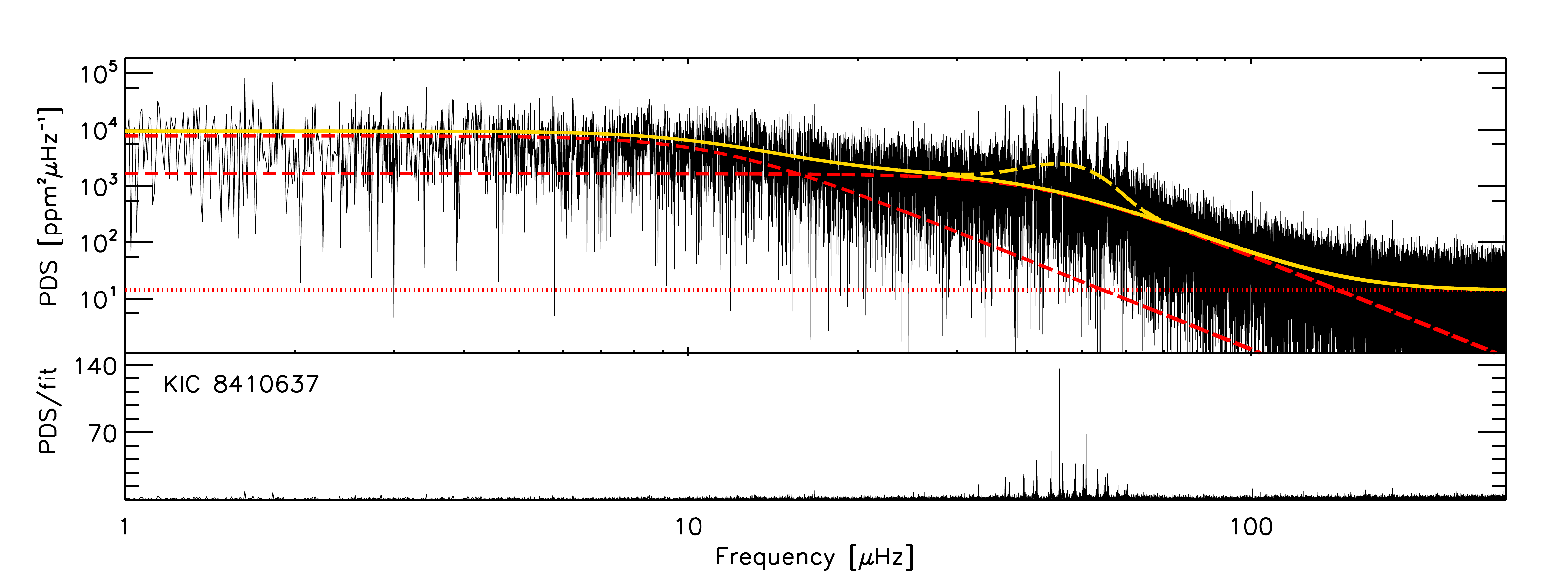}
	\includegraphics[width=2\columnwidth]{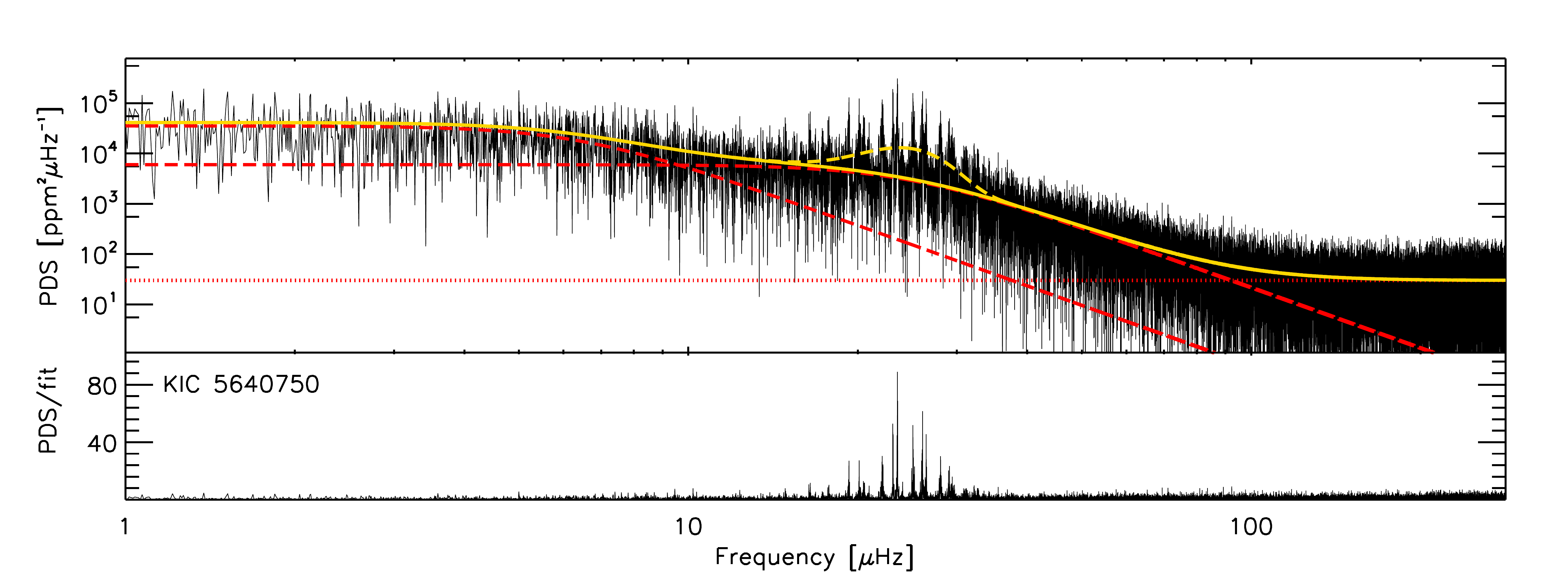}
	\includegraphics[width=2\columnwidth]{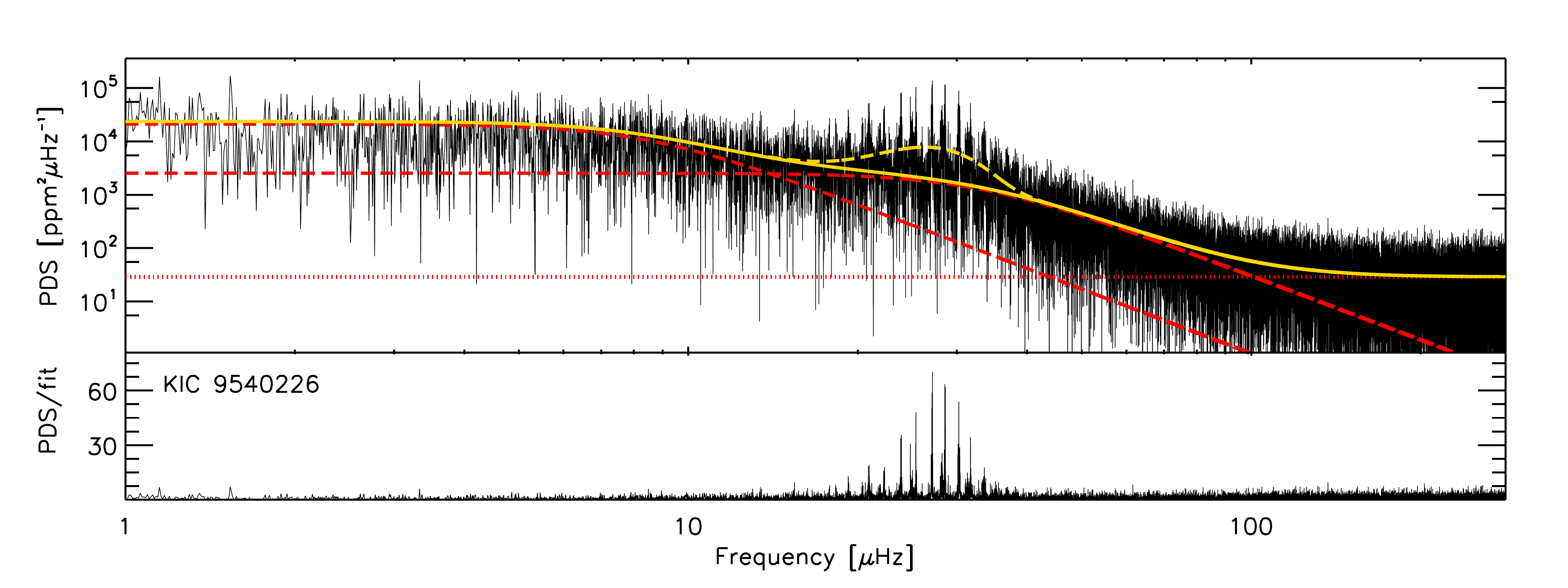}
    \caption{Fourier power density spectra of KIC\,8410637 (top), KIC\,5640750 (middle) and  KIC\,9540226 (bottom) in black. The yellow solid lines represent the best global background fits to the data. The granulation background components are indicated by the red dashed lines, while the red horizontal dotted lines depict the constant white noise components. The oscillation excesses are modelled by Gaussians (yellow dashed lines, see Eq.~\ref{eq:gauss}, Section~\ref{sec:osc}). The smaller panels below each power density spectrum show the background normalized spectra.}
    \label{fig:fig07}
\end{figure*}

Figure~\ref{fig:fig07} shows the global background fits to the Fourier power density spectra. The oscillation power excesses, distinct for pulsating red-giant stars, are clearly visible in each spectrum. For illustrative purposes, we also present the background normalized spectra in the lower panels, which reveal that after correcting for the background only the oscillations are left in the spectra.

\begin{table}
	\centering
	\caption{Ranges of uniform prior distributions used for the global background and Gaussian parameters.}
	\label{tab:bg_par}
	\begin{tabular}{ll}
		\hline\hline
		Parameter & Ranges of the uniform priors\\
		\hline
		Noise $w_{\rm noise}$ & $<10\,\times$\,mean power($0.75\times\nu_{\rm nyq}$ to $\nu_{\rm nyq}$)\\
		$Rms$ amplitude $a_1$, $a_2$ & $<\sqrt{\rm maximum~power}$\\
		Frequency $b_1$, $b_2$ & 1 to $\nu_{\rm nyq}$ with $b_1<b_2$\\
		\hline
		Height of Gaussian $\Lambda_{\rm g}$ & $<$ maximum power\\
		Frequency $\nu_{\rm max}$ & $\nu_{\rm max,guess} \pm 1.5\times\Delta\nu_{\rm guess}$\\
		Standard deviation $\sigma_{\rm g}$ & $0.05\times\nu_{\rm max,guess}$ to $0.5\times\nu_{\rm max,guess}$\\
		\hline
	\end{tabular}
\end{table}

\subsection{Solar-like oscillations}
\label{sec:osc}

\begin{figure}
	\includegraphics[width=\columnwidth]{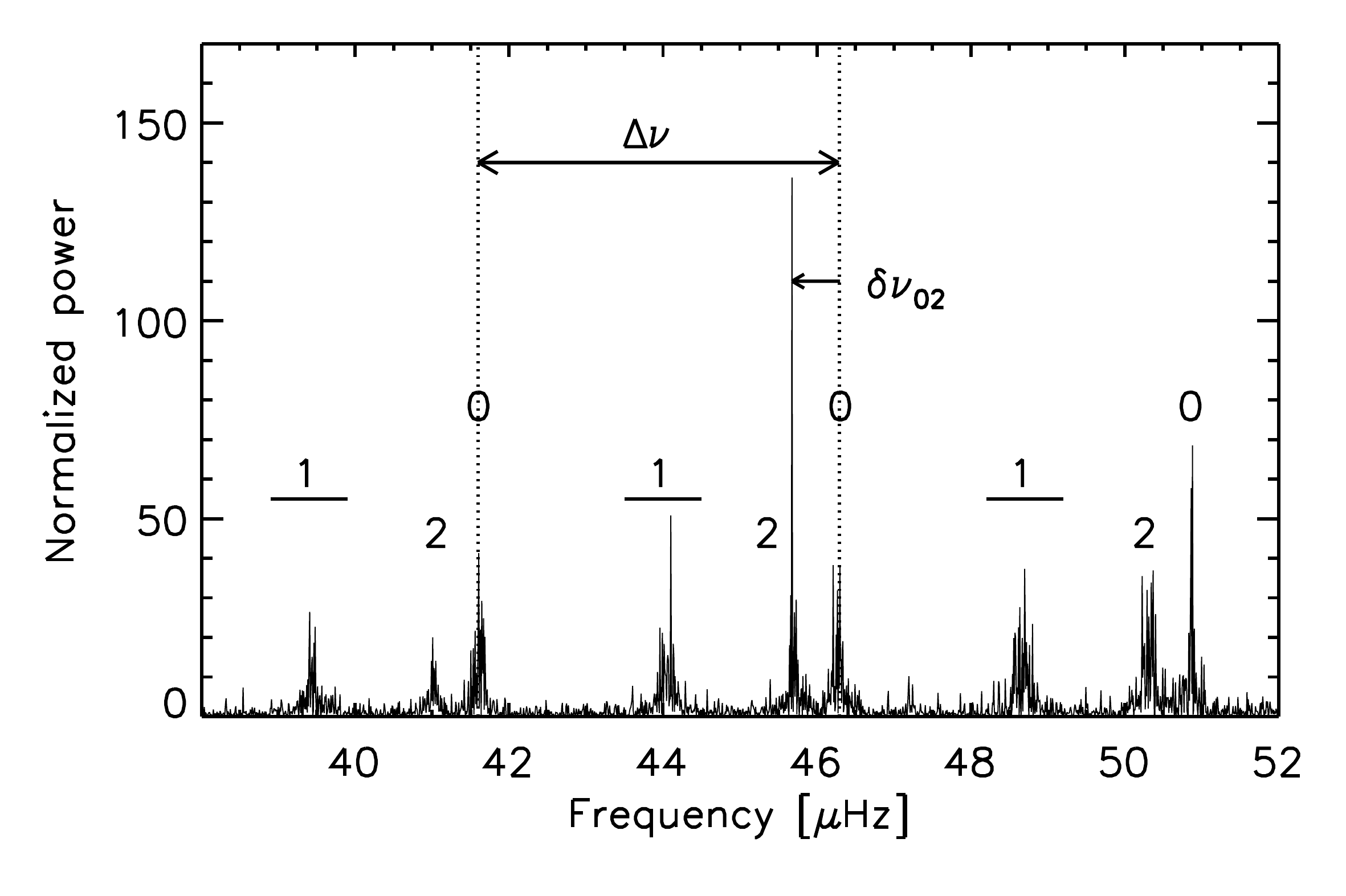}
    \caption{Normalized (by background, see Section~\ref{sec:bg}) Fourier power density spectrum of KIC\,8410637 centered around the frequency of maximum oscillation power. The large frequency separation $\Delta\nu$ and the small frequency separation $\delta\nu_{02}$ are indicated for some modes only. The dominant peaks represent modes of spherical degree $\ell=0,1$, and 2, as indicated.}
    \label{fig:fig08}
\end{figure}

The Fourier spectrum of solar-like oscillators consists of several overtones of radial order ($n$) and spherical degree ($\ell$) modes. A zoom of the individual oscillation modes is shown in Figure~\ref{fig:fig08}. The dominant peaks are arranged in a well-defined sequence, which forms in an asymptotic approximation a so-called universal pattern \citep{1980tas,2011mos}. This pattern reveals the structure of the radial and non-radial modes. In stellar time-series observations only low spherical degree modes ($\ell\leq3$) are observable. Due to cancellation effects higher degree modes are not visible in observations of the whole stellar disk.

\subsubsection{The frequency of maximum oscillation power $\nu_{\rm max}$}

The oscillation region of red giants is visible as excess power in the PDS (e.g. for KIC\,8410637 at $\sim45\,\mu$Hz as shown in the top panel of Figure~\ref{fig:fig07}). The centre of this power excess is known as the frequency of maximum oscillation power $\nu_{\rm max}$. This global seismic parameter is one of the direct observables used for deriving the asteroseismic mean density, mass, radius and surface gravity of the red giants that we study and thus has to be obtained accurately. We derived $\nu_{\rm max}$ from a Gaussian fit to the power excess according to:
\begin{equation}
    P_{\rm g}(\nu)=P_{\rm bg}(\nu)+\eta(\nu)^2\left[\Lambda_{\rm g}\exp\left({\frac{-(\nu-\nu_{\rm max})^2}{2\sigma_{\rm g}^2}}\right)\right].
	\label{eq:gauss}
\end{equation}
Here, $\Lambda_{\rm g}$ and $\sigma_{\rm g}$ indicate the height and the standard deviation of the Gaussian.
To estimate the free parameters we applied the same Bayesian MCMC method including Metropolis-Hastings sampling which we described before in Section~\ref{sec:bg}. 
The ranges for the uniform prior distributions of the Gaussian parameters are defined in the bottom part of Table~\ref{tab:bg_par}. We used the frequency peak with the highest amplitude in the oscillation region as an initial guess for $\nu_{\rm max}$ ($\nu_{\rm max,guess}$) and we computed a first estimate of the large frequency separation ($\Delta\nu_{\rm guess}$, see Section~\ref{sec:sig_from_f}) from the relation between the frequency of maximum oscillation power and the large frequency spacing \citep{2009hek,2009stel,2010mos}. Since the global background was determined in a preceding step, we kept the parameters of $P_{\rm bg}(\nu)$ (Eq.~\ref{eq:granbg}) in the Gaussian model (Eq.~\ref{eq:gauss}) fixed. 
The Gaussian fits to the power excesses of KIC\,8410637, KIC\,5640750 and KIC\,9540226 are shown in Figure~\ref{fig:fig07}. These are based on the model parameters that are reported in Table~\ref{tab:bg_results}, as computed from the MCMC algorithm. 

\subsubsection{Determination of individual frequencies}
\label{sec:pb}

\begin{figure*}
	\includegraphics[width=2\columnwidth]{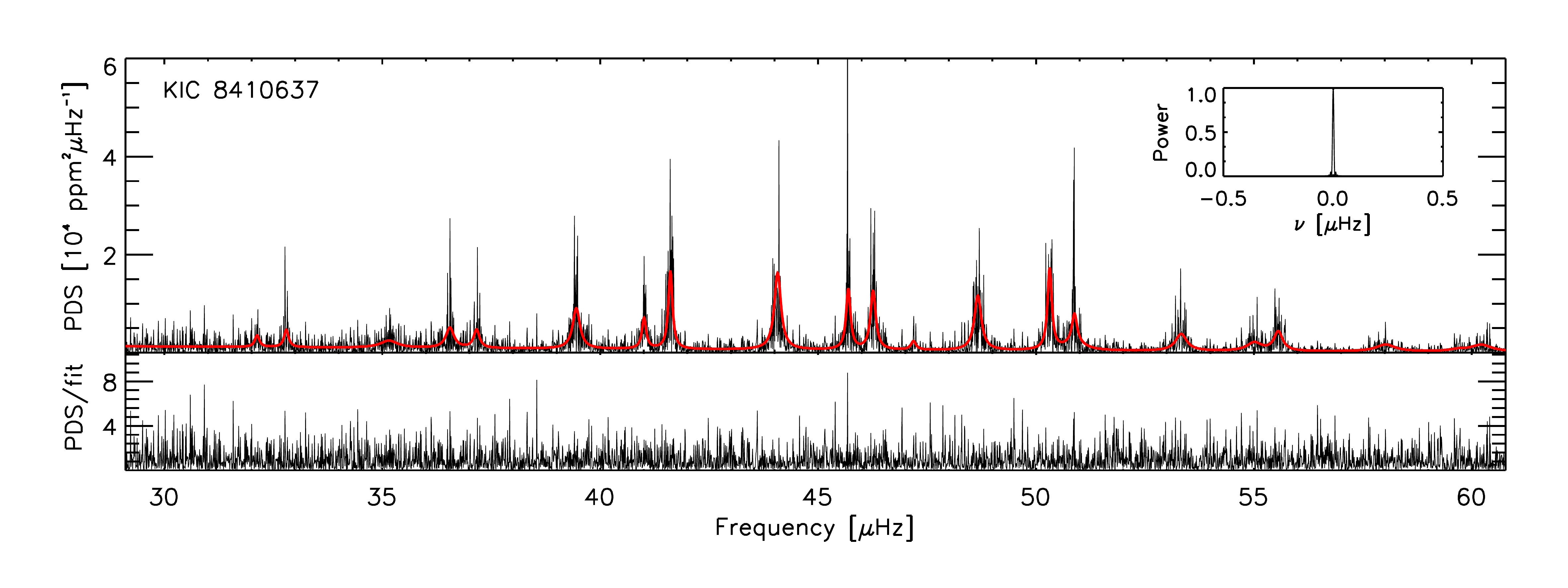}
	\includegraphics[width=2\columnwidth]{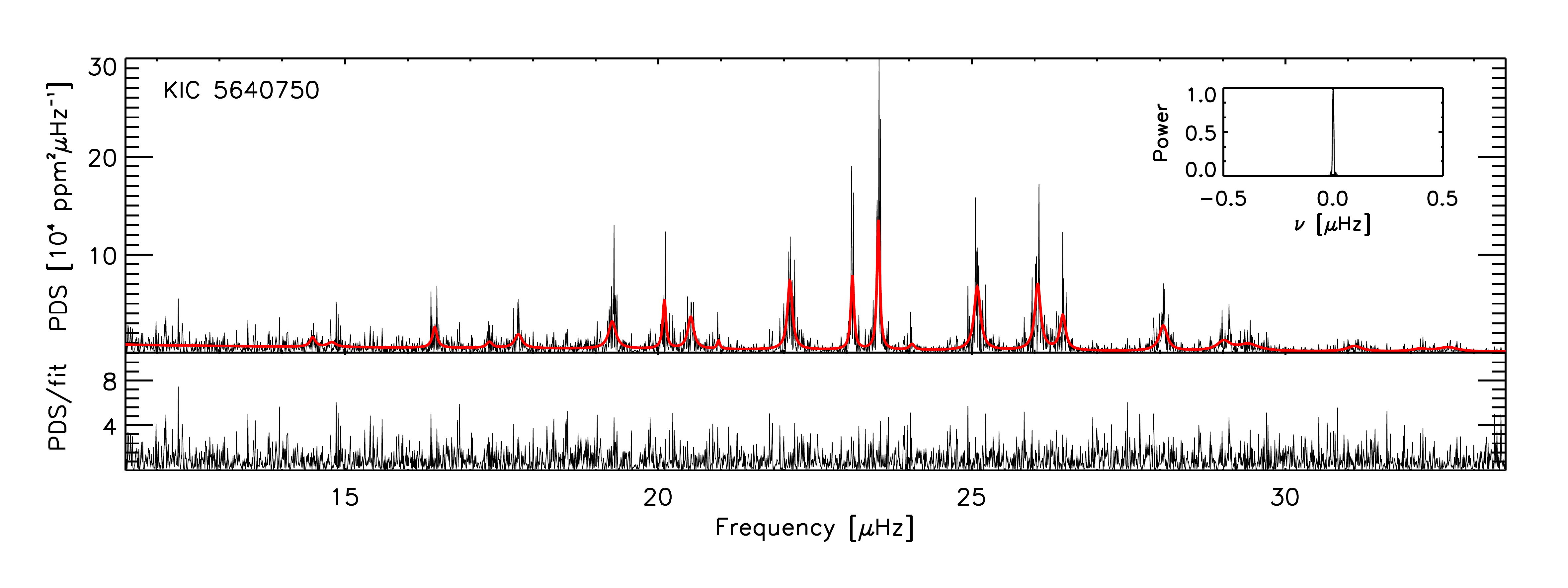}
	\includegraphics[width=2\columnwidth]{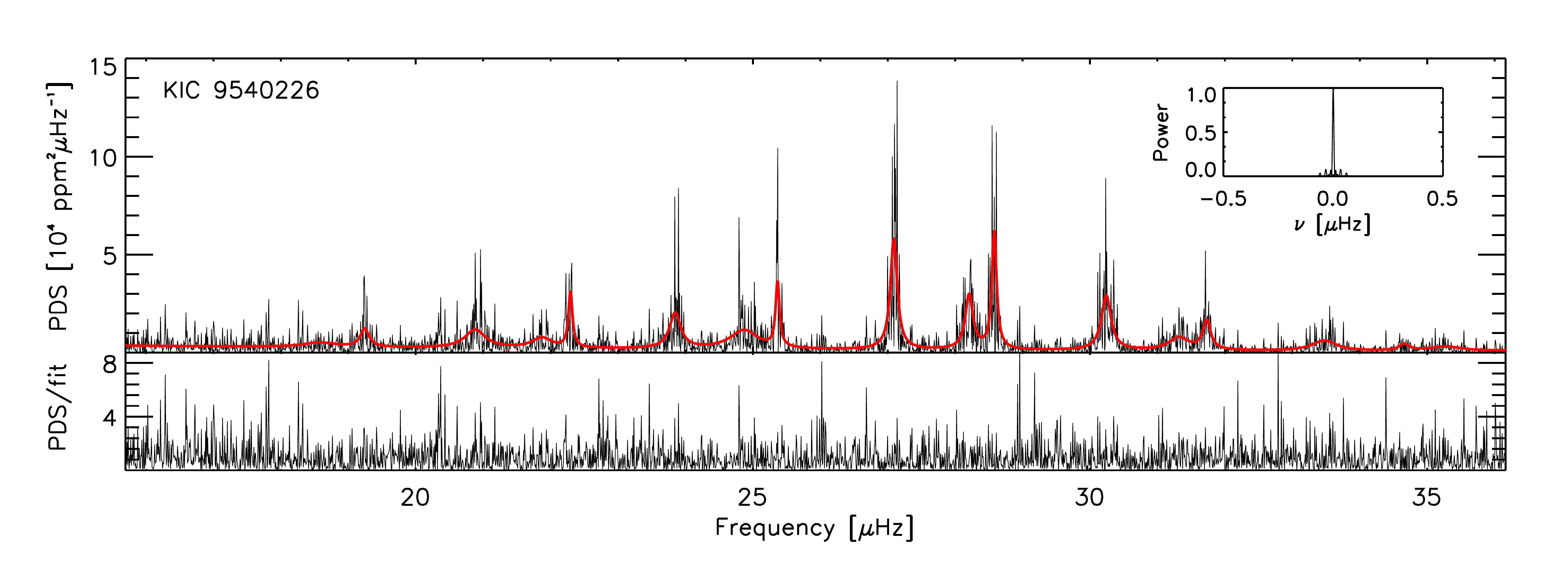}
    \caption{Fourier power density spectra (in black) of KIC\,8410637 (top), KIC\,5640750 (middle) and KIC\,9540226 (bottom) in the frequency range of the oscillations. The red solid lines represent the fits to the modes. The spectral window functions are shown in the insets in each panel. The smaller panels below each spectrum show the residuals of the peakbagging fits.}
    \label{fig:fig09}
\end{figure*}

Individual frequencies of oscillation modes contain valuable information about the stellar properties and provide essential constraints for detailed stellar modelling. In asteroseismology, the extraction of frequencies is often referred to as ``peakbagging'' analysis. Our aim is to extract all significant oscillation modes from the power density spectrum to calculate the mean large frequency spacing, which in combination with $\nu_{\rm max}$ and $T_{\rm eff}$ provides access to the stellar parameters of red-giant stars through so-called scaling relations \citep{1986ulr,1991bro,1995kje}.
Since frequencies with large power are found around the frequency of maximum oscillation power, we restricted the peakbagging analysis to the frequency range covering $\nu_{\rm max}\pm\,4\,\Delta\nu$. In this region, we used the asymptotic relation \citep{1980tas,2011mos} to obtain the spherical degree and initial frequencies of the modes. We only included the dominant peak of each degree per (acoustic) radial order without incorporating mixed or rotationally-split modes explicitly. These p-dominated mode frequencies are necessary to compute the mean large and small frequency separations. 
The resulting set of modes were simultaneously fit with Lorentzian profiles \citep[e.g.][]{1990and,2015cor}:
\begin{equation}
    P_{\rm peaks}(\nu)=P_{\rm bg}(\nu)+\eta(\nu)^2\left[\sum^{\rm n}_{\rm i=1}\frac{A_{\rm i}^2/(\pi\Gamma_{\rm i})}{1+4(\frac{\nu-\nu_{\rm i}}{\Gamma_{\rm i}})^2}\right].
	\label{eq:lorentzian}
\end{equation}
Each Lorentzian $i$ consists of a central mode frequency $\nu_{\rm i}$, mode amplitude $A_{\rm i}$ and mode linewidth $\Gamma_{\rm i}$. For the peakbagging we kept the global background parameters (Eq.~\ref{eq:granbg}) fixed and considered uniform prior distributions for the variables representing the Lorentzian profiles.
By using Metropolis-Hastings sampling in our framework of a MCMC simulation (see Section~\ref{sec:bg} for more details), we explored the parameter space of about 54 free parameters on average per star. Figure~\ref{fig:fig09} shows the global peakbagging fits to the frequency range of the oscillations as well as the spectral window functions and the residuals of the fits. Unresolved frequency peaks were also excluded from this analysis since Lorentzian profiles are not appropriate for fitting them.
Due to low S/N, we also omitted some of the outermost modes which achieved poor fits and ambiguous posterior probability distributions for the sampled parameters. Median values for all significant mode frequencies, mode widths and mode amplitudes (Eq.~\ref{eq:lorentzian}) that were computed with our fitting method are listed in Tables~\ref{tab:pb84}, \ref{tab:pb56} and \ref{tab:pb95} for the three red giants studied here.

\begin{table*}
	\centering
		\caption{Median values and corresponding 68~per cent credible interval for the global background (Eq.~\ref{eq:granbg}) and Gaussian (Eq.~\ref{eq:gauss}) parameters for the three red giants investigated here.}
	\label{tab:bg_results}
	\begin{tabular}{c l r l r l r l c c} 
		\hline\hline
		\multicolumn{1}{c}{KIC} & \multicolumn{1}{c}{$a_1$ [ppm]} & \multicolumn{1}{c}{$b_1$ [$\mu$Hz]} & \multicolumn{1}{c}{$a_2$ [ppm]} & \multicolumn{1}{c}{$b_2$ [$\mu$Hz]} & \multicolumn{3}{c}{$\Lambda_{\rm g}$ [ppm$^2\mu$Hz$^{-1}$]} & \multicolumn{1}{c}{$\nu_{\rm max}$ [$\mu$Hz]} & \multicolumn{1}{c}{$\sigma_{\rm g}$ [$\mu$Hz]}\\
		\hline
8410637 & $295 \pm 6$ & $11.3 \pm 0.4$ & $269 \pm 7$ & $43.5 \pm 0.9$ & &1743 & \HT $\pm$  67 & $46.4 \pm 0.3$ & $7.1 \pm 0.2$\\
5640750 & $477 \pm 22$ & $6.5 \pm 0.4$ & $380 \pm 36$ & $24.7 \pm 1.5$ & & 9732 & \HT $\pm$ 483 & $24.1 \pm 0.2$ & $3.6 \pm 0.1$\\
9540226 & $416\pm 17$ & $8.2\pm 0.5$ & $303 \pm 30$ & $32.1 \pm 2.3$ & & 5806 & \HT $\pm$ 257 & $26.7 \pm 0.2$ & $4.6 \pm 0.1$\\
		\hline
	\end{tabular}
\end{table*}

\subsubsection{Signatures derived from individual frequencies}
\label{sec:sig_from_f}

In the current study, we use the individual mode frequencies to derive the mean large and small frequency separations. The large frequency separation $\Delta\nu_{\rm n,\ell}$ is the spacing between oscillation modes of the same spherical degree ($\ell$) and consecutive radial order ($n$). The large frequency spacing is related to the sound travel time across the stellar diameter and thus to the mean density of the star. 
We computed the global mean large frequency separation ($\Delta\nu$) from a linear weighted fit to the frequencies of all fitted $\ell=0$ modes versus radial order that are reported in Tables~\ref{tab:pb84}, \ref{tab:pb56} and \ref{tab:pb95}. Each fit was weighted according to the uncertainties of the individual frequencies as derived from the peakbagging analysis. The slope of this linear fit corresponds to $\Delta\nu$ and the intercept refers to the offset $\epsilon$ in the asymptotic relation \citep{1980tas} multiplied with $\Delta\nu$. Based on the central three radial $\ell=0$ modes we also calculated local values of $\Delta\nu_{\rm c}$ and $\epsilon_{\rm c}$ that can be used as an indicator for the evolutionary stage of red-giant stars (see Section~\ref{sec:evo}).

Other parameters of interest are the mean small frequency separations $\delta\nu_{02}$, i.e. the frequency difference between $\ell=0$ and $\ell=2$ modes, and $\delta\nu_{01}$, i.e. the offset of the $\ell=1$ modes from the midpoint between consecutive $\ell=0$ modes. The small frequency spacings have some sensitivity in the stellar core of main-sequence stars and possibly for red giants they provide some information about their evolutionary state \citep[e.g.][]{2012cor,2017han}. For each couple of modes we obtained estimates of these frequency spacings by using all significant frequencies and we adopted the weighted mean of these measurements as mean small frequency separations $\delta\nu_{02}$ and $\delta\nu_{01}$.

The large and small frequency separations change with evolution and can be used to infer stellar properties of stars showing solar-like oscillations. We report the global seismic parameters for KIC\,8410637, KIC\,5640750 and KIC\,9540226 in Table~\ref{tab:dnu_results}. We note that radial and non-radial modes are used to compute the small frequency separations, hence these measurements can also be perturbed by mixed modes. With $\Delta\nu$ known, we constructed so-called \'{e}chelle diagrams \citep{1983gre}, in which we detect three clear ridges of $\ell=0,1,2$ modes and several detections of $\ell=3$ modes (see Appendix~\ref{sec:ech} and Figure~\ref{fig:fig13}). These diagrams are consistent with the mode identification of the asymptotic relation.

In addition to the large frequency separation, which represents the first frequency difference, we also investigated the second frequency difference for acoustic glitch signatures (see Appendix~\ref{sec:d2nu}).

In Figure~\ref{fig:fig10} we show the comparison between $\nu_{\rm max}$ and $\Delta\nu$ derived from our analysis procedures with the results from previous asteroseismic studies. For all three red giants we observe small variations of the order of a few per cent in the derived parameter estimates, which are partly caused by different analysis procedures and different datasets. We find the local mean large frequency separations ($\Delta\nu_{\rm c}$) to have a larger value than the global mean large frequency separations ($\Delta\nu$). The difference in their computations is the frequency range that is used, which can cause a change in the $\Delta\nu$ value and can be linked to stellar structure changes that occur over longer scales \citep{2016hek}.
In Figure~\ref{fig:fig10} we also show different $\Delta\nu - \nu_{\rm max}$ relations that were observed for field and cluster giants \citep{2011hek}. 
KIC\,8410637, KIC\,5640750 and KIC\,9540226 follow such relations and their stellar parameters are in line with the mass ranges observed for the cluster stars.

\begin{table*}
	\centering
	\caption{Weighted mean $\Delta\nu$ (global and local), small frequency separations, $\delta\nu_{02}$ and $\delta\nu_{01}$, and local offset $\epsilon_{\rm c}$ computed from the frequencies obtained from the peakbagging analysis (see Section~\ref{sec:pb}).}
	\label{tab:dnu_results}
	\begin{tabular}{lccccc} 
		\hline\hline
		KIC & $\Delta\nu_{\ell=0}$ [$\mu$Hz] & $\delta\nu_{02}$ [$\mu$Hz] & $\delta\nu_{01}$ [$\mu$Hz] & $\Delta\nu_{\rm c}$ [$\mu$Hz] & $\epsilon_{\rm c}$\\
		\hline
8410637 & 4.564 $\pm$ 0.004 & 0.583 $\pm$ 0.014 & -0.109 $\pm$ 0.016 & 4.620 $\pm$ 0.008 & 1.02 $\pm$ 0.02 \\
5640750 & 2.969 $\pm$ 0.006 & 0.429 $\pm$ 0.013 & -0.098 $\pm$ 0.016 & 2.978 $\pm$ 0.015 & 0.90 $\pm$ 0.02 \\
9540226 & 3.153 $\pm$ 0.006 & 0.517 $\pm$ 0.019 & -0.095 $\pm$ 0.014 & 3.192 $\pm$ 0.010 & 1.01 $\pm$ 0.02 \\
		\hline
	\end{tabular}
\end{table*}

\begin{figure}
		\includegraphics[width=\columnwidth]{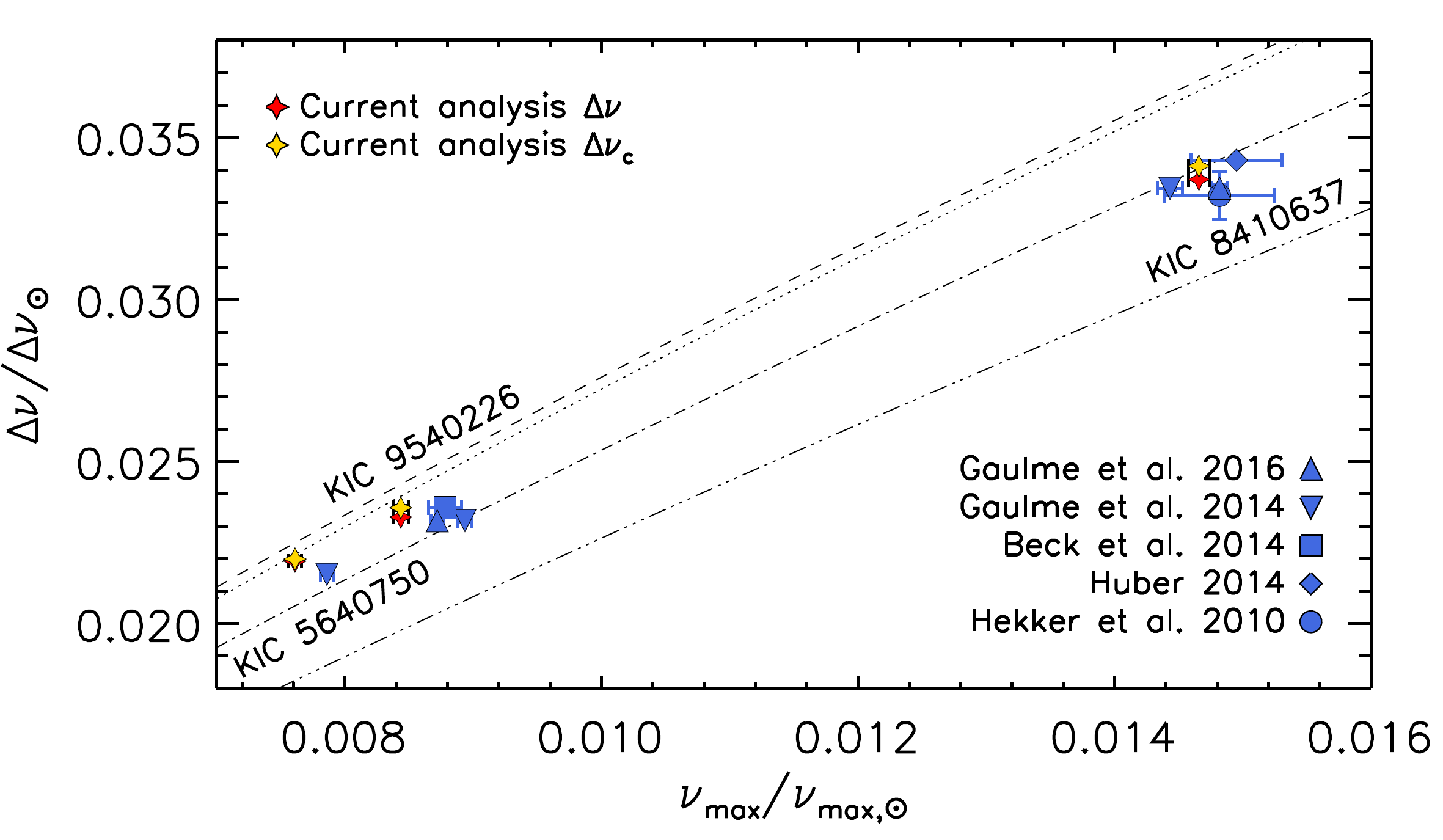}
    \caption{Global oscillation parameters $\nu_{\rm max}$ and $\Delta\nu$ for KIC\,8410637, KIC\,5640750 and KIC\,9540226. Different symbols correspond to different asteroseismic studies that were performed (see legend). The red and gold symbols correspond to estimates of the global ($\Delta\nu$) and local ($\Delta\nu_{\rm c}$) large frequency separation that we discuss in Section~\ref{sec:sig_from_f}. Different linestyles indicate the $\Delta\nu - \nu_{\rm max}$ relations derived by \citet{2011hek} for three open clusters with different masses (NGC\,6791: dash, NGC\,6819: dash dot, NGC\,6811: dash triple dot) and field stars (dotted line).}
    \label{fig:fig10}
\end{figure}

\subsubsection{Comparison with other asteroseismic fitting methods}

To check for consistency with other analysis methods, the three red giants under study were independently fit by several co-authors and their respective methods. Since these methods have been thoroughly tested on red-giant stars, they provide the means to probe the fitting procedures that we used (see Section~\ref{sec:osc} for more details).
The comparison between the global seismic parameters reported in Tables~\ref{tab:bg_results} and \ref{tab:dnu_results} with those calculated from the methods developed by \citet{2009mos,2014kal,2014cor,2015cor} are presented in Appendix \ref{subsec:diff} and Figure~\ref{fig:fig14}. In this Figure we show that the derived $\nu_{\rm max}$ and $\Delta\nu$ values from different methods are in line for the three red giants investigated here.
In addition, we checked the individual frequencies of oscillation modes (Tables~\ref{tab:pb84} -- \ref{tab:pb95}) that we obtained based on the fitting algorithm described in Section~\ref{sec:pb} with independent sets of frequencies that were extracted according to the methods developed by \citet{2014kal,2014cor,2015cor}. For each red-giant star we only report  frequencies that were independently detected by different analysis methods.

\subsection{Derivation of the stellar parameters}
\label{sec:stellarparameters}

\subsubsection{Evolutionary state of red giants}
\label{sec:evo}

Asteroseismology allows us to differentiate between red-giant branch and red-clump stars by using different oscillation features \citep[e.g.][]{2011mos,2011bed,2012kal,2014mos,2017els} that are discussed here.

All non-radial modes in red giants are mixed pressure-gravity modes, which carry information of the outer layers of the star as well as from the core \citep[e.g.][]{2011bec,2012mos}. These mixed modes can be used to distinguish between less evolved red-giant branch and more evolved red-clump stars through a study of their period spacings. \citet{2011mos} and \citet{2011bed} considered observed (bumped) period spacings of mixed dipole ($\ell=1$) modes which give an estimate of the spacings of the g-dominated modes. For hydrogen-shell burning stars on the red-giant branch they observed period spacings of the order of about 50\,s, while typical spacings of red-clump stars reached values around 100 to 300\,s. The observed period spacing is generally smaller than the so-called asymptotic period spacing which is directly related to the core size of the star. In more recent studies, \citet{2014mos,2015mos,2016vra} developed a method to measure this asymptotic period spacing and they found values of about 40 to 100\,s for red-giant-branch stars and of roughly 200 to 350\,s for more evolved stars in the red clump. 
Based on the technique described by \citet{2015mos} and implemented by \citet{2016vra}, we derived asymptotic period spacings of $58\pm3$\,s and $55\pm5$\,s for two of the red giants under study, KIC\,5640750 and KIC\,9540226, which suggests that these stars belong to the red-giant branch.

A clear advantage of this method is that it does not require individual frequencies of g-dominated mixed modes. For the three red giants investigated here only the p-dominated non-radial modes are pronounced. \citet{2014gau} found red-giant components in close binary systems where tidal interactions caused extra mode damping and even complete mode suppression. In our red giants, the lack of distinct mixed modes can also be an indication for some binary influence. In a preliminary study we investigated the presence of only p-dominated mixed modes in a small number of known red giants in binary systems as well as in a larger number of stars from the APOKASC \citep{2014pin} sample. We observed mainly p-dominated mixed modes in a large fraction of known binaries, while we detected the same feature in a significantly smaller fraction of red giants in the APOKASC sample \citep{2017the}.

For stars without distinct g-dominated mixed modes, \citet{2012kal} proposed another method to determine their evolutionary stage, which is based on the local offset ($\epsilon_{\rm c}$) of the asymptotic relation. By plotting the local large frequency separation ($\Delta\nu_{\rm c}$) against this offset, non-helium (red giant branch) and helium-burning (clump) stellar populations occupy two different parts in this $\Delta\nu_{\rm c}$ versus $\epsilon_{\rm c}$ space as shown in Figure 4 of \citet{2012kal}. 
The theoretical explanation for this relation was provided by \citet{2014jcd} and additional observational evidence was found by \citet{2015vra} based on the study of the acoustic glitches due to the second-helium ionization. They both note that the separation between red-giant branch and red-clump stars does not only relate to the different structures in their cores. The differences in the cores also cause a change in the outer stellar layers. The observed effect of this is a shift in the acoustic glitch of the helium second ionization zone that affects the oscillations. According to the $\Delta\nu_{\rm c}-\epsilon_{\rm c}$ diagram, the three red-giant stars KIC\,8410637, KIC\,5640750 and KIC\,9540226 are located on the red-giant branch. For KIC\,9540226, the identification of the evolutionary state, from both mixed modes and the local offset of the asymptotic relation, agrees with the findings of \citet{2014bec}. They found that nearly all heartbeat stars in their sample are unambiguously hydrogen-shell burning stars. Therefore they identified this as a selection effect through stellar evolution as those systems are likely to undergo a common-envelope phase and eventually eject the outer envelope before they can reach the tip of the red giant branch.

\subsubsection{Asteroseismology: Direct method}
\label{sec:sr}

\begin{table}
	\centering
	\caption{Stellar parameters obtained from asteroseismic scaling relations. Whenever the solar symbol is shown, we used solar values derived in this work and presented in Section \ref{sec:sr}.}
	\label{tab:sr_par}
	\begin{tabular}{lcccc} 
		\hline\hline
		KIC & $M$ [$\rm M_{\sun}$] & $R$ [$\rm R_{\sun}$] & $\bar{\rho}$ [$\rm \bar{\rho}_{\sun}\times10^{-3}$] & $\log g$ (cgs)\\
		\hline
		\multicolumn{5}{|l|}{Scaling relations (SR) + $\nu_{\rm max,ref}=\nu_{\rm max,\sun}$ and $\Delta\nu_{\rm ref}=\Delta\nu_{\sun}$ (SR$_{\rm a}$)}\\
		\hline
		8410637 & $1.74\pm 0.06$ & $11.53\pm 0.15$ & $1.136\pm 0.006$ & $2.555\pm 0.005$\\
		5640750 & $1.33 \pm 0.05$ & $14.02 \pm 0.20$ & $0.481 \pm 0.003$ & $2.267 \pm 0.005$\\
		9540226 & $1.45\pm 0.06$ & $13.87\pm 0.20$ & $0.542\pm 0.003$ & $2.314\pm 0.006$\\
		\hline
		\multicolumn{5}{|l|}{SR + $\nu_{\rm max,ref}=\nu_{\rm max,\sun}$ and $\Delta\nu_{\rm ref}$ from Gug16 (SR$_{\rm b}$)}\\
		\hline
		8410637 & $1.62\pm 0.06$ & $11.12\pm 0.13$ & $1.178\pm 0.002$ & $2.555\pm 0.005$\\		
		5640750 & $1.19\pm 0.05$  & $13.31\pm 0.17$ & $0.506\pm 0.002$ & $2.267\pm 0.005$\\
		9540226 & $1.31\pm 0.05$ & $13.22\pm 0.18$ & $0.569\pm 0.002$ & $2.314\pm 0.006$\\
		\hline
		\multicolumn{5}{|l|}{SR + $\nu_{\rm max,ref}=\nu_{\rm max,\sun}$ and $\Delta\nu_{\rm ref}$ from Gug17 (SR$_{\rm c}$)}\\
		\hline
		8410637 & $1.61\pm 0.06$ & $11.08\pm 0.13$ & $1.182\pm 0.002$ & $2.555\pm 0.005$\\		
		5640750 & $1.17\pm 0.04$  & $13.20\pm 0.17$ & $0.511\pm 0.002$ & $2.267\pm 0.005$\\
		9540226 & $1.30\pm 0.05$ & $13.14\pm 0.17$ & $0.572\pm 0.002$ & $2.314\pm 0.006$\\
		\hline
		\multicolumn{5}{|l|}{SR + $\nu_{\rm max,ref}=\nu_{\rm max,\sun}$ and $\Delta\nu_{\rm ref}=\Delta\nu_{\rm ref,emp}\sim 131\,\mu$Hz (SR$_{\rm emp}$, Section~\ref{sec:dnu_emp})}\\
		\hline
		8410637 & $1.51 \pm 0.07$ & $10.75 \pm 0.20$ & $1.218\pm 0.018$ & $2.555\pm 0.005$\\		
		5640750 & $1.15 \pm 0.06$ & $13.08 \pm 0.26$ & $0.515\pm 0.008$ & $2.267\pm 0.005$\\
		9540226 & $1.26 \pm 0.06$ & $12.94 \pm 0.25$ & $0.581\pm 0.009$ & $2.314\pm 0.006$\\
		\hline
	\end{tabular}
\end{table}

To apply asteroseismic methods to calculate the stellar parameters of a solar-like oscillating star, we need to know the star's effective temperature and metallicity.  In Section~\ref{sec:atmos} we accurately estimated these parameters from the analysis of the disentangled spectra. The effective temperatures and metallicities for the red-giant components are summarized in Table~\ref{tab:atmos}.

The frequency of maximum oscillation power is proportional to the acoustic cut-off frequency with $\nu_{\rm max}$ defined as \citep{1991bro,1995kje}:
\begin{equation}
\begin{aligned}
\nu_{\rm max}&\simeq\frac{M/\rm M_{\sun}}{(R/{\rm R}_{ \sun})^2\sqrt{T_{\rm eff}/T_{\rm eff, ref}}}\nu_{\rm max,ref} \\
&\simeq\frac{g/\rm g_{\sun}}{\sqrt{T_{\rm eff}/T_{\rm eff,ref}}}\nu_{\rm max,ref}.
	\label{eq:numax_rel}
	\end{aligned}
\end{equation}
The large frequency separation scales with the sound travel time across the stellar diameter and is therefore a measure of the mean density of the star \citep{1986ulr}:
\begin{equation}
\begin{aligned}
\Delta\nu&\simeq\left(\frac{M}{\rm M_{ \sun}}\right)^{0.5}\left(\frac{R}{\rm R_{ \sun}}\right)^{-1.5}\Delta\nu_{\rm ref} \\
&\simeq\sqrt{\bar{\rho}/\rm \bar{\rho}_{\sun}} \Delta\nu_{\rm ref}.
	\label{eq:dnu_rel}
\end{aligned}
\end{equation}
These equations represent the well-known asteroseismic scaling relations. Parameters $M,\,R,\,g$ and $\bar{\rho}$ are given in solar units and ``ref'' refers to a reference value. By combining these two relations (Eq.~\ref{eq:numax_rel} and \ref{eq:dnu_rel}) we can compute an estimate of the asteroseismic stellar parameters:
\begin{equation}
\begin{aligned}
        \left(\frac{\bar{\rho}}{\rm \bar{\rho}_{\sun}}\right)&\simeq\left(\frac{\Delta\nu}{\Delta\nu_{\rm ref}}\right)^2,\\
        \left(\frac{g}{\rm g_{\sun}}\right)&\simeq\left(\frac{\nu_{\rm max}}{\nu_{\rm max,ref}}\right)\left(\frac{T_{\rm eff}}{T_{\rm eff,ref}}\right)^{0.5},\\
    \left(\frac{R}{\rm R_{\sun}}\right)&\simeq\left(\frac{\nu_{\rm max}}{\nu_{\rm max,ref}}\right)\left(\frac{\Delta\nu}{\Delta\nu_{\rm ref}}\right)^{-2}\left(\frac{T_{\rm eff}}{T_{\rm eff,ref}}\right)^{0.5},\\
        \left(\frac{M}{\rm M_{\sun}}\right)&\simeq\left(\frac{\nu_{\rm max}}{\nu_{\rm max,ref}}\right)^3\left(\frac{\Delta\nu}{\Delta\nu_{\rm ref}}\right)^{-4}\left(\frac{T_{\rm eff}}{T_{\rm eff,ref}}\right)^{1.5}.\\
\end{aligned}
	\label{eq:scalrel}
\end{equation}
For these relations the Sun is often used as reference star. To obtain solar values in a consistent way we determined the global seismic parameters of the Sun with the same procedures as applied to the stellar data (see Sections~\ref{sec:bg} and \ref{sec:osc}). The only difference was the number of granulation background components that were used for the solar background model. We find a better fit to the Fourier spectrum of the Sun if we include a third granulation component. We used full-disk integrated light measurements of the Sun from the VIRGO experiment \citep{1995fro} onboard the ESA/NASA Solar and Heliospheric Observatory (SOHO).
To analyze the oscillations of the Sun, we used red and green data from 12~years of continuous solar observations with a cadence of one minute. The time span of these observations covers a whole solar cycle during which the p-mode parameters are known to vary with the solar activity \citep[e.g.][]{1990lib,2002jim}. We divided the data into subsets of four years with a one year step to mimic the observation time span of the nominal {\it Kepler} mission. Then, we analysed each subset separately. From the results of all subsets we computed the following mean solar values of $\nu_{\rm max,\sun}=3166\pm6\,\mu$Hz and $\Delta\nu_{\rm \sun}=135.4\pm0.3\,\mu$Hz. In terms of spectroscopic measurements of the Sun, we make use of the nominal solar effective temperature of $\mathcal{T}^{\rm N}_{\rm eff,\sun}=5771.8\pm0.7$\,K \citep{2015mam,2016prs}.
Using these solar values as a reference, we directly determined the stellar parameters of our three red-giant stars (Eq.~\ref{eq:scalrel}, SR$_{\rm a}$ in Table~\ref{tab:sr_par}). By using a single star as a reference it is implicitly assumed that the internal properties of stars change in a homologous way with stellar evolution for all stars of different masses and metallicities \citep{2011bel,2013bel}. However, from theoretical predictions and observations we know that many structural changes occur in stars when they pass through different evolutionary stages during their lives and hence the assumption of homology does not strictly hold.
Due to these known difficulties connected with the asteroseismic scaling relations, many studies \citep[e.g.][]{2011whi,2012mig,2013mos,2013hek,2016gug,2016sha,2017rod,2017via,2017gug} tried to improve the results obtained from these relations. \citet{2016gau} tested different scaling-relation corrections for their study of 10 red-giant stars in eclipsing binary systems and found that they lead to similar results. \citet[][hereafter Gug16]{2016gug} derived a metallicity and effective temperature dependent reference function applicable to red giant branch stars in the mass and $\nu_{\rm max}$ range that we are investigating here. They showed that their reference improves the precision of mass and radius estimates by a factor of two, which translates to an accuracy of 5 per cent in mass and 2 per cent in radius. In \citet[][hereafter Gug17]{2017gug}, they expanded their method by including a mass dependence in their formulation of the $\Delta\nu$ reference. We adopted both of their methods to calculate the stellar parameters for KIC\,8410637, KIC\,5640750 and KIC\,9540226, which we present in Table~\ref{tab:sr_par} (SR$_{\rm b}$ and SR$_{\rm c}$). 

\subsubsection{Asteroseismology: Grid-based modelling}
\label{sec:gbm}

Interpolation through a precomputed grid of stellar models and finding the best fit to the observational data is another method by which stellar parameters can be determined (grid-based modelling, see \citealt{2011gai}). For the grid-based modelling (GBM) we used the canonical BASTI grid\footnote{\url{http://albione.oa-teramo.inaf.it/}} \citep{2004pie}, which spans masses from 0.5~$\rm M_{\sun}$ to 3.5~$\rm M_{\rm\sun}$ in steps of 0.05~$\rm M_{\sun}$ and metallicities of $Z=0.0001$, $0.0003$, $0.0006$, $0.001$, $0.002$, $0.004$, $0.008$, $0.01$, $0.0198$, $0.03$, and $0.04$. (The corresponding helium abundances are $Y=0.245$, $0.245$, $0.246$, $0.246$, $0.248$, $0.251$, $0.256$, $0.259$, $0.2734$, $0.288$, and $0.303$.) The BASTI grid includes models from the zero-age main sequence all the way to the asymptotic giant branch phase. The models were computed using an updated version of the code described by \citet{1997cas} and \citet{1998sal}.

We extracted stellar parameters for the stars under study from this grid using an independent implementation of the likelihood method described by \citet{2010bas}. In short, the likelihood of each model was computed given the values of some chosen set of observed parameters, in this case $\nu_{\rm max}$, $\Delta\nu$, $T_{\rm eff}$ and [M/H]. To obtain a reliable uncertainty for the derived parameters a Monte Carlo analysis was performed, in which the observed values were perturbed within their uncertainties and a new likelihood was determined. The final answer was derived from the centre and width of a Gaussian fit through the total likelihood distribution of 1\,000 perturbations.
Furthermore, we used the temperature and metallicity dependent reference function developed by Gug16 for the $\Delta\nu$ scaling relation (Eq.~\ref{eq:dnu_rel}).
Additionally, we included the fractional solar uncertainties on $\Delta\nu$ and $\nu_{\rm max}$ in the uncertainties of the $\Delta\nu$ and $\nu_{\rm max}$ values derived for the three giants investigated here to account for uncertainties in the reference values. The grid-based modelling was carried out twice once using only models on the red-giant branch and once using models in the helium-core burning phase. We only report here the results from the red giant branch models as this is the evolutionary stage of the stars according to the present asteroseismic analysis (see Section~\ref{sec:evo}).

For the grid-based modelling we used the effective temperatures given in Table~\ref{tab:atmos}, which were derived from the atmospheric analysis of the disentangled spectra. The resulting stellar parameters are listed in Table~\ref{tab:gbm_par} and shown in Figure~\ref{fig:fig11}. One advantage of grid-based modelling is that it also provides age estimates for the stars. According to our grid-based analysis KIC\,8410637, KIC\,5640750 and KIC\,9540226 have approximate ages of $1.2\pm0.6,2.8\pm0.8$, and $2.1\pm0.8$\,Gyr, respectively.

\begin{table*}
	\centering
	\caption{Asteroseismic stellar parameters obtained from grid-based modelling. Whenever the solar symbol is shown, we used solar values derived in this work and presented in Section \ref{sec:sr}.}
	\label{tab:gbm_par}
	\begin{tabular}{lccccc} 
		\hline\hline
		KIC & $M$ [$\rm M_{\sun}$] & $R$ [$\rm R_{\sun}$] & $\bar{\rho}$ [$\rm \bar{\rho}_{\sun}\times10^{-3}$] & $\log g$ (cgs) & age [Gyr]\\
		\hline
		\multicolumn{6}{|l|}{Grid-based modelling + $\nu_{\rm max,ref}=\nu_{\rm max,\sun}$ and $\Delta\nu_{\rm ref}$ from Gug16 (GBM)}\\
		\hline
		8410637 & $1.91 \pm 0.06$ & $11.83 \pm 0.19$ & $1.148\pm 0.003$ & $2.571\pm 0.003$ & $1.2\pm 0.6$\\		
		5640750 & $1.39 \pm 0.05$ & $14.10 \pm 0.53$ & $0.496\pm 0.005$ & $2.282\pm 0.005$ & $2.8\pm 0.8$\\
		9540226 & $1.52 \pm 0.06$ & $14.03 \pm 0.42$ & $0.557\pm 0.006$ & $2.329\pm 0.005$ & $2.1\pm 0.8$\\
		\hline
		\multicolumn{6}{|l|}{Grid-based modelling + $\nu_{\rm max,ref}=\nu_{\rm max,\sun}$ and $\Delta\nu_{\rm ref}=\Delta\nu_{\rm ref,emp}\sim 131\,\mu$Hz (GBM$_{\rm emp}$, Section~\ref{sec:dnu_emp})}\\
		\hline
		8410637 & $1.48 \pm 0.06$ & $10.65 \pm 0.52$ & $1.223\pm 0.035$ & $2.555\pm 0.010$ & $2.8\pm 1.2$\\		
		5640750 & $1.22 \pm 0.10$ & $13.31 \pm 0.85$ & $0.515\pm 0.016$ & $2.274\pm 0.011$ & $4.6\pm 1.9$\\
		9540226 & $1.32 \pm 0.10$ & $13.14 \pm 0.88$ & $0.581\pm 0.017$ & $2.321\pm 0.011$ & $3.5\pm 1.6$\\
		\hline
	\end{tabular}
\end{table*}

\section{Comparison between asteroseismic and dynamical stellar parameters}
\label{sec:comp}

The aim of this complementary study is to explore consistencies between stellar parameters derived from asteroseismology, i.e. the asteroseismic scaling relations and grid-based modelling, and from binary analyses.

\subsection{Comparison}

We first consider the mass, radius, mean density and logarithmic surface gravity values obtained from the binary analyses. The results are shown as the shaded boxes in Figure~\ref{fig:fig11}. The size of the box is given by the $1\sigma$ uncertainties of the individual parameters.
For KIC\,8410637 we note agreement between the different estimates with the exception of the mass and the mean density from the current study which is lower than the other values.
As previously indicated, for KIC\,5640750 there are two possible binary solutions at inconsistent values of the parameters. This ambiguity can be attributed to insufficient phase coverage of the spectroscopic observations (Section~\ref{sec:orbital}). We will shortly use the asteroseismic values to guide us in making a choice between these two options.
For KIC\,9540226 there is agreement within $1\sigma$ for the stellar masses and within $2\sigma$ for the stellar radii. Differences in these results can partly be attributed to the use of different datasets, analysis methods and possibly some signal that is still left in the orbital phase.

We now consider the asteroseismic results that are shown as the open boxes in Figure~\ref{fig:fig11}.
Results from grid-based modelling (GBM) and the application of scaling laws (SR) are shown.
Because of underlying physical principles used to compute asteroseismic estimates of mass and radius, mass and radius are correlated and the use of different $\Delta\nu$ values naturally leads to the obvious trends seen in the left panels of Figure~\ref{fig:fig11}.
We note that the uncertainties in $\bar{\rho}$ and $\log g$ for the asteroseismic results are smaller than the corresponding uncertainties of the values that are computed directly from the derived dynamical masses and radii.

From GBM we obtained larger values for the stellar masses and radii than the parameter estimates calculated from the asteroseismic scaling relations. \citet{2017tay} pointed out that GBM may modify the inferred effective temperatures in order to find a better match between observed parameters and stellar models. This is the case here. We were not able to find a matching model for the $T_{\rm eff}$ derived from the binary analyses (Table~\ref{tab:atmos}). Instead, GBM favored models with $\sim200$\,K higher effective temperatures. As a consequence, we find larger values for the logarithmic surface gravities as well as stellar masses and radii from GBM. 
We note that these results have relatively large uncertainties, because their computations also take metallicities with uncertainties into account. 
Furthermore, we obtained lower stellar masses and radii and larger mean densities when using asteroseismic scaling relations that also take the metallicity, temperature and mass dependence of the stars into account (SR$_{\rm c}$).

For KIC\,5640750 we now compare the dynamical stellar parameters from both orbital solutions with the asteroseismic values. We see that the second option (denoted as $\rm BA_2$) with $M_{\rm B}=1.158\pm0.014\,\rm M_{\sun}$, $R_{\rm B}=13.119\pm0.090\,\rm R_{\sun}$ and $\log g_{\rm B}=2.266\pm0.006$\,(cgs) is in line with what we observe for the other two red giants where the dynamical masses and radii are lower than the asteroseismic values. Despite the fact that the statistical significance of both binary solutions is almost the same, we can now give more weight to the second option which suits the overall picture of the three red giants studied here. In the further analysis, we consider only the second set of dynamical stellar parameters for KIC\,5640750.

In Figure~\ref{fig:fig11}, in addition to the metallicity-independent mass-radius correlation already alluded to, we see that there is a lack of agreement between the asteroseismic and dynamical stellar parameters for the three red giants under study. To investigate these discrepancies further, we examine the influence of different asteroseismic observables and the asteroseismic scaling relations since these are important for inferring the stellar properties. Among the parameters of interest are the global seismic parameters ($\nu_{\rm max}$, $\Delta\nu$) as well as the reference values ($\nu_{\rm max,ref}$, $\Delta\nu_{\rm ref}$).

\subsection{Empirically derived $\Delta\nu_{\rm ref, emp}$}
\label{sec:dnu_emp}

Following the determination of the asteroseismic stellar parameters in Section~\ref{sec:sr}, we now reverse the scaling relations (Eq.~\ref{eq:numax_rel}--\ref{eq:dnu_rel}) to obtain estimates for the global seismic parameters ($\nu_{\rm max},\Delta\nu$) and for the reference values ($\nu_{\rm max,ref},\Delta\nu_{\rm ref}$) to check for coherency. In both cases we use the dynamical stellar masses and radii and the spectroscopic effective temperatures as input. 

In combination with the observed reference values, i.e. from the Sun, we calculated the global seismic parameters, which for all three red-giant stars agree with the observed values within uncertainties. We subsequently computed the reference values by using the observed global seismic parameters of the red-giant stars together with the dynamical $M$, $R$ and $T_{\rm eff}$. We obtained consistent reference values for the frequency of maximum oscillation power with a mean value of $3137\pm45\,\mu$Hz which agrees with the observed solar reference reported in Section~\ref{sec:sr}.

Based on the same approach we calculated $\Delta\nu_{\rm ref}$ for the three red giants investigated here. We consistently derived lower values around a mean value of $\Delta\nu_{\rm ref,emp}=130.8\pm0.9\,\mu$Hz, which is inconsistent with the observed solar value of $135.4\pm0.3\,\mu$Hz (Section~\ref{sec:sr}).

We now consider the different $\Delta\nu_{\rm ref}$ references that we used throughout our asteroseismic analysis. 
We adopted either the observed solar value of $\sim135\,\mu$Hz or
we included the temperature, mass and metallicity dependence of the stars using corrections based on models. This latter approach led to a $\Delta\nu_{\rm ref} \sim132\,\mu$Hz based on the formulations provided by \citet{2017gug}. In the latter reference, the so-called surface effect \citep[e.g.][]{2017bal} is not included, yet it is present in the models on which the $\Delta\nu_{\rm ref}$ is based. This effect arises due to improper modelling of the near surface layers and it causes a shift in the p-mode frequencies which then also changes the value of the mean large frequency separation.

In short, if we consider a star with one solar mass, an effective temperature of 5772\,K and $\rm [Fe/H]=0.0$, i.e. the Sun, we would derive a large frequency separation of about $136\,\mu$Hz from a solar model. The reason why this is different from the observed solar value of about $135\,\mu$Hz is due to the surface effect. Thus, we may still have to decrease $\Delta\nu_{\rm ref}$ from Gug17 by $\sim1\,\mu$Hz, i.e. $\sim1$ per cent, because we would expect a shift of this magnitude for the reference value. This would then be very close to our empirically determined $\Delta\nu_{\rm ref,emp}$ value. A detailed analysis of the scale of the surface effect in KIC\,8410637, KIC\,5640750 and KIC\,9540226 is presented in \citet{2017bbal}. It is worth noting that due to the surface effect a shift in frequency between $\sim 0.1$ and $0.3\,\mu$Hz at $\nu_{\rm max}$ is observed for the three red-giant stars studied here, which show oscillations in the range between $\sim 24$ and $46\,\mu$Hz. The magnitude of these frequency differences is similar to the one percent that was found for the model of the Sun.

Most recently, \citet{2018bro} also reported consistencies between asteroseismic and dynamical stellar parameters when using a model-dependent theoretical correction factor that was proposed by \citet{2017rod} instead of the usual solar reference values. Their correction of $\Delta\nu_{\rm ref}$ is of the same order of magnitude as the one that we present in the current study.

\begin{figure*}
	\includegraphics[width=2\columnwidth]{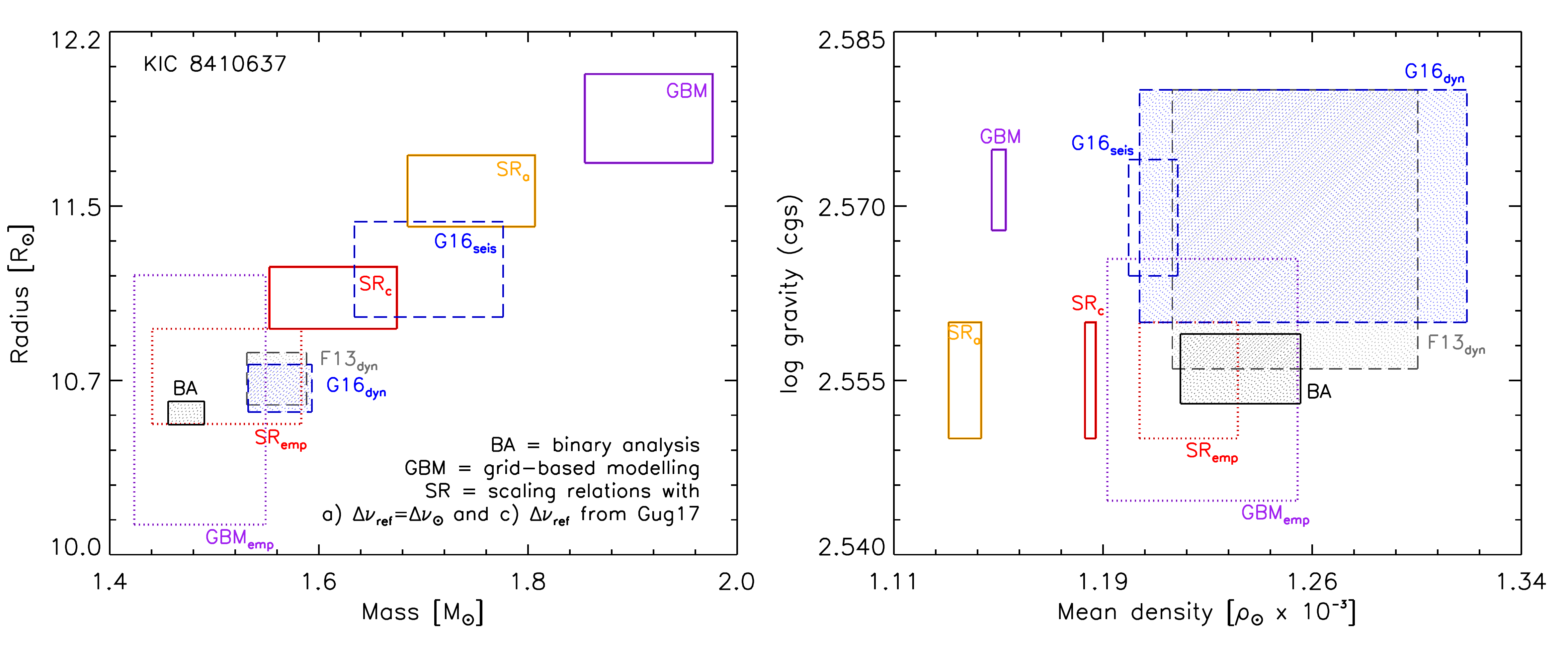}
	\includegraphics[width=2\columnwidth]{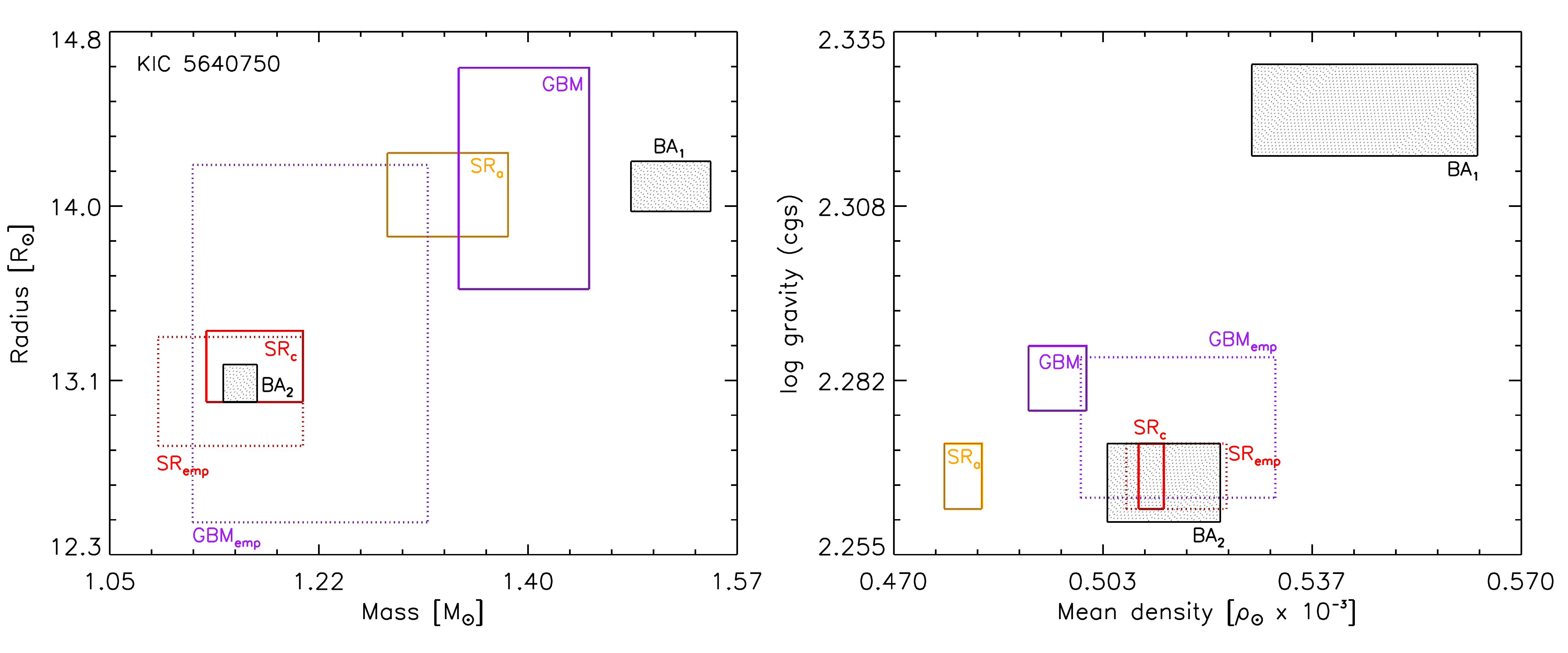}
	\includegraphics[width=2\columnwidth]{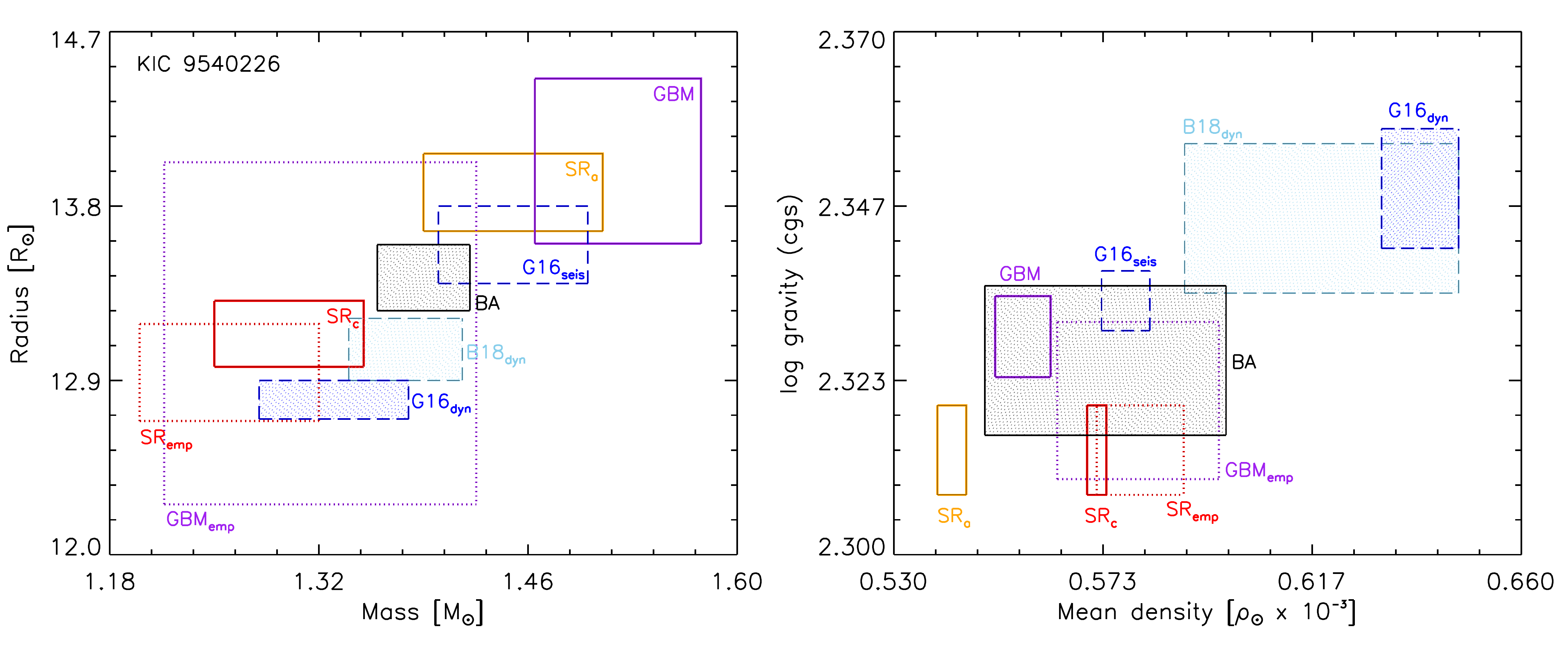}
    \caption{Stellar mass versus radius (left) and mean density versus logarithmic surface gravity (right) for KIC\,8410637 (top), KIC\,5640750 (middle) and KIC\,9540226 (bottom) derived from binary orbit analyses (filled boxes) and from asteroseismology (open boxes). Results from the current analysis are shown with solid and dotted lines. The latter refers to stellar parameters determined on the basis of an empirical $\Delta\nu_{\rm ref,emp}$ reference (Section~\ref{sec:dnu_emp}). Dashed lines represent already published values of the stellar parameters (\citet[][F13$_{\rm dyn}$]{2013fra}, \citet[][G16$_{\rm dyn;seis}$]{2016gau}, \citet[][B18$_{\rm dyn}$]{2018bro}). For different colours see legend and Tables~\ref{tab:sysprop}, \ref{tab:sr_par} and \ref{tab:gbm_par}.}
    \label{fig:fig11}
\end{figure*}

From the combined asteroseismic and binary analysis we derived an empirical $\Delta\nu_{\rm ref,emp}$ reference that seems to be more appropriate for these three specific red-giant stars instead of the commonly used solar reference.
We used this in the asteroseismic scaling relations (Eq.~\ref{eq:scalrel}) and determined revised stellar parameters for KIC\,8410637, KIC\,5640750 and KIC\,9540226. 
The asteroseismic stellar masses and radii from both the scaling relations and GBM are lower and they are now in line with the dynamical stellar parameters. We show the revised stellar parameters with dotted boxes in Figure~\ref{fig:fig11}. The uncertainties for these stellar parameters are larger due to a larger uncertainty in the $\Delta\nu_{\rm ref,emp}$ reference. In the GBM analysis, these lower masses have an impact on the ages of these red-giant stars which are now on average 1.5\,Gyr older (Table~\ref{tab:gbm_par}).
In addition we note that our conclusions do not change when using slightly different $\Delta\nu$ values (such as the ones given in Figure~\ref{fig:fig14}) for the determination of the stellar parameters of the three red giants studied here.

\begin{table*}
	\centering
	\caption{Relative differences in per cent of the asteroseismic stellar parameters (Section~\ref{sec:sr} and \ref{sec:gbm}) with respect to the binary solution (Section~\ref{sec:ecl_model}). In the case of KIC\,5640750 we compare the dynamical stellar parameters that are based on the second binary solution.}
	\label{tab:percent_diff}
	\begin{tabular}{lcccccccccccccc}
		\hline\hline
		Method & \multicolumn{4}{c}{KIC\,8410637} & \multicolumn{6}{c}{KIC\,5640750} & \multicolumn{4}{c}{KIC\,9540226}\\
		\hline
		& $M$ & $R$ & $\bar{\rho}$ & $\log g$ & & $M$ & $R$ & $\bar{\rho}$ & $\log g$ & & $M$ & $R$ & $\bar{\rho}$ & $\log g$\\
		\hline
	SR$_{\rm a}$ & 18\% & 9\% & 8\% & <1\% & & 15\% & 7\% & 6\% & <1\% & & 4\% & 3\% & 6\% & <1\%\\
	SR$_{\rm b}$ & 10\% & 5\% & 5\% & <1\% & & 3\% & 2\% & 1\% & <1\% & & 6\% & 2\% & <1\% & <1\%\\
	SR$_{\rm c}$ & 9\% & 5\% & 4\% & <1\% & & 1\% & <1\% & <1\% & <1\% & & 7\% & 2\% & <1\% & <1\%\\
	SR$_{\rm emp}$ & 3\% & 2\% & 2\% & <1\% & & <1\% & <1\% & <1\% & <1\% & & 9\% & 4\% & 1\% & <1\%\\
		\hline
	GBM &  30\% & 12\% & 7\% & <1\% & & 20\% & 8\% & 3\% & 1\% & &  9\% & 5\% & 3\% & <1\% \\
	GBM$_{\rm emp}$ & <1\% & <1\% & 1\% & <1\% & & 5\% & 2\% & <1\% & <1\% & & 5\% & 2\% & 1\% & <1\%\\
		\hline
	\end{tabular}
\end{table*}

\subsubsection{Consistency check with a larger sample of red giants}
\label{sec:dnu_app}

To further test the empirically derived reference value, we applied it to a number of known oscillating red-giant-branch components in eclipsing binary systems. By using published $\nu_{\rm max},\Delta\nu$ and $T_{\rm eff}$ we recomputed the asteroseismic stellar parameters for nine stars that show solar-like oscillations in the range between $\sim 20$ and $\sim77\,\mu$Hz, i.e. KIC\,9540226 \citep{2014bec,2016gau,2018bro}, KIC\,8410637 \citep{2013fra}, KIC\,4663623, KIC\,9970396, KIC\,7037405, KIC\,5786154, KIC\,10001167, KIC\,7377422 and KIC\,8430105 \citep{2016gau}.

\begin{figure*}
	\includegraphics[width=2\columnwidth]{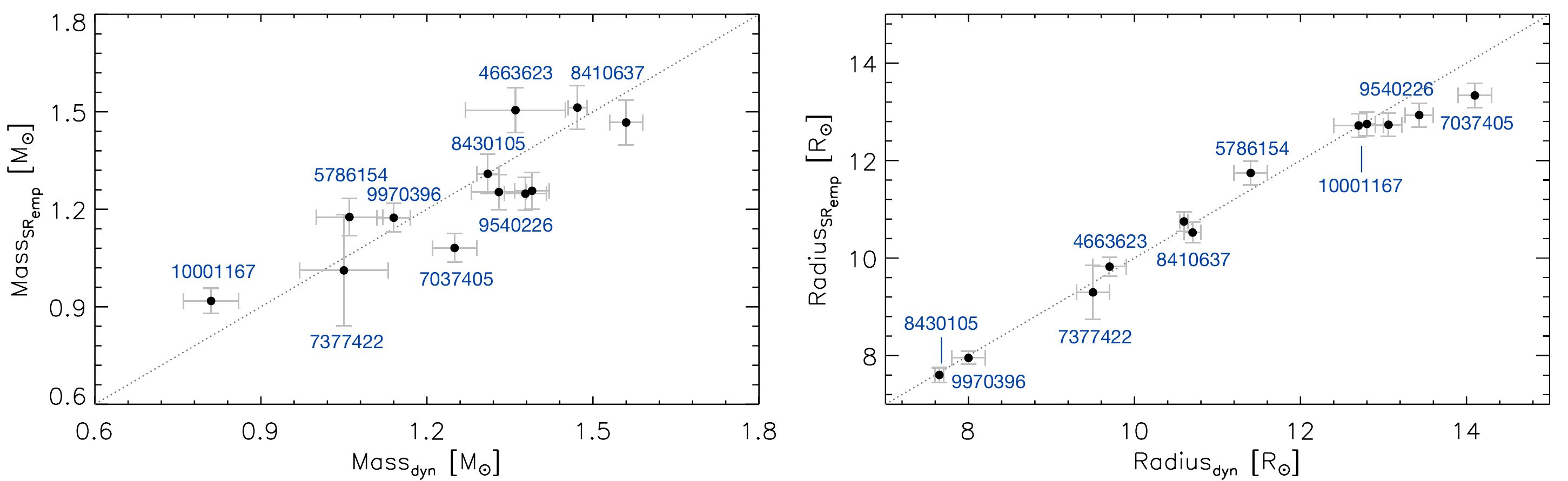}
    \caption{Asteroseismic versus dynamical masses (left) and radii (right) for nine RGB components in known eclipsing binary systems with KIC numbers indicated. Asteroseismic stellar parameters (SR$_{\rm emp}$) were determined by using $\Delta\nu_{\rm ref,emp}\sim 131\,\mu$Hz (Section~\ref{sec:dnu_emp}).}
    \label{fig:fig16}
\end{figure*}

Although we find consistencies between the derived asteroseismic and published dynamical stellar parameters, we see some scatter in the results (see Figure~\ref{fig:fig16}). This scatter is partly caused by combining global seismic parameters, effective temperatures, RVs and/or disentangled spectra that were obtained from different methods and specific calibrations. The systematics and biases that are introduced by asteroseismic techniques have been thoroughly discussed in literature \citep[e.g.][]{2011bhek,2011ver,2012hek,2014kal}.

The observed scatter in the asteroseismic results is larger than the quoted uncertainties of the dynamical stellar parameters. This shows the importance of a homogeneous analysis for all stars under study, which is expected to increase the consistency between the asteroseismic and dynamical stellar parameters. We note that \citet{2018kal} present a careful study where they test the scaling relations with dynamical stellar parameters of eclipsing binary systems and they only found six stars, for which the stellar properties are known with sufficient precision for solid conclusions. 

Despite these issues we determined a consistent mean value for the $\Delta\nu$ reference (albeit with large uncertainties) when using the published properties of these nine red-giant-branch stars, which cover a larger stellar parameter space than our original sample. This is a further indication that a $\Delta\nu_{\rm ref,emp}$ of about $131\,\mu$Hz is appropriate for stars on the red giant branch.

\section{Conclusions}
\label{sec:con}

We find agreement between the stellar parameters determined using asteroseismic scaling relations and eclipse modelling only when using a $\Delta\nu$ reference that we derived empirically and that can be justified by known physical parameters such as mass, effective temperature, metallicity and the surface effect. In Table~\ref{tab:percent_diff} we show the relative differences of the asteroseismic stellar parameters for the red-giant components of KIC\,8410637, KIC\,5640750 and KIC\,9540226 with respect to the binary solutions. As we look at a small sample of stars here, we cannot investigate the global differences between the asteroseismic scaling relations and orbital parameters. Yet, we performed a careful in-depth observational analysis for each of the red giants and we found consistent results for the three systems. 

Calculations of detailed stellar models of the three red giants studied here are on the way. The highly-precise and accurate {\it Kepler} data can be used to fit oscillation frequencies in a large number of red-giant stars, which provide additional information for stellar models. The oscillation frequencies, however, have to be corrected for surface effects before they can be compared to modelled frequencies \citep[e.g.][]{2017bal}. The surface effects in red giants are not yet fully understood, even though 3-D simulations are insightful \citep[][for instance]{2015son}. The results from detailed modelling of the surface effects of these three red giants in eclipsing binary systems using individual frequencies is presented in \citet{2017bbal}. We note however that for the red giants investigated here we require a lower $\Delta\nu$ reference value of the order of $131\,\mu$Hz when mass, temperature and metallicity dependence as well as surface effect are taken into account. This supports the idea of using $\Delta\nu_{\rm ref,emp} =130.8\pm0.9\,\mu$Hz which in our case provides consistencies between asteroseismic and dynamical stellar parameters for KIC\,8410637, KIC\,5640750 and KIC\,9540226 which are all located close to the red giant branch bump.

The current study shows the importance of $\Delta\nu_{\rm ref}$ in the asteroseismic scaling relations. Ideally, we would extend this work to more pulsating red-giant stars in eclipsing binary systems that cover a wide range in stellar parameters. In future, stars with different properties of mass, metallicity and evolutionary stages (e.g. red-giant components in eclipsing binaries that are in the red clump phase) need to be studied in order to identify appropriate reference values and to investigate the presence of other sources that could influence the stellar parameter measurements.

\section*{Acknowledgements}

We are very grateful to B. Mosser for his contribution and his valuable comments on earlier drafts of the manuscript. We also thank the anonymous referees for their constructive and useful suggestions and comments, which helped to improve the manuscript.
NT and SH received funding from the European Research Council under the European Community's Seventh Framework Programme (FP7/2007-2013) / ERC grant agreement no 338251 (StellarAges). This research was undertaken in the context of the International Max Planck Research School for Solar System Science at the University of G{\"o}ttingen. PGB acknowledges the ANR (Agence Nationale de la Recherche, France) program IDEE (n$^\circ$ ANR-12-BS05-0008) ``Interaction Des Etoiles et des Exoplanetes'' and has also received funding from the CNES grants at CEA. PGB acknowledges the support of the Spanish Minstry of Economy and Competitiveness (MINECO) under the programme 'Juan de la Cierva' (IJCI-2015-26034). EC is funded by the European Community's Seventh Framework Programme (FP7/2007-2013) under grant agreement N$^\circ$312844 (SPACEINN) and the European Union's Horizon 2020 research and innovation programme under the Marie Sklodowska-Curie grant agreement n$^\circ$ 664931. YE and WHB acknowledge the support of the UK Science and Technology Facilities Council (STFC). JS acknowledges financial support from the Leverhulme Trust in the form of a Philip Leverhulme Prize. KP is supported by the Croatian Science Foundation grant  2014-09-8656. AT acknowledges the support of the Fonds Wetenschappelijk Onderzoek - Vlaanderen (FWO) under the grant agreement G0H5416N (ERC Opvangproject). Funding for the Stellar Astrophysics Centre (SAC) is provided by The Danish National Research Foundation (Grant agreement no.: DNRF106). The research leading to these results has (partially) received funding from the European Research Council (ERC) under the European Union's Horizon 2020 research and innovation programme (grant agreement N$^\circ$670519: MAMSIE) and from the Belgian Science Policy Office (Belspo) under ESA/PRODEX grant ``PLATO mission development''. We thank S{\o}ren Frandsen and Patrick Gaulme for discussions about the RV measurements of KIC\,8410637. We also thank the technical team as well as the observers of the \textsc{Hermes} spectrograph and Mercator Telescope, operated on the island of La Palma by the Flemish Community, at the Spanish Observatorio del Roque de los Muchachos of the Instituto de Astrof\'{i}sica de Canarias.



\bibliographystyle{mnras}
\bibliography{example}



\appendix

\section{Radial velocity measurements for KIC\,8410637, KIC\,5640750 and KIC\,9540226}

We applied the cross-correlation method to all available spectra of KIC\,8410637, KIC\,5640750 and KIC\,9540226 to determine the radial velocities of the giants and their corresponding uncertainties, which are given in Tables~\ref{tab:rv84}, \ref{tab:rv56} and \ref{tab:rv95}. For completeness, we also report the RVs for the dwarf components of these systems, which were obtained by previous studies.

\begin{table}
\centering
\caption{Radial velocity measurements of the two components of KIC\,8410637. These are the measurements used in the analysis in Section\,\ref{sec:ecl_model}. They were not used in the determination of the physical properties of the system (which was done with the velocity amplitudes from the spectral disentangling analysis). The RVs in this table are those used by co-author JS in his analysis in Section\,\ref{sec:ecl_model} and also in \citet{2013fra}. However, it was retrospectively discovered that they are not the same as those published in \citet{2013fra}, and source of the discrepancy has resisted our attempts to find it (S.\ Frandsen, 2015, priv.\ comm.). In future we advise that analyses of this binary system use the results from spectral disentangling, or as a second choice the RVs published in \citet{2013fra}. We report the RVs in this table only for completeness.}
\label{tab:rv84}
\begin{tabular}[t]{crlrlr}
\hline
\hline
KIC\,8410637 & & & & &\\
\hline
HJD-2450000 & \multicolumn{2}{c}{Giant RV [\kms]}        & \multicolumn{2}{c}{Dwarf RV [\kms]}        & Source \\
\hline
5660.7131 & -54.69 & \HT $\pm$ 0.05 & -30.68 & \HT $\pm$  2.59 & {\sc Fies}    \\
5660.7362 & -54.69 & \HT $\pm$ 0.04 & -30.23 & \HT $\pm$  2.31 & {\sc Fies}    \\
5733.6205 & -60.91 & \HT $\pm$ 0.05 & -27.79 & \HT $\pm$  0.76 & {\sc Fies}    \\
5749.5119 & -62.18 & \HT $\pm$  0.05 & -27.39 & \HT $\pm$  0.74 & {\sc Fies}    \\
5762.6434 & -62.98 & \HT $\pm$  0.06 & -26.92 & \HT $\pm$  0.88 & {\sc Fies}    \\
5795.4991 & -56.90 & \HT $\pm$  0.05 & -29.78 & \HT $\pm$  1.04 & {\sc Fies}    \\
5810.4756 & -28.70 & \HT $\pm$  0.05 & -64.60 & \HT $\pm$  1.38 & {\sc Fies}    \\
5825.3478 & -11.75 & \HT $\pm$  0.05 & -85.47 & \HT $\pm$  1.21 & {\sc Fies}    \\
5828.3417 & -12.90 & \HT $\pm$  0.05 & -84.60 & \HT $\pm$  1.30 & {\sc Fies}    \\
5834.4285 & -16.01 & \HT $\pm$  0.06 & -81.75 & \HT $\pm$  1.21 & {\sc Fies}    \\
5844.3966 & -21.13 & \HT $\pm$  0.05 & -76.04 & \HT $\pm$  1.79 & {\sc Fies}    \\
5855.3398 & -25.64 & \HT $\pm$  0.05 & -68.46 & \HT $\pm$  1.59 & {\sc Fies}    \\
5886.3003 & -34.33 & \HT $\pm$  0.05 & -58.98 & \HT $\pm$  1.15 & {\sc Fies}    \\
5903.3175 & -37.68 & \HT $\pm$  0.06 & -58.22 & \HT $\pm$  1.18 & {\sc Fies}    \\
5903.3444 & -37.80 & \HT $\pm$  0.07 & -57.31 & \HT $\pm$  1.55 & {\sc Fies}    \\
5700.4998 & -58.00 & \HT $\pm$  0.05 & -29.28 & \HT $\pm$  1.14 & {\sc Ces}    \\
5726.4642 & -60.47 & \HT $\pm$  0.06 & -28.01 & \HT $\pm$  1.27 & {\sc Ces}    \\
5734.4230 & -60.87 & \HT $\pm$  0.06 & -26.41 & \HT $\pm$  2.28 & {\sc Ces}    \\
5734.5069 & -60.84 & \HT $\pm$  0.07 & -27.56 & \HT $\pm$  1.29 & {\sc Ces}    \\
5754.4370 & -62.55 & \HT $\pm$  0.06 & -26.19 & \HT $\pm$  1.19 & {\sc Ces}    \\
5793.3535 & -58.62 & \HT $\pm$  0.05 & -29.30 & \HT $\pm$  1.48 & {\sc Ces}    \\
5799.4801 & -53.04 & \HT $\pm$  0.10 & -30.89 & \HT $\pm$  1.52 & {\sc Ces}    \\
5810.4495 & -28.76 & \HT $\pm$  0.06 & -66.44 & \HT $\pm$  2.19 & {\sc Ces}    \\
5817.3353 & -14.40 & \HT $\pm$  0.08 & -83.10 & \HT $\pm$  1.85 & {\sc Ces}    \\
5850.3024 & -23.98 & \HT $\pm$  0.06 & -76.39 & \HT $\pm$  2.48 & {\sc Ces}    \\
5852.2713 & -24.85 & \HT $\pm$  0.06 & -76.17 & \HT $\pm$  3.31 & {\sc Ces}    \\
5880.2040 & -33.14 & \HT $\pm$  0.05 & -59.91 & \HT $\pm$  1.48 & {\sc Ces}    \\
5334.5342 & -61.94 & \HT $\pm$  0.04 & -27.85 & \HT $\pm$  0.82 & {\sc Hermes} \\
5334.6737 & -61.53 & \HT $\pm$  0.05 & -27.59 & \HT $\pm$  0.84 & {\sc Hermes} \\
5335.5792 & -61.90 & \HT $\pm$  0.04 & -27.90 & \HT $\pm$  0.74 & {\sc Hermes} \\
5335.6967 & -61.58 & \HT $\pm$  0.05 & -27.48 & \HT $\pm$  0.88 & {\sc Hermes} \\
5336.5676 & -61.88 & \HT $\pm$  0.05 & -27.69 & \HT $\pm$  0.78 & {\sc Hermes} \\
5336.7291 & -61.47 & \HT $\pm$  0.05 & -27.26 & \HT $\pm$  1.04 & {\sc Hermes} \\
5609.7553 & -50.64 & \HT $\pm$  0.04 & -30.81 & \HT $\pm$  1.69 & {\sc Hermes} \\
5715.4610 & -59.72 & \HT $\pm$  0.05 & -28.80 & \HT $\pm$  1.68 & {\sc Hermes} \\
5765.4437 & -63.07 & \HT $\pm$  0.05 & -26.82 & \HT $\pm$  0.98 & {\sc Hermes} \\
5779.4483 & -63.58 & \HT $\pm$  0.04 & -27.28 & \HT $\pm$  0.88 & {\sc Hermes} \\
5801.4416 & -49.85 & \HT $\pm$  0.05 & -41.61 & \HT $\pm$  2.13 & {\sc Hermes} \\
5835.3802 & -16.82 & \HT $\pm$  0.05 & -82.05 & \HT $\pm$  1.41 & {\sc Hermes} \\
5871.3083 & -30.75 & \HT $\pm$  0.05 & -62.92 & \HT $\pm$  1.61 & {\sc Hermes} \\
5888.3093 & -35.35 & \HT $\pm$  0.05 & -58.77 & \HT $\pm$  1.00 & {\sc Hermes} \\
5888.3273 & -34.80 & \HT $\pm$  0.05 & -59.05 & \HT $\pm$  1.21 & {\sc Hermes} \\
5965.7723 & -45.72 & \HT $\pm$  0.05 & -34.82 & \HT $\pm$  1.41 & {\sc Hermes} \\
\hline
\end{tabular}
\end{table}

\begin{table}
\centering
\caption{Radial velocity measurements of the red-giant component of KIC\,5640750 derived with the cross-correlation method ({\sc ccf}) from {\sc Hermes} spectra.}
\label{tab:rv56}
\begin{tabular}[t]{crlcr}
\hline
\hline
KIC\,5640750 & & & &\\
\hline
HJD-2450000 &\multicolumn{2}{c}{Giant RV [\kms]}  & S/N (\ion{Mg}{i} triplet) & Source \\
\hline
5623.7602 & -23.34 & \HT $\pm$ 0.04 & 41.80 & {\sc Hermes} \\
5765.5211 & -27.56 & \HT $\pm$ 0.03 & 32.00 & {\sc Hermes} \\
5778.6673 & -28.13 & \HT $\pm$ 0.03 & 31.10 & {\sc Hermes} \\
5835.4160 & -30.93 & \HT $\pm$ 0.03 & 32.00 & {\sc Hermes} \\
5867.3840 & -32.67 & \HT $\pm$ 0.03 & 26.20 & {\sc Hermes} \\
5870.3324 & -32.83 & \HT $\pm$ 0.03 & 27.10 & {\sc Hermes} \\
6011.7324 & -43.42 & \HT $\pm$ 0.04 & 29.30 & {\sc Hermes} \\
6101.5477 & -50.77 & \HT $\pm$ 0.03 & 37.00 & {\sc Hermes} \\
6101.5691 & -50.76 & \HT $\pm$ 0.03 & 38.10 & {\sc Hermes} \\
6119.4962 & -51.43 & \HT $\pm$ 0.04 & 36.80 & {\sc Hermes} \\
6119.5199 & -51.43 & \HT $\pm$ 0.04 & 37.50 & {\sc Hermes} \\
6138.4671 & -51.53 & \HT $\pm$ 0.03 & 25.70 & {\sc Hermes} \\
6138.4908 & -51.53 & \HT $\pm$ 0.04 & 24.90 & {\sc Hermes} \\
6176.4195 & -49.03 & \HT $\pm$ 0.03 & 33.90 & {\sc Hermes} \\
6182.4333 & -48.36 & \HT $\pm$ 0.03 & 35.70 & {\sc Hermes} \\
6183.4953 & -48.24 & \HT $\pm$ 0.03 & 38.70 & {\sc Hermes} \\
6183.5190 & -48.22 & \HT $\pm$ 0.04 & 37.50 & {\sc Hermes} \\
6196.5769 & -46.42 & \HT $\pm$ 0.03 & 26.80 & {\sc Hermes} \\
6215.3964 & -43.59 & \HT $\pm$ 0.04 & 38.30 & {\sc Hermes} \\
6479.6168 & -22.34 & \HT $\pm$ 0.03 & 20.10 & {\sc Hermes} \\
6484.4643 & -22.42 & \HT $\pm$ 0.04 & 20.80 & {\sc Hermes} \\
6488.5134 & -22.32 & \HT $\pm$ 0.04 & 20.30 & {\sc Hermes} \\
\hline
\end{tabular}
\end{table}

\begin{table}
\centering
\caption{Radial velocity measurements of the red-giant component of KIC\,9540226 derived with {\sc ccf} from {\sc Hermes} spectra.}
\label{tab:rv95}
\begin{tabular}[t]{crlcr}
\hline
\hline
KIC\,9540226 & & & &\\
\hline
HJD-2450000 & \multicolumn{2}{c}{Giant RV [\kms]} & S/N (\ion{Mg}{i} triplet) & Source \\
\hline
5765.4977 & -25.85 & \HT $\pm$ 0.03 & 29.60 & {\sc Hermes}\\
5783.5060 & -22.50 & \HT $\pm$ 0.03 & 32.80 & {\sc Hermes}\\
5872.3865 & -12.13 & \HT $\pm$ 0.04 & 27.10 & {\sc Hermes}\\
5884.3425 & -19.18 & \HT $\pm$ 0.04 & 21.80 & {\sc Hermes}\\
5884.3570 & -19.19 & \HT $\pm$ 0.05 & 19.90 & {\sc Hermes}\\
5889.3317 & -21.16 & \HT $\pm$ 0.05 & 16.70 & {\sc Hermes}\\
5889.3459 & -21.30 & \HT $\pm$ 0.04 & 10.90 & {\sc Hermes}\\
5889.3662 & -21.18 & \HT $\pm$ 0.04 & 18.40 & {\sc Hermes}\\
5990.7423 & -5.01 & \HT $\pm$ 0.04 & 22.90 & {\sc Hermes}\\
6106.4950 & -26.43 & \HT $\pm$ 0.03 & 25.40 & {\sc Hermes}\\
6106.5165 & -26.41 & \HT $\pm$ 0.04 & 26.90 & {\sc Hermes}\\
6126.6582 & -24.17 & \HT $\pm$ 0.04 & 24.40 & {\sc Hermes}\\
6126.6796 & -24.22 & \HT $\pm$ 0.03 & 24.10 & {\sc Hermes}\\
6132.5252 & -22.91 & \HT $\pm$ 0.04 & 27.60 & {\sc Hermes}\\
6132.5471 & -22.84 & \HT $\pm$ 0.04 & 25.60 & {\sc Hermes}\\
6136.6113 & -21.65 & \HT $\pm$ 0.03 & 32.20 & {\sc Hermes}\\
6136.6327 & -21.61 & \HT $\pm$ 0.03 & 31.70 & {\sc Hermes}\\
6139.4500 & -21.00 & \HT $\pm$ 0.04 & 31.50 & {\sc Hermes}\\
6148.5050 & -17.38 & \HT $\pm$ 0.04 & 32.10 & {\sc Hermes}\\
6148.5264 & -17.36 & \HT $\pm$ 0.04 & 32.10 & {\sc Hermes}\\
6158.4969 & -11.59 & \HT $\pm$ 0.04 & 34.20 & {\sc Hermes}\\
6158.5183 & -11.58 & \HT $\pm$ 0.04 & 35.00 & {\sc Hermes}\\
6176.4456 & 6.03 & \HT $\pm$ 0.04 & 30.10 & {\sc Hermes}\\
6182.4104 & 13.44 & \HT $\pm$ 0.04 & 33.40 & {\sc Hermes}\\
6184.5669 & 15.87 & \HT $\pm$ 0.03 & 31.40 & {\sc Hermes}\\
6184.5895 & 15.92 & \HT $\pm$ 0.03 & 29.50 & {\sc Hermes}\\
6195.5294 & 18.81 & \HT $\pm$ 0.04 & 30.60 & {\sc Hermes}\\
6506.4110 & -13.52 & \HT $\pm$ 0.03 & 21.90 & {\sc Hermes}\\
6506.4324 & -13.55 & \HT $\pm$ 0.04 & 20.70 & {\sc Hermes}\\
6510.5030 & -10.81 & \HT $\pm$ 0.03 & 16.50 & {\sc Hermes}\\
6518.3925 & -4.06 & \HT $\pm$ 0.04 & 26.10 & {\sc Hermes}\\
6519.4668 & -2.94 & \HT $\pm$ 0.04 & 26.70 & {\sc Hermes}\\
\hline
\end{tabular}
\end{table}

\section{Frequencies}

\subsection{Peakbagging results}

\begin{figure*}
	\includegraphics[width=2\columnwidth]{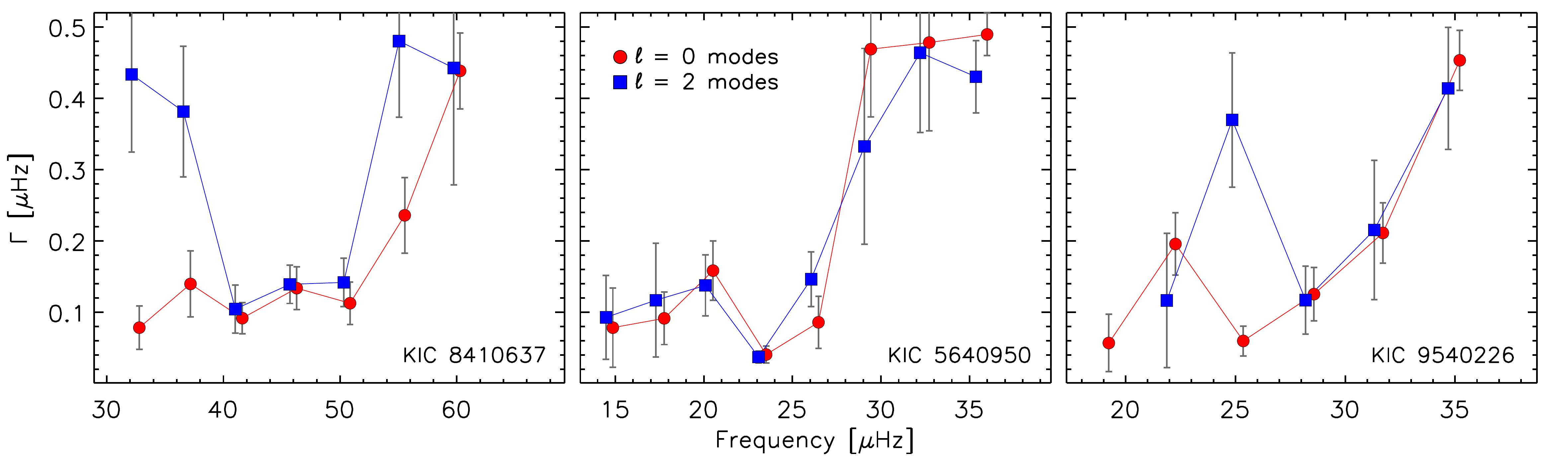}
    \caption{Mode linewidths of significant frequencies of KIC\,8410637 (left), KIC\,5640750 (middle) and KIC\,9540226 (right). Red circles and blue squares represent $\ell=0$ and $\ell=2$ modes, respectively.}
    \label{fig:fig12}
\end{figure*}

Based on the methods described in Section~\ref{sec:pb}, we extracted all the significant frequencies from the oscillation spectra of the three stars. The complete list of central mode frequencies, amplitudes and linewidths are reported in Tables~\ref{tab:pb84}, \ref{tab:pb56} and \ref{tab:pb95}. In addition, we show the linewidths of all significant $\ell=0$ and $\ell=2$ modes in Figure~\ref{fig:fig12}. It is clear that the extracted non-radial modes incorporate contributions from unresolved mixed and rotationally-split modes. For KIC\,8410637, KIC\,5640750 and KIC\,9540226 we derived linewidths of the same order of magnitude as reported by \citet{2015cor}. They performed an extensive peakbagging analysis of 19 red-giant stars and they pointed out a linewidth depression close to the frequency of maximum oscillation power which is also visible in Figure~\ref{fig:fig12}.

\begin{table}
\centering
\caption{Median values and 68~per cent confidence interval for central mode frequencies, mode amplitudes and mode heights for KIC\,8410637.}
\label{tab:pb84}
\begin{tabular}{ccrrr}
\hline\hline
KIC\,8410637 & & & & \\
\hline
$n$ & $\ell$ & $\nu$ [$\mu$Hz] & A [ppm] & $\Gamma$ [$\mu$Hz]\\
\hline
5 &2&32.13$\pm$0.05&47.6$\pm$6.3&0.43$\pm$0.11\\
6 &0&32.79$\pm$0.02&40.6$\pm$3.6&0.08$\pm$0.03\\
6 &1&35.16$\pm$0.04&52.1$\pm$4.9&0.45$\pm$0.11\\
6 &2&36.56$\pm$0.03&66.7$\pm$7.1&0.38$\pm$0.09\\
7 &0&37.18$\pm$0.02&46.1$\pm$5.4&0.14$\pm$0.05\\
6 &3&37.70$\pm$0.09&12.3$\pm$6.4&0.33$\pm$0.13\\
7 &1&39.44$\pm$0.02&87.4$\pm$6.8&0.23$\pm$0.04\\
7 &2&41.02$\pm$0.01&62.5$\pm$6.2&0.10$\pm$0.03\\
8 &0&41.62$\pm$0.01&88.0$\pm$5.6&0.09$\pm$0.02\\
8 &1&44.06$\pm$0.02&90.4$\pm$6.3&0.26$\pm$0.04\\
8 &2&45.69$\pm$0.01&87.4$\pm$5.0&0.14$\pm$0.03\\
9 &0&46.28$\pm$0.02&94.5$\pm$6.3&0.13$\pm$0.03\\
8 &3&47.19$\pm$0.01&29.6$\pm$4.2&0.07$\pm$0.03\\
9 &1&48.70$\pm$0.02&93.3$\pm$5.7&0.21$\pm$0.03\\
9 &2&50.31$\pm$0.02&96.4$\pm$7.1&0.14$\pm$0.03\\
10 &0&50.85$\pm$0.01&87.3$\pm$5.0&0.11$\pm$0.03\\
9 &3&51.71$\pm$0.06&37.2$\pm$5.1&0.48$\pm$0.13\\
10 &1&53.32$\pm$0.02&81.4$\pm$4.0&0.30$\pm$0.04\\
10 &2&55.06$\pm$0.04&75.1$\pm$8.6&0.48$\pm$0.11\\
11 &0&55.54$\pm$0.03&61.4$\pm$7.7&0.24$\pm$0.05\\
11 &1&58.04$\pm$0.04&64.5$\pm$3.7&0.46$\pm$0.04\\
11 &2&59.74$\pm$0.05&47.6$\pm$8.4&0.44$\pm$0.16\\
12 &0&60.28$\pm$0.05&62.0$\pm$7.0&0.44$\pm$0.05\\
12 &1&62.95$\pm$0.02&82.7$\pm$4.6&0.50$\pm$0.01\\
\hline
\end{tabular}
\end{table}

\begin{table}
\centering
\caption{Same as Table \ref{tab:pb84} for KIC\,5640750.}
\label{tab:pb56}
\begin{tabular}{cccrc}
\hline\hline
KIC\,5640750 & & & & \\
\hline
$n$ & $\ell$ & $\nu$ [$\mu$Hz] & A [ppm] & $\Gamma$ [$\mu$Hz]\\
\hline
3 &2&14.47$\pm$0.06&18.9$\pm$17.0&0.09$\pm$0.05\\
4 &0&14.86$\pm$0.06&48.1$\pm$19.1&0.08$\pm$0.05\\
4 &1&16.46$\pm$0.03&86.2$\pm$11.9&0.10$\pm$0.06\\
4 &2&17.29$\pm$0.03&53.5$\pm$11.8&0.12$\pm$0.08\\
5 &0&17.76$\pm$0.02&75.1$\pm$12.7&0.09$\pm$0.04\\
5 &1&19.27$\pm$0.01&142.2$\pm$14.9&0.06$\pm$0.03\\
5 &2&20.09$\pm$0.01&107.6$\pm$11.1&0.14$\pm$0.04\\
6 &0&20.52$\pm$0.02&99.8$\pm$08.4&0.16$\pm$0.04\\
5 &3&20.94$\pm$0.07&51.8$\pm$14.6&0.06$\pm$0.02\\
6 &1&22.12$\pm$0.01&140.6$\pm$12.0&0.12$\pm$0.03\\
6 &2&23.09$\pm$0.01&141.1$\pm$15.7&0.04$\pm$0.01\\
7 &0&23.52$\pm$0.01&187.4$\pm$16.1&0.04$\pm$0.01\\
6 &3&24.02$\pm$0.02&57.4$\pm$10.7&0.13$\pm$0.06\\
7 &1&25.09$\pm$0.01&182.4$\pm$11.7&0.15$\pm$0.02\\
7 &2&26.06$\pm$0.02&178.5$\pm$12.1&0.15$\pm$0.04\\
8 &0&26.47$\pm$0.02&128.8$\pm$13.2&0.09$\pm$0.04\\
8 &1&28.02$\pm$0.02&147.2$\pm$10.3&0.19$\pm$0.03\\
8 &2&29.07$\pm$0.03&127.2$\pm$24.2&0.33$\pm$0.14\\
9 &0&29.43$\pm$0.08&99.9$\pm$23.3&0.47$\pm$0.09\\
9 &1&31.08$\pm$0.04&93.1$\pm$07.1&0.38$\pm$0.09\\
9 &2&32.21$\pm$0.05&70.0$\pm$13.4&0.46$\pm$0.11\\
10 &0&32.72$\pm$0.06&66.9$\pm$12.8&0.48$\pm$0.12\\
10 &1&34.23$\pm$0.06&58.4$\pm$06.7&0.42$\pm$0.07\\
10 &2&35.36$\pm$0.08&60.3$\pm$06.2&0.43$\pm$0.05\\
11 &0&35.99$\pm$0.05&64.4$\pm$11.5&0.49$\pm$0.03\\ \hline
\end{tabular}
\end{table}

\begin{table}
\centering
\caption{Same as Table \ref{tab:pb84} for KIC\,9540226.}
\label{tab:pb95}
\begin{tabular}{ccrrr}
\hline\hline
KIC\,9540226 & & & & \\
\hline
$n$ & $\ell$ & $\nu$ [$\mu$Hz] & A [ppm] & $\Gamma$ [$\mu$Hz]\\
\hline
5 &0&19.24$\pm$0.02&66.1$\pm$10.2&0.06$\pm$0.03\\
5 &1&20.90$\pm$0.04&140.8$\pm$09.0&0.47$\pm$0.09\\
5 &2&21.88$\pm$0.04&38.3$\pm$10.1&0.12$\pm$0.09\\
6 &0&22.27$\pm$0.02&97.3$\pm$09.5&0.20$\pm$0.04\\
6 &1&23.84$\pm$0.02&130.6$\pm$09.1&0.23$\pm$0.06\\
6 &2&24.83$\pm$0.03&129.7$\pm$08.7&0.37$\pm$0.09\\
7 &0&25.36$\pm$0.01&99.8$\pm$11.8&0.06$\pm$0.02\\
7 &1&27.09$\pm$0.01&172.0$\pm$12.2&0.11$\pm$0.03\\
7 &2&28.19$\pm$0.02&135.6$\pm$12.6&0.12$\pm$0.05\\
8 &0&28.58$\pm$0.01&144.1$\pm$13.0&0.13$\pm$0.04\\
8 &1&30.24$\pm$0.01&154.8$\pm$10.9&0.17$\pm$0.03\\
8 &2&31.32$\pm$0.04&97.2$\pm$14.8&0.22$\pm$0.10\\
9 &0&31.71$\pm$0.02&127.2$\pm$13.2&0.21$\pm$0.04\\
9 &1&33.46$\pm$0.03&111.7$\pm$05.9&0.48$\pm$0.06\\
9 &2&34.69$\pm$0.05&53.4$\pm$09.7&0.41$\pm$0.08\\
10 &0&35.21$\pm$0.04&83.5$\pm$10.3&0.45$\pm$0.04\\
\hline
\end{tabular}
\end{table}

\begin{figure*}
	\includegraphics[height=2\columnwidth,angle=90]{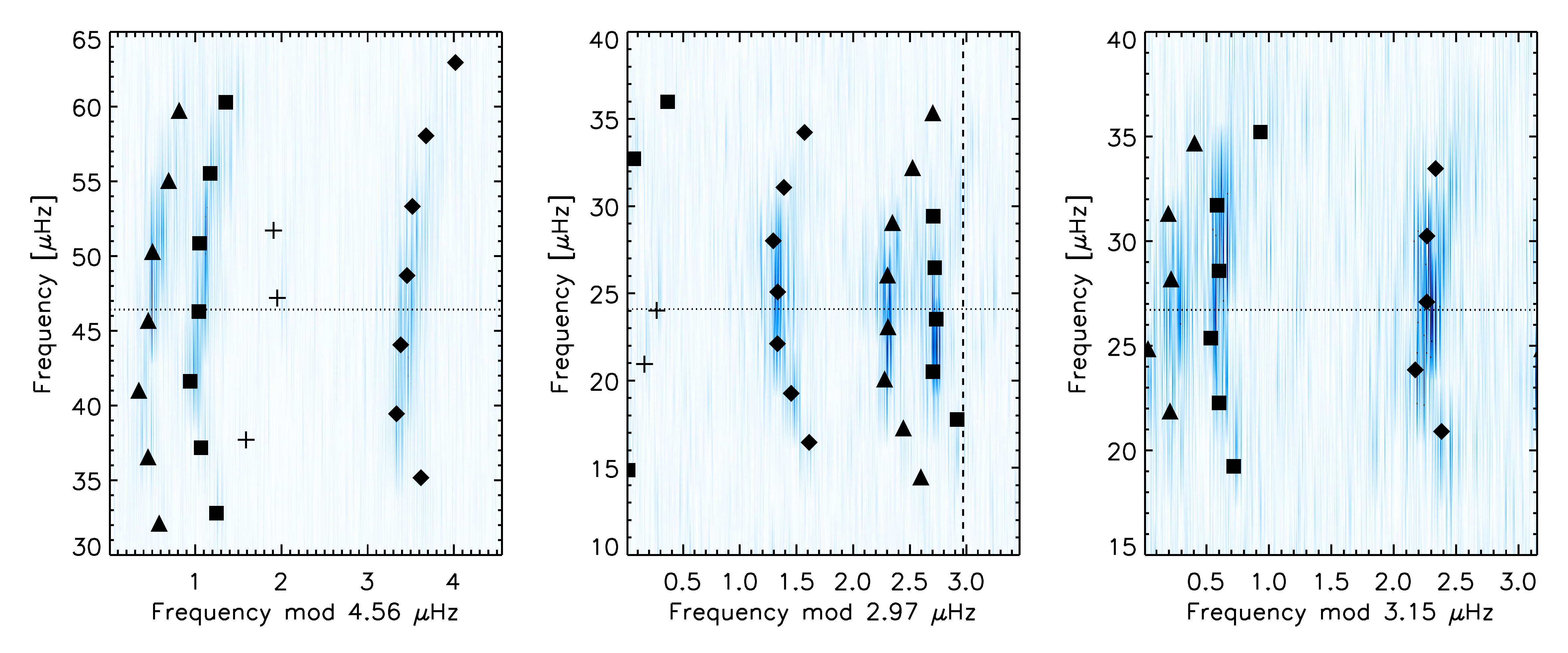}
    \caption{\'{E}chelle diagrams for KIC\,8410637 (left), KIC\,5640750 (middle) and KIC\,9540226 (right). Darker blue tones correspond to higher power. Extracted mode frequencies are indicated by symbols (square: $\ell=0$, triangle: $\ell=2$, diamond: $\ell=1$, plus: $\ell=3$) and the horizontal dotted lines show measured $\nu_{\rm max}$ values.}
    \label{fig:fig13}
\end{figure*}

\subsection{\'{E}chelle diagrams}
\label{sec:ech}

\'{E}chelle diagrams \citep{1983gre} present an alternative technique to identify modes of different spherical degree $\ell$. To construct such a diagram the Fourier power density spectrum is divided into segments of equal lengths ($\Delta\nu$). Then these segments are stacked on top of each other. The modes of the same spherical degree line up as near vertical ridges in the diagram. Figure~\ref{fig:fig13} shows the \'{e}chelle diagrams for KIC\,8410637, KIC\,5640750 and KIC\,9540226. Three ridges are clearly visible, which correspond to spherical degrees $\ell=0,1,2$. The strongest mode frequencies that were extracted through the peakbagging analysis are marked with symbols.

\subsection{Differences between asteroseismic methods}
\label{subsec:diff}

\begin{figure*}
	\includegraphics[width=2\columnwidth]{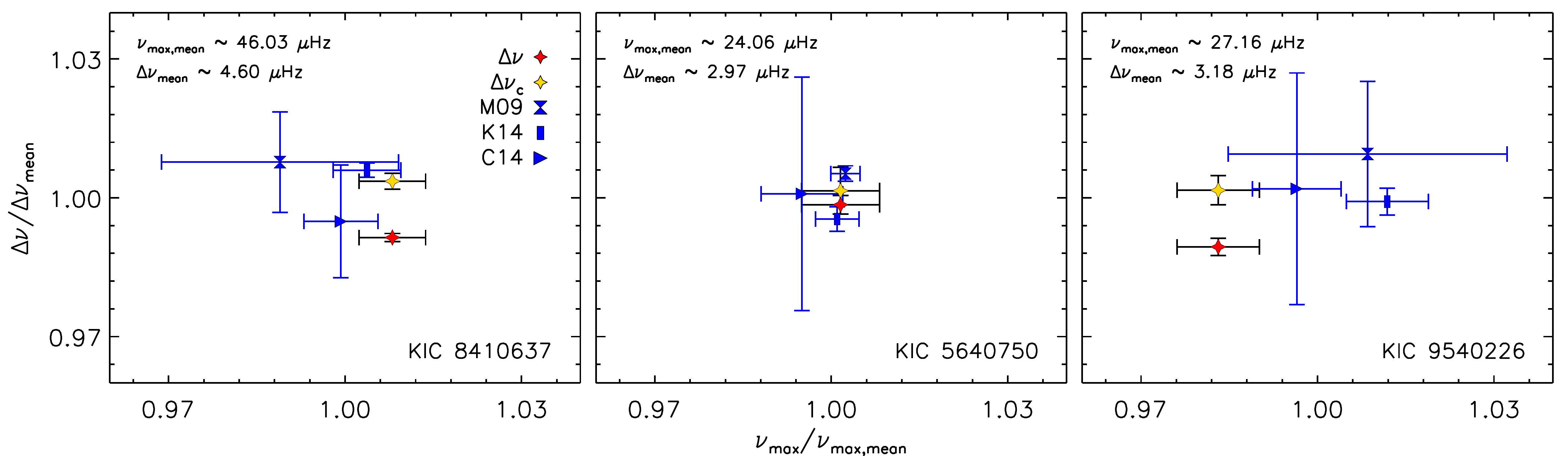}
    \caption{Estimates of the global oscillation parameters from different methods for KIC\,8410637 (left), KIC\,5640750 (middle) and KIC\,9540226 (right). The red and gold star symbols represent the global ($\Delta\nu$) and local ($\Delta\nu_{\rm c}$) mean large frequency separations that were used to determine the stellar parameters and evolutionary stages of the red giants. The KIC number is indicated in each panel.}
    \label{fig:fig14}
\end{figure*}

The global oscillation parameters and frequencies of oscillation modes were derived independently from several co-authors using different approaches. We refer the interested reader to \citet{2009mos} (EACF, hereafter M09), \citet{2014kal} (hereafter K14) and \citet{2014cor,2015cor} (\textsc{D\large{iamonds}}, hereafter C14), for detailed descriptions of these methods. Figure~\ref{fig:fig14} shows that the seismic parameter estimates, $\nu_{\rm max}$ and $\Delta\nu$, are of the same order of magnitude and that the use of different procedures does not substantially affect the results.  
Based on our analysis (Section~\ref{sec:sig_from_f}) we determined a local ($\Delta\nu_{\rm c}$) and global ($\Delta\nu$) mean large frequency separation, which are shown in the same Figure.

\subsection{Second frequency differences and acoustic glitches}
\label{sec:d2nu}

\begin{figure}
	\includegraphics[width=\columnwidth]{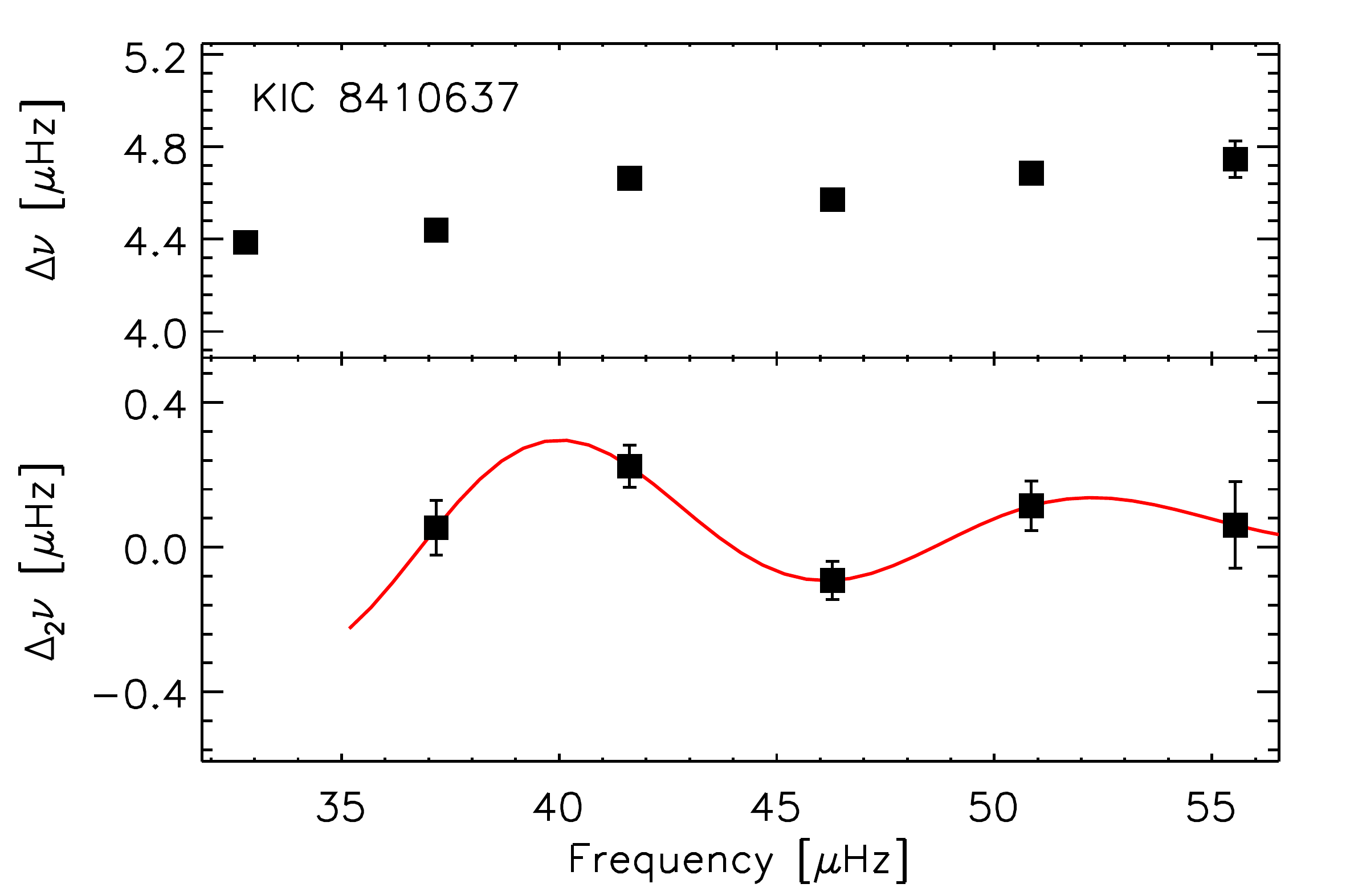}
	\includegraphics[width=\columnwidth]{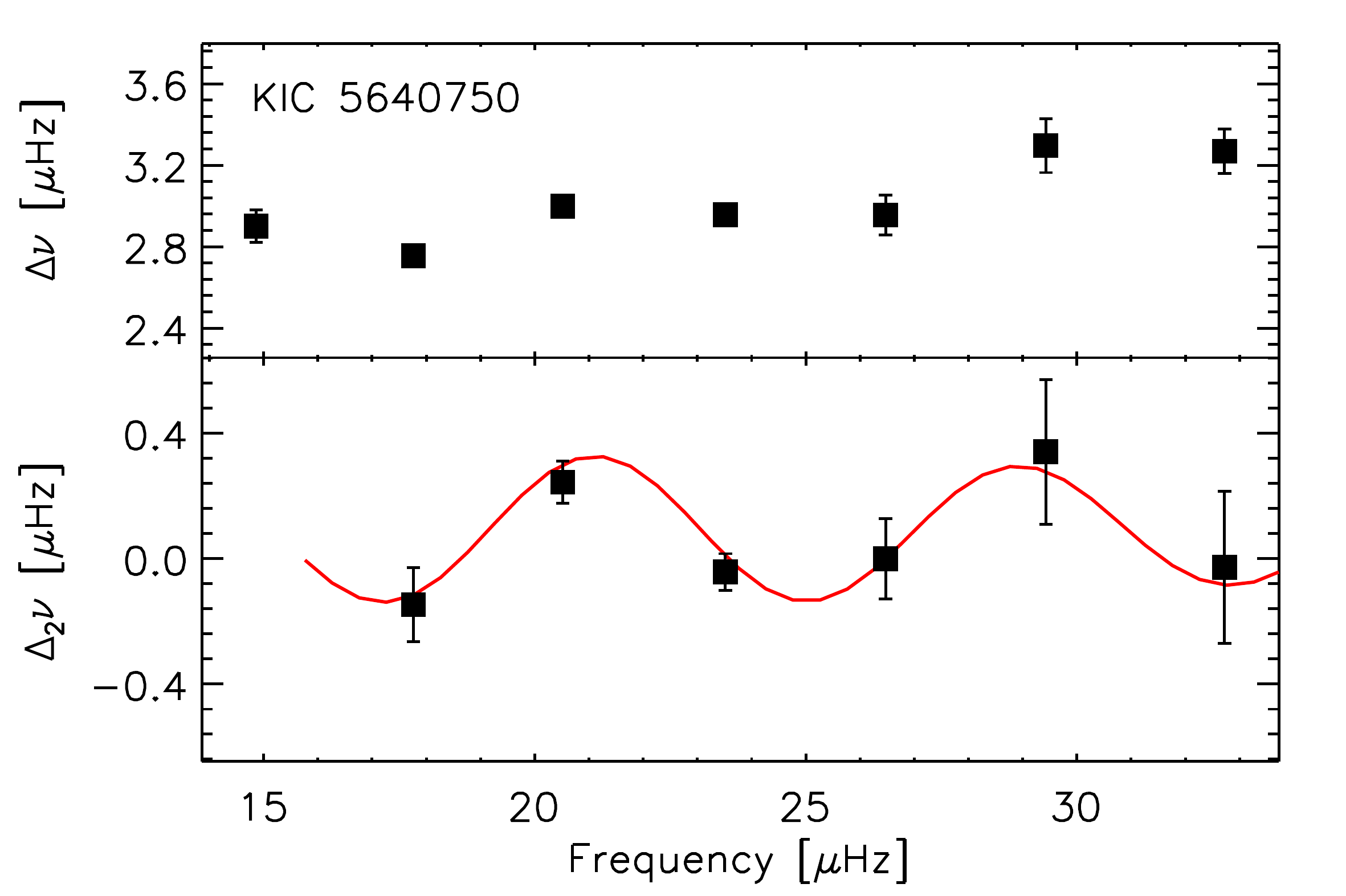}
	\includegraphics[width=\columnwidth]{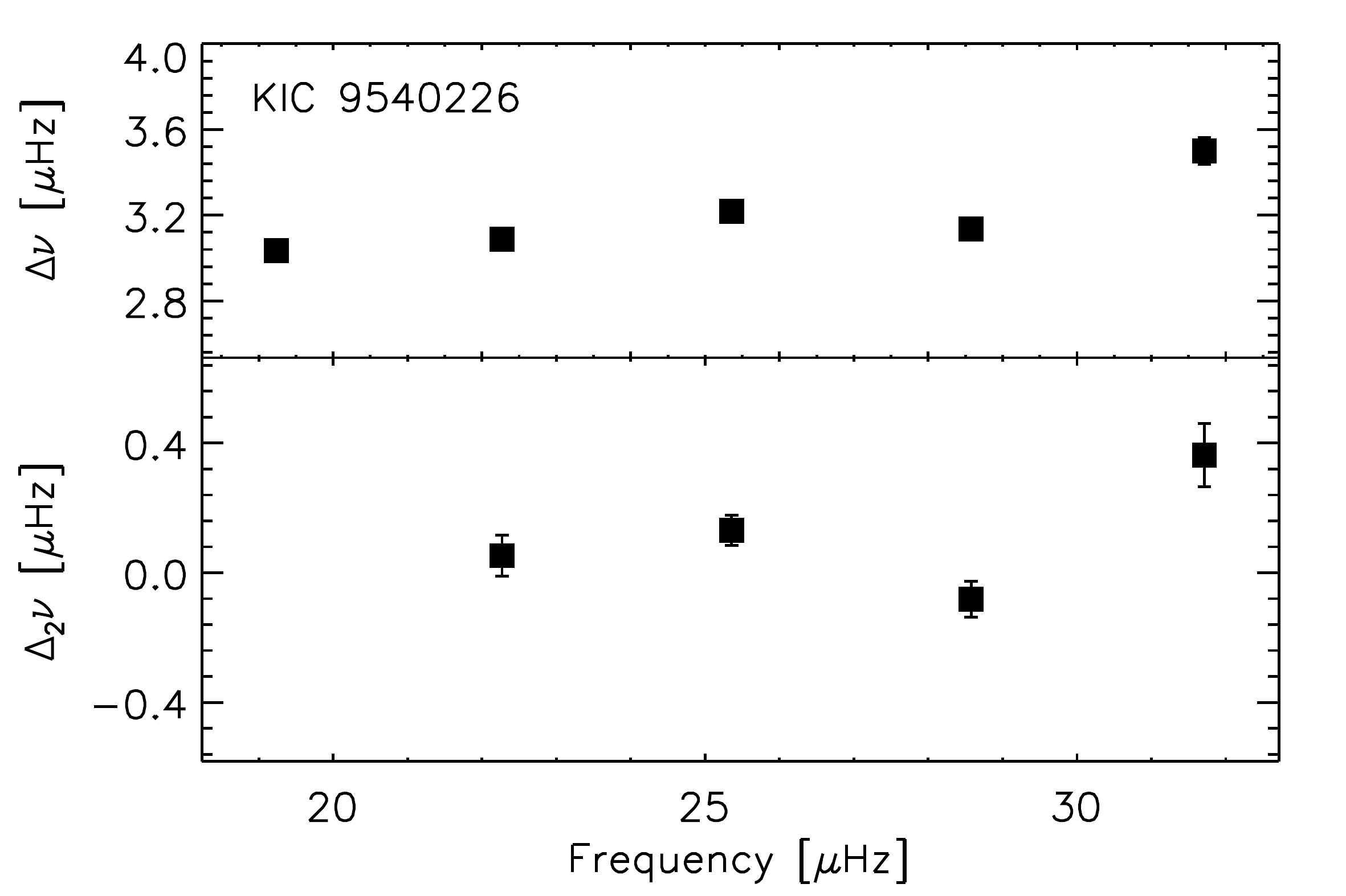}
    \caption{First (top panels, Eq.~\ref{eq:1st_f_diff}) and second (bottom panels, Eq. \ref{eq:2nd_f_diff}) frequency differences for KIC\,8410637, KIC\,5640750 and KIC\,9540226. The red solid lines indicate the fits of the acoustic glitches (Eq.~\ref{eq:2nd_diff} and Table~\ref{tab:acoustic_r}) to the second differences. Uncertainties are shown on all points. In some cases the error bars are smaller than the symbol size.}
    \label{fig:fig15}
\end{figure}

In addition to the first frequency difference i.e. the large frequency separation $\Delta\nu_{\rm n,\ell}$ which is given by:
\begin{equation}
    \Delta\nu_{\rm n,\ell}\equiv\nu_{\rm n+1,\ell}-\nu_{\rm n,\ell},
	\label{eq:1st_f_diff}
\end{equation}
we also computed the second frequency difference $\Delta_{\rm 2}\nu_{\rm n,\ell}$ \citep{1990gou}:
\begin{equation}
\Delta_{\rm 2}\nu_{\rm n,\ell}\equiv\Delta\nu_{\rm n,\ell}-\Delta\nu_{\rm n-1,\ell}=\nu_{\rm n-1,\ell}-2\nu_{\rm n,\ell}+\nu_{\rm n+1,\ell}.
\label{eq:2nd_f_diff}
\end{equation}
In both Equations (\ref{eq:1st_f_diff}--\ref{eq:2nd_f_diff}) $\nu_{\rm n,\ell}$ represents the frequency of a mode with given radial order $n$ and spherical degree $\ell$. The first and second frequency differences of the radial ($\ell=0$) modes for KIC\,8410637, KIC\,5640750 and KIC\,9540226 are shown in Figure~\ref{fig:fig15}.
The uncertainties in $\Delta\nu_{\rm n,\ell}$ and $\Delta_2\nu_{\rm n,\ell}$ were derived from a standard error propagation of the uncertainties of the individual mode frequencies. The oscillatory signal which is visible in the second frequency difference is caused by a so-called acoustic glitch. This observable feature appears due to changes in the stellar interior that occur at scales shorter than the local wavelengths of the oscillations. In red giants, such a change is caused by the helium second ionization zone. Recently, acoustic glitch signatures could be examined in modelled and observed oscillation frequencies of stars exhibiting solar-like oscillations \citep[e.g.][]{2010mig,2014maz,2014bro,2015acor,2015vra,2016per}.

In the current analysis we show that it is possible to perform a basic study of the acoustic glitches of two of the red giants (KIC\,8410637 and KIC\,5640750) for which we have 4 years of nearly uninterrupted {\it Kepler} data without any larger gaps available. Since this periodic signature caused by the helium second ionization zone is observable in pure p modes \citep[e.g.][]{1988vor,1990gou}, we can measure the acoustic depth and width of the helium second ionization zone \citep{2013chek} through the analysis of the second frequency differences of $\ell=0$ modes. The modulation due to the glitch can be described by the following model \citep{2007hou}:
\begin{equation}
\Delta_{\rm 2}\omega_{\rm n,\ell}=A_{\rm osc}\omega_{\rm n,\ell}\exp{(-2b_{\rm osc}^2\omega^2_{\rm n,\ell})}\cos{[2(\tau_{\rm HeII}\omega_{\rm n,\ell}+\phi)]}+c,
\label{eq:2nd_diff}
\end{equation}
where $\omega_{\rm n,\ell}$ and $\Delta_2\omega_{\rm n,\ell}$ define the angular versions of $\nu_{\rm n,\ell}$ and $\Delta_2\nu_{\rm n,\ell}$:
\begin{equation}
\omega_{\rm n,\ell}\equiv2\pi\nu_{\rm n,\ell},\\
\Delta_{\rm 2}\omega_{\rm n,\ell}\equiv2\pi\Delta_2\nu_{\rm n,\ell}.
\label{eq:angular_ver}
\end{equation}
This model is composed of a dimensionless amplitude $A_{\rm osc}$, an acoustic depth $\tau_{\rm HeII}$ and characteristic width $b_{\rm osc}$ of the second ionization zone, a constant phase shift $\phi$ and an offset $c$.
Based on the Bayesian MCMC method with Metropolis-Hastings sampling which is described in more detail in Section~\ref{sec:bg}, we estimated the model parameters in Equation~\ref{eq:2nd_diff} and their 68 per cent credible intervals. In this case we assumed that the parameters follow a Gaussian distribution and therefore used a normal likelihood function. We did not account for correlations between the individual second frequency differences. For the free parameters, we chose uniform priors and a strict constraint on the period $\tau_{\rm HeII}$ with the convention that at least two measurements of the second frequency differences should cover one period of the acoustic glitch.
The bottom panels in Figure~\ref{fig:fig15} show a fit of the acoustic glitch model to the second frequency differences of KIC\,8410637 and KIC\,5640750 and are based on the  best-fit parameters presented in Table~\ref{tab:acoustic_r}. Due to a low number of available second frequency differences (less than five) we omitted KIC\,9540226 (Fig.~\ref{fig:fig15}) from this part of the analysis.

\begin{table*}
	\centering
	\caption{Median values and corresponding 68~per cent credible interval as derived for the fit parameters of the glitch model (Eq.~\ref{eq:2nd_diff}) using the Metropolis-Hastings MCMC algorithm.}
	\label{tab:acoustic_r}
	\begin{tabular}{lcllcc}
		\hline\hline
		KIC & $A_{\rm osc}$ & $\tau_{\rm HeII}$ [s] & $b_{\rm osc}$ [s] & $\phi$ [rad] & $c$ [$\mu$Hz] \\
		\hline
		8410637 & $0.041\pm 0.003$ & $40437\pm 16$ & $3875\pm 23$ & $2.33\pm 0.01$ & $0.06 \pm 0.01$ \\
		5640750 & $0.020\pm 0.007$ & $62901\pm 4281$ & $4118\pm 656$ & $1.08\pm0.67$ & $0.09\pm 0.04$ \\
		\hline
	\end{tabular}
\end{table*}


\bsp	
\label{lastpage}
\end{document}